\pdfoutput=1

\documentclass[leqno]{article}

\usepackage{aoa}
\usepackage{eucal}
\usepackage{amsmath}
\usepackage{amssymb}
\usepackage{amsthm}
\usepackage{tipa}
\usepackage{tabularx}
\usepackage{graphicx}
\usepackage{hyperref}
\usepackage{subfigure}
\usepackage{mathtools}
\usepackage[labelfont={bf,small},font=small,labelsep=period]{caption}
\usepackage{array}
\usepackage{pbox}
\usepackage{booktabs}
\usepackage[nottoc]{tocbibind}
\usepackage{appendix}
\newcolumntype{M}[1]{>{\centering\arraybackslash}m{#1}}
\newcolumntype{N}{@{}m{0pt}@{}}
\usepackage{pdfpages}
\usepackage[margin=1.25in]{geometry}

\renewcommand\P{\mathbb{P}}

\newcommand\E{\mathbb{E}}
\newcommand\G{\mathcal{G}}
\let\Sec\S
\renewcommand\S{\mathcal{S}}
\newcommand\T{\mathcal{T}}
\newcommand\smexp[1]{\mathrm{e}^{#1}}
\newcommand\lgexp[1]{\exp{\left({#1}\right)}}

\newcommand\F{\mathcal{F}}

\newcommand\n[1]{\left\vert#1\right\vert}
\newcommand\eps{\varepsilon}

\newcommand\sq{\text{\fontsize{4}{1}$\square$}}
\newcommand\bsq{\text{\fontsize{4}{1}$\blacksquare$}}
\newcommand{\comment}[1]{}
\newcommand\A{\mathcal{A}}
\newcommand\Av{\A^{\bullet}}
\newcommand\Avn{\A_n^{\bullet}}

\newcommand\Au{\A^{\bullet-\bullet}}
\newcommand\Aun{\A_n^{\bullet-\bullet}}

\newcommand\Ad{\A^{\bullet\ra\bullet}}
\newcommand\Adn{\A_n^{\bullet\ra\bullet}}
\DeclareRobustCommand\cyc{\tiny\rotatebox[origin=c]{270}{$\circlearrowright$}}
\newcommand\cycs{\raisebox{-1.5pt}{\includegraphics[width=2mm]{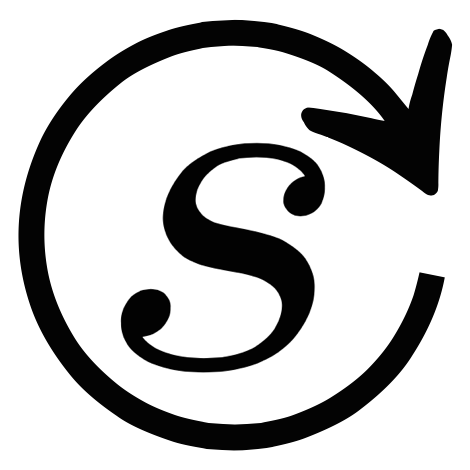}}}
\newcommand\cp[1]{#1^{\cyc}}
\newcommand\scp[1]{#1^{\cycs}}
\newcommand\Acp{\cp{\A}}
\newcommand\Bcp{\cp{\B}}
\newcommand\Bscp{\scp{\B}}
\newcommand\Ccp{\C^{\,\cyc}}
\newcommand\Acpn{\cp{\A_n}}
\newcommand\Ascp{\scp{\A}}
\newcommand\Ascpv{\scp{\A_v}}
\newcommand\Ascpe{\scp{\A_e}}
\newcommand\Setcp{\Set^{\,\cyc}}
\newcommand\Setscp{\scp{\Set}}
\renewcommand\DH{\mathcal{DH}}
\newcommand\TLP{3\mathcal{LP}}
\newcommand\DHcp{\DH^{\,\cyc}}
\newcommand\TLPcp{\TLP^{\,\cyc}}
\newcommand\B{\mathcal{B}}
\newcommand\C{\mathcal{C}}
\newcommand\Z{\mathcal{Z}}
\newcommand\Zcp{\Z^{\bullet}}
\newcommand\Zat{\textsf{Z}}
\newcommand\Zcpat{\Zat^{\bullet}}
\newcommand\SX{\mathcal{S_X}}
\newcommand\SC{\mathcal{S_C}}
\newcommand\K{\mathcal{K}}
\newcommand\Tcp{\cp{\T}}
\newcommand\Tscp{\scp{\T}}
\newcommand\Tscpv{\scp{\T_v}}
\newcommand\Tscpe{\scp{\T_e}}
\newcommand\Tv{\T^{\bullet}}
\newcommand\Tvn{\T_n^{\bullet}}
\newcommand\Td{\T^{\bullet\ra\bullet}}
\newcommand\Tu{\T^{\bullet-\bullet}}
\newcommand\Scp{\cp{\S}}
\newcommand\Sym{\mathrm{Sym}}
\newcommand\Rcsym{\mathrm{RSym}}
\newcommand\Fix{\mathrm{Fix}}
\newcommand\sub{\circledcirc}
\newcommand{\TextUnderscore}{\underline{\hspace{2mm}}}
\newcommand\rhosub[1]{\rho_{\scalebox{0.5}{$#1$}}}
\newcommand\rhoA{\rhosub{\A}}

\newtheorem*{thm*}{Theorem}
\newtheorem{thm}{Theorem}[section]

\newtheorem{lem}[thm]{Lemma}
\newtheorem{cor}[thm]{Corollary}
\theoremstyle{definition}

\newtheorem{definition}[thm]{Definition}

\theoremstyle{remark}
\newtheorem{remark}[thm]{Remark}

\renewcommand\l{\ell}

\newcommand\vb{\,\vert\,}

\newcommand\ra{\rightarrow}

\newcommand\etal{\emph{et~al.~}}
      
\everymath{\displaystyle}

\begin{document}

\author{Alexander Iriza}
\title{Enumeration and random generation of unlabeled classes of graphs:\\ A practical study of cycle pointing and the dissymmetry theorem}
\date{September 2015}
\includepdf{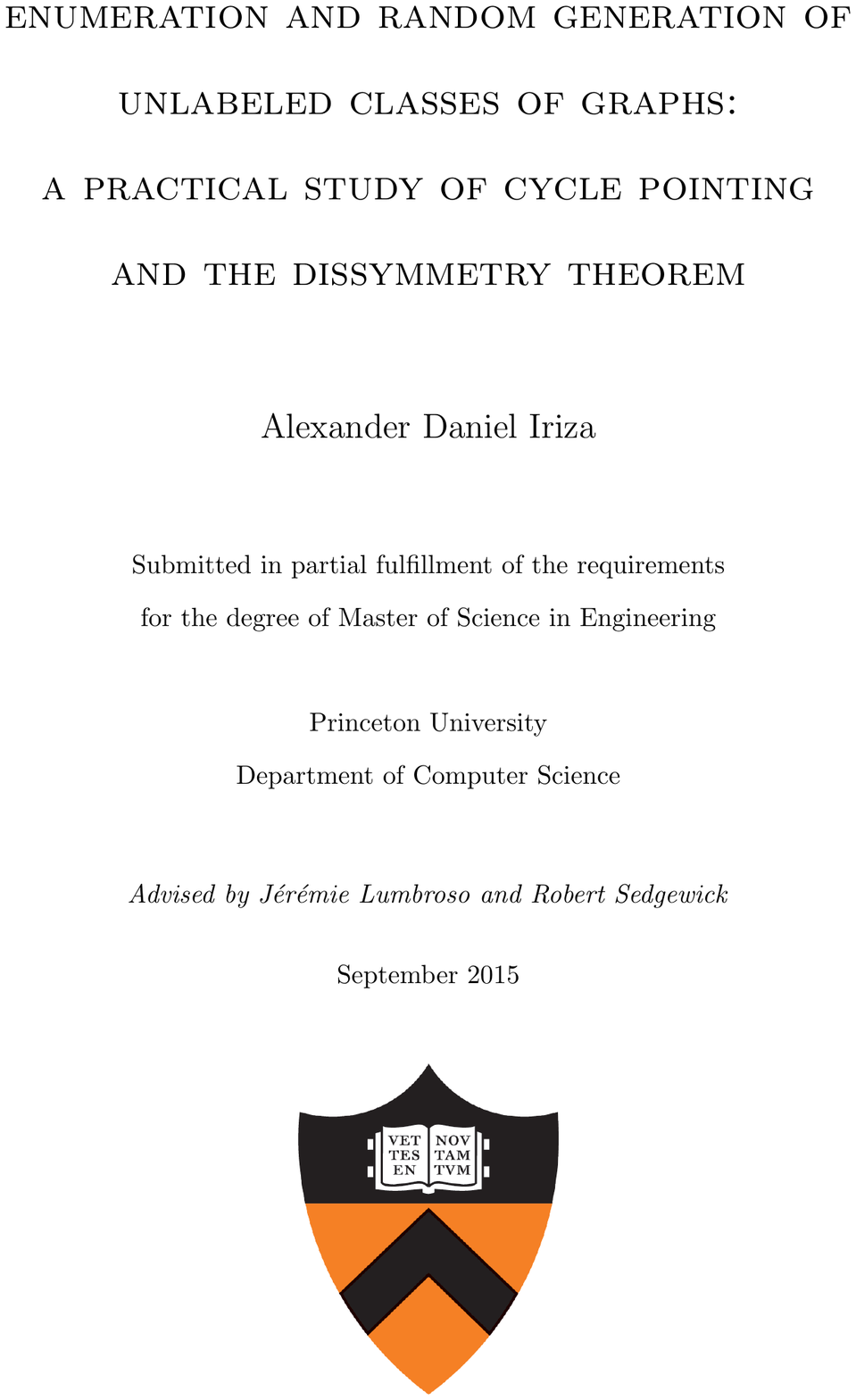}
\tableofcontents
\newpage

\begin{abstract}
Our work studies the enumeration and random generation of unlabeled combinatorial classes of unrooted graphs. While the technique of vertex pointing provides a straightforward procedure for analyzing a labeled class of unrooted graphs by first studying its rooted counterpart, the existence of nontrivial symmetries in the unlabeled case causes this technique to break down. Instead, techniques such as the dissymmetry theorem (of Otter~\cite{otter}) and cycle pointing (of Bodirsky~\etal\cite{cpshort, cplong}) have emerged in the unlabeled case, with the former providing an enumeration of the class and the latter providing both an enumeration and an unbiased sampler. In this work, we extend the power of the dissymmetry theorem by showing that it in fact provides a Boltzmann sampler for the class in question. We then present an exposition of the cycle pointing technique, with a focus on the enumeration and random generation of the underlying unpointed class. Finally, we apply cycle pointing to enumerate and implement samplers for the classes of distance-hereditary graphs and three-leaf power graphs.
\end{abstract}

\section{Introduction}

The study of families of graphs -- enumerating them, randomly generating them, and analyzing their parameters and asymptotics -- has been one of the many great success stories of analytic combinatorics (see, for example, the discussion by Flajolet and Sedgewick~\cite[Sec.~II.5.]{purplebook}). These graphs may come in many varieties -- labeled or unlabeled, rooted or unrooted, plane or non-plane, and with or without cycles -- and the techniques for analyzing these variations are just as numerous.

The typical first step in the analysis is to build a combinatorial specification for the class in question. In the case of graphs this specification is usually recursive, expressing a graph as a combination of smaller elements of the same class or related classes. If the class happens to be rooted, then the root of a graph, whether a vertex, edge, or other feature, provides a distinguished location at which it can be decomposed and by which a specification of the class can be written. However if the graphs in the class are unrooted, no such distinguished location exists.

To address this difficulty, techniques have arisen in both the labeled and unlabeled cases which analyze an unrooted class by relating it to different rooted (also sometimes called \emph{marked} or \emph{pointed}) versions of itself. If the class is labeled, meaning that the vertices of each graph of size $n$ are given distinct labels from $1$ to $n$, then there are $n$ distinct vertex-rooted graphs for each unrooted graph of size $n$, and this fact allows for an analysis of the vertex-rooted class to be easily translated into an analysis of the unrooted class~\cite{purplebook}, and even a random sampler~\cite[\Sec 2.2.1]{DaPaRoSo12}. However if the class is unlabeled, meaning that the vertices of a graph are distinguished not by distinct labels but rather only by their adjacencies to other vertices in the graph, then two different graphs of the same size may have two different sets of non-trivial symmetries (reflections, rotations, etc.) and therefore two different numbers of rooted graphs corresponding to them.

In order then to analyze an unlabeled, unrooted class of graphs, a common strategy is to study not only the corresponding vertex-rooted class but also to study corresponding classes that are rooted at features other than vertices. One example of such a strategy is the dissymmetry theorem, introduced by Otter in 1948~\cite{otter} and popularized by Bergeron~\etal in 1997~\cite{bergeron}, which allows one to compute the enumeration of various unlabeled, unrooted classes of \emph{trees} by enumerating the corresponding classes that are rooted at a vertex, an edge, and a directed edge. A second example is the technique of cycle pointing, introduced by Bodirsky~\etal\cite{cpshort, cplong}, in which a graph is rooted at a cycle on some of its vertices satisfying a certain property, and this property is chosen so that there are $n$ distinct cycle-pointed graphs corresponding to each unrooted graph of size $n$. Cycle pointing allows one both to compute the enumerations of unlabeled, unrooted classes of graphs, and to construct unbiased random generators for these classes.

This work has three main contributions. In Section~\ref{dissymmetry}, we introduce a new Boltzmann sampler for an arbitrary class of trees that is specified by the dissymmetry theorem, assuming that there exist samplers for the corresponding vertex-rooted and edge-rooted classes. This allows the dissymmetry theorem to be used not only for enumeration but also for sampling. The sampler relies on the concept of the \emph{center} of a tree -- informally, the vertex or edge of the tree that is farthest from its leaves -- and it works by repeatedly drawing a tree from the union of the vertex-rooted and edge-rooted classes until the drawn tree is rooted at its center.

In Section~\ref{cyclepointing} we provide an exposition of the cycle pointing technique, complete with diagrams to visually convey the important concepts. We tailor our exposition to our aims of enumerating and sampling from unlabeled, unrooted classes of graphs, and omit certain aspects of the theory for the sake of clarity.

In Sections~\ref{example} and \ref{implementation}, we apply the cycle pointing technique to analyze two unlabeled classes of graphs -- distance-hereditary graphs and three-leaf power graphs. Using cycle pointing, we compute exact enumerations for these classes that agree with the ones developed with the dissymmetry theorem by Chauve~\etal\cite{chauvelumbrosofusy}. We then build unbiased samplers for these two classes of graphs using cycle pointing. A full implementation in Maple is provided, along with a description of some of its features, empirical results, and drawings of randomly generated graphs.

\section{Analysis of unrooted graph classes}
\label{analysis}

\subsection{Enumeration}
\label{analysis-enumeration}

When studying a combinatorial class $\A$, one of the first and most fundamental challenges to address is to determine how many objects of a given size exist in the class. The sequence $$\A_n = \#\{\gamma\in\A\vert\n{\gamma} = n\}$$ that answers this question is called the \emph{enumeration} of $\A$, and the formal power series $$\A(z) = \sum_{n = 0}^{\infty}\A_nz^n\qquad \A(z) = \sum_{n = 0}^{\infty}\frac{\A_n}{n!}z^n$$ are called, respectively, the \emph{ordinary generating function (OGF)} and \emph{exponential generating function (EGF)} of $\A$. The former is used in the case when $\A$ is unlabeled, and the latter in the case when $\A$ is labeled, so no confusion should arise from this overloaded notation.

If $\A$ is \emph{decomposable}, meaning that it can be specified recursively in terms of basic classes ($\eps$, $\Z$, $\Set$, $\Seq$, $\Cyc$, etc.), itself, other decomposable classes, and operators (disjoint union, product, substitution, etc.), then by the theory of symbolic transfer theorems \cite{purplebook} its combinatorial specification immediately gives a generating function equation  that can often be solved in order to recover the coefficients $\A_n$. 

For example, let $\C$ be the class of Cayley trees, which are labeled, rooted, non-plane trees. An element $\gamma\in\C$ consists of a root connected to a set of $0$ or more elements of $\C$, so we have the recursive specification $$\C = \Z\times\Set(\C)$$ for the class. This results in the exponential generating function equation $$\C(z) = z\cdot\lgexp{\C(z)},$$ and by the Lagrange Inversion theorem \cite{purplebook} it follows that $$\C_n = n!\cdot\frac{1}{n}[u^{n-1}]\smexp{nu} = (n-1)!\frac{n^{n-1}}{(n-1)!} = n^{n-1}.$$ When $\A$ is not decomposable, however, this method is not sufficient for computing its enumeration, because its key tool - a symbolic specification for $\A$ - is missing. In this case a variety of other techniques may apply, depending on the particular nature of $\A$, and in this work we will focus on the techniques used for one important family of non-decomposable combinatorial classes: classes of unrooted graphs.

\subsection{Boltzmann sampling}
\label{analysis-sampling}

After studying the enumeration of a class, a natural next step is to investigate potential methods of randomly generating objects from this class. This can be useful in order to visualize large random objects in the class, and to study the behavior of parameters of the objects as their size grows.

One of the first proposed methods for randomly generating objects from a combinatorial class was the \emph{recursive method} of Flajolet~\etal\cite{recursive}, which uses the enumeration of the class (\textit{i.e.} the coefficients of its generating function) to sample an object of a specified size uniformly at random. More recently, \emph{Boltzmann samplers} have been introduced by Duchon~\etal as a general technique to sample objects from an arbitrary \emph{decomposable} combinatorial class \cite{duchon, flfupi, cpshort}. Indeed, the rules outlined in these articles allow for an automatic, algorithmic translation of the combinatorial specification of the class into a Boltzmann sampler for that class. Boltzmann samplers are particularly attractive because they are more efficient than the recursive method, running in linear time in the size of the output and not requiring the linear-time precomputation of the recursive method, and they rely not on the individual terms of the enumeration but instead on basic constructs from probability theory and the ability to evaluate the generating function of the class.

\begin{definition}
Let $\A$ be an unlabeled combinatorial class, and let $\A(z)$ be its OGF. For a fixed parameter value $z > 0$ at which $\A(z)$ converges, an \emph{ordinary Boltzmann sampler} $\Gamma\A(z)$ is a random generator that draws an object $\gamma\in\A$ with probability $$\P_z[\gamma] = \frac{z^{\n{\gamma}}}{\A(z)}.$$
\end{definition}
\noindent Since the only property of $\gamma$ upon which $\P_z[\gamma]$ depends is its size, we see that a Boltzmann sampler is \emph{unbiased}, in the sense that it draws all objects of a given size in $\A$ with equal probability. However unlike in recursive sampling, it is not possible to specify at the outset the size of the object that will be returned. Instead, this size is a random variable $S$ whose distribution depends on the parameter $z$ as follows: $$\P_z[S = n] = \sum_{\substack{\gamma\in\A,\\ \n{\gamma} = n}}\frac{z^{\n{\gamma}}}{\A(z)} = \frac{\A_nz^n}{\A(z)},$$ $$\E_z[S] = \sum_{n = 0}^{\infty}\frac{n\A_nz^n}{\A(z)} = \frac{z\A'(z)}{\A(z)}.$$  A parallel definition holds when $\A$ is a labeled class, except that $\A(z)$ must be the EGF of the class and the expressions for $\P_z[\gamma]$ and $\P_z[S = n]$ must be scaled by $1/n!$.

Duchon~\etal\cite{duchon} provide a set of rules for automatically building a Boltzmann sampler for an arbitrary labeled decomposable class, and this theory was extended to unlabeled classes by Flajolet~\etal\cite{flfupi} and Bodirsky~\etal\cite{cpshort}. Some of the basic rules in the labeled case are shown in Table~\ref{analysis-labeledboltzmanntable}, where the Poisson, geometric, and logarithmic distributions are the power series distributions for the functions $\smexp{z}$, $1/(1-z)$, and $\log(1/(1-z))$, respectively. The first four rules apply in the unlabeled case as well, while the rules for unlabeled sets, sequences, and cycles will be discussed in Section~\ref{cyclepointing-sampler-theory}.
\begin{table}[!htb]
\begin{center}
\begin{tabular} {cl}
\toprule
Class & Boltzmann sampler \\
\midrule\\[-8pt]
$\C = \eps$ & $\Gamma\C(z) = \circ\text{ (atom of size }0)$ \\[7pt]
$\C = \Z$ & $\Gamma\C(z) = \bullet\text{ (atom of size }1)$ \\[10pt]
$\C = \A + \B$ & $\Gamma\C(z) =$ \textbf{if} Bern$\left(\frac{\A(z)}{\A(z) + \B(z)}\right)$ \textbf{then} $\Gamma\A(z)$ \textbf{else} $\Gamma\B(z)$ \\[16pt]
$\C = \A\times\B$ & $\Gamma\C(z) = (\Gamma\A(z), \Gamma\B(z))$ \\[10pt]
$\C = \Set(\A)$ & $\Gamma\C(z) = \underbrace{(\Gamma\A(z), \ldots, \Gamma\A(z))}_{\text{Pois}(\A(z))}$ \\[20pt]
$\C = \Seq(\A)$ & $\Gamma\C(z) = \underbrace{(\Gamma\A(z), \ldots, \Gamma\A(z))}_{\text{Geom}(\A(z))}$ \\[20pt]
$\C = \Cyc(\A)$ & $\Gamma\C(z) = \underbrace{(\Gamma\A(z), \ldots, \Gamma\A(z))}_{\text{Loga}(\A(z))}$ \\[10pt]
\bottomrule
\end{tabular}
\caption{Boltzmann sampler rules for labeled classes.}
\label{analysis-labeledboltzmanntable}
\end{center}
\end{table}

To see a concrete example (which we will return to in Section~\ref{analysis-challenges-labeled}), we consider the class $\C$ of Cayley trees. This class can be specified by $$\C = \Z\times\Set(\C),$$ and by the rules above a Boltzmann sampler for this class is given by: $$\Gamma\C(z) = (T\leftarrow (v = \bullet; \underbrace{\Gamma\C(z), \ldots, \Gamma\C(z)}_{\text{Pois}(\C(z))});\textbf{ return }(\text{label}(T), v));$$ where the first element $v$ of the tuple is an atom denoting the root of the tree, the subtrees that appear after the semicolon in the tuple are its children, and the \emph{label} function assigns a random permutation of the labels $1, 2, \ldots, \n{T}$ to the atoms of $T$.

As is the case with enumeration, this technique does not provide Boltzmann samplers for classes that are not decomposable, because it relies on a recursive specification for the class in terms of classes whose Boltzmann samplers have already been constructed.

\subsection{The challenges of unrooted graphs}
\label{analysis-challenges}

Our aim in this work is to study techniques for enumerating and sampling from various classes of unrooted graphs. One issue that arises here is that such graphs have no ``distinguished'' vertex, edge, or other feature at which they can be recursively decomposed into smaller elements of the same class or other classes; instead, they are simply a set of vertices, together with a set of edges connecting certain pairs of those vertices. Thus, techniques beyond the ones described in Sections~\ref{analysis-enumeration} and \ref{analysis-sampling} are needed. We now introduce some of these techniques, first in the case when the class is labeled, and then in the more challenging case when it is unlabeled.

\subsubsection{The labeled case}
\label{analysis-challenges-labeled}

In order to study a class of labeled, unrooted graphs, a useful technique is to begin by studying the corresponding class of graphs that are rooted at a vertex. This is known as \emph{vertex-rooting} or \emph{vertex-pointing}, and the intuition behind it is straightforward: for any labeled, unrooted graph with $n$ nodes, there are exactly $n$ vertex-rooted graphs corresponding to it (since the root can be chosen as the vertex labeled $1$, the vertex labeled $2$, \ldots, or the vertex labeled $n$), so there is a $1$-to-$n$ correspondence between the size-$n$ elements of the unrooted class and the size-$n$ elements of the rooted class. 

If the rooted class is decomposable (as was the class of Cayley trees in Section~\ref{analysis-enumeration}, for instance), the standard techniques can be employed to develop an enumeration and Boltzmann sampler for it. As we will see below, this enumeration and sampler, together with the $1$-to-$n$ correspondence, can be used to enumerate and sample from the unrooted class.

\begin{definition}
For a class $\A$ of unrooted objects, the \emph{vertex-rooted} class corresponding to $\A$ is the class $\Av$ defined by $$\Av = \{(\gamma, v) \vb \gamma\in\A\text{ and } v \text{ is a node of } \gamma\},$$ where the size of an element $(\gamma, v)$ in $\Av$ is defined as the size of $\gamma$ in $\A$.
\end{definition}

\begin{lem}
For a class $\A$ of labeled, unrooted objects with EGF $\A(z)$, the EGF for $\Av$ is given by $$\Av(z) = z\A'(z).$$
\end{lem}
\begin{proof}
There are $n$ objects of size $n$ in $\Av$ for each graph of size $n$ in $\A$, so $\Avn = n\A_n$. Thus the EGF for $\Av$ is $$\Av(z) = \sum_{n = 0}^{\infty}\frac{\Avn}{n!}z^n = \sum_{n = 0}^{\infty}\frac{n\A_n}{n!}z^n = z\sum_{n = 0}^{\infty}\frac{\A_n}{n!}nz^{n-1} = z\A'(z).$$
\end{proof}

\begin{lem}
\label{analysis-challenges-correspondencelemma}
For a class $\A$ of labeled objects, $$\A_n = \frac{1}{n}\Avn.$$ Furthermore, if $\Gamma\Av(z)$ is a Boltzmann sampler for $\Av$, then $$\tilde{\Gamma}\A(z) = \{(\gamma, v)\leftarrow\Gamma\Av(z);\textbf{ return }\gamma;\}$$ is an unbiased sampler for $\A$, in the sense that $$\P_z[\,\gamma\,\vb\n{\gamma} = n] = \frac{1}{\A_n}.$$
\end{lem}
\begin{proof}
By the discussion in the first paragraph of this section, we see that $\A_n$ is the correct enumeration for $\A$.  For the sampler, the probability of drawing $\gamma$ from $\tilde{\Gamma}\A(z)$ is equal to the probability of drawing $(\gamma, v)$ from $\Gamma\Av(z)$ for some vertex $v$ of $\gamma$, and since $\Gamma\Av(z)$ is a Boltzmann sampler for $\Av$, this probability is $$\P_z[\gamma] = \sum_{v \text{ is a vertex of } \gamma}\frac{z^{\n{\gamma}}}{\n{\gamma}!\Av(z)} = \frac{\n{\gamma}z^{\n{\gamma}}}{\n{\gamma}!z\A'(z)} = \frac{z^{\n{\gamma} - 1}}{(\n{\gamma} - 1)!\A'(z)}.$$ Since the only property of $\gamma$ on which this expression depends is $\n{\gamma}$, it follows that $\tilde{\Gamma}\A(z)$ draws all objects of a given size from $\A$ with equal probability, and hence is unbiased.
\end{proof}
\noindent We note, however, that $\P_z[\gamma]$ is not equal to $$\frac{z^{\n{\gamma}}}{\n{\gamma}!\A(z)},$$ so $\tilde{\Gamma}\A(z)$ is not in fact a Boltzmann sampler for $\A$. In order to obtain a Boltzmann sampler, the technique of rejection may be employed as follows:

\begin{lem}
The following is a Boltzmann sampler for $\A$: $$\Gamma\A(z) = \{\textbf{do }(\gamma, v)\leftarrow\Gamma\Av(z)\textbf{ while }\text{label}(v)\neq 1;\textbf{ return }\gamma;\}$$
\end{lem}
\noindent This simple rejection solution is one that has been suggested before, for instance by Bousquet-M\'{e}lou and Weller~\cite[\Sec 11.1]{BoWe14} to draw random minor-closed classes of graphs. It has been improved upon by Darrasse~\etal~\cite[\Sec 2.2.1]{DaPaRoSo12}, who, instead of fixing the parameter $z$, draw it according to a certain differentiated probability distribution which biases the exponential Boltzmann sampler in order to mimic an unrooted distribution. While their technique, which avoids rejection altogether, is suitable for labeled objects, it is unclear how to apply it to the unlabeled objects we will study beginning in Section~\ref{analysis-challenges-unlabeled}. Indeed it seems that it does not address how to obtain an enumeration of the unrooted class, but instead assumes that such an enumeration is available (of course, by Lemma~\ref{analysis-challenges-correspondencelemma}, this enumeration is trivially available in the labeled case).
\begin{proof}
The probability of drawing $\gamma$ from $\Gamma\A(z)$ is equal to the probability of drawing $(\gamma, v)$ from $\Gamma\Av(z)$ conditioned on the event that $\text{label}(v) = 1$, which is 
\begin{align*}
\frac{\displaystyle\frac{z^{\n{\gamma}}}{\n{\gamma}!\Av(z)} }{\P^{\Av}_z[\,\text{label}(v) = 1]} &= \frac{\displaystyle\frac{z^{\n{\gamma}}}{\n{\gamma}!\Av(z)} }{\displaystyle\sum_{n = 1}^{\infty}\P^{\Av}_z[\,\text{label}(v) = 1\vb \n{\gamma} = n]\cdot\P^{\Av}_z[\n{\gamma} = n]}\\
&= \frac{\displaystyle\frac{z^{\n{\gamma}}}{\n{\gamma}!\Av(z)} }{\displaystyle\sum_{n = 1}^{\infty}\frac{1}{n}\frac{\Avn z^n}{n!\Av(z)}} \\
&= \frac{z^{\n{\gamma}}}{\n{\gamma}!\displaystyle\sum_{n = 1}^{\infty}\frac{1}{n}\frac{n\A_n}{n!}z^n} \\
&= \frac{z^{\n{\gamma}}}{\n{\gamma}!\A(z)}.
\end{align*}
\end{proof}
\noindent To see an example of this in action, let $\T$ be the class of labeled non-plane trees, which are connected graphs with no cycles. Then $\Tv$ is the class of labeled rooted non-plane trees, \textit{i.e.} Cayley trees, whose enumeration is given (from Section~\ref{analysis-enumeration}) by $$\Tvn = n^{n-1}.$$ Thus the number of labeled trees with $n$ vertices is $$\T_n = \frac{1}{n}\Tvn = n^{n-2}.$$ Furthermore, the class $\Tv$ has a Boltzmann sampler $$\Gamma\Tv(z) = \{T\leftarrow (v; \underbrace{\Gamma\Tv(z), \ldots, \Gamma\Tv(z)}_{\text{Pois}(\Tv(z))});\textbf{ return }(\text{label}(T), v);\}$$  so a Boltzmann sampler for $\T$ is given by $$\Gamma\T(z) = \{\textbf{do }(T, v) \leftarrow\Gamma\Tv(z) \textbf{ while }\text{label}(v)\neq 1; \textbf{ return }T;\}.$$

\subsubsection{The unlabeled case}
\label{analysis-challenges-unlabeled}

The analysis of unlabeled, unrooted classes of graphs poses a greater challenge than the analysis of their labeled counterparts, and the techniques described in the previous section do not suffice in general. The difficulty here arises from the existence of symmetries: without labels on the vertices, a graph may have internal symmetries that cause some of its vertices to be indistinguishable from each other, and rooting at two indistinguishable vertices will give rise to identical rooted graphs. These symmetries are of course not the same for all graphs of a given class and size. Thus while it is still possible to build the vertex-rooted class for a given unrooted class, it is no longer the case that each unrooted graph of size $n$ gives rise to the same number of rooted graphs of size $n$.

\begin{figure}[!htb]
\begin{center}
\includegraphics[width=0.4\linewidth]{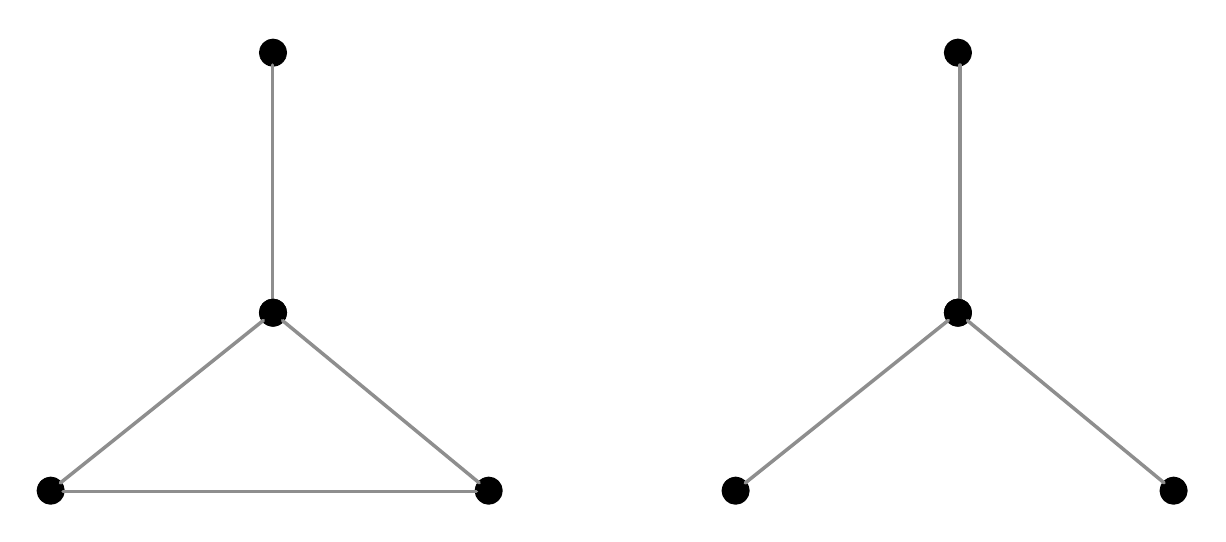}
\caption{Two graphs of size $4$.}
\label{analysis-graphsexample}
\end{center}
\end{figure}

\begin{figure}[!htb]
\begin{center}
\includegraphics[width=\linewidth]{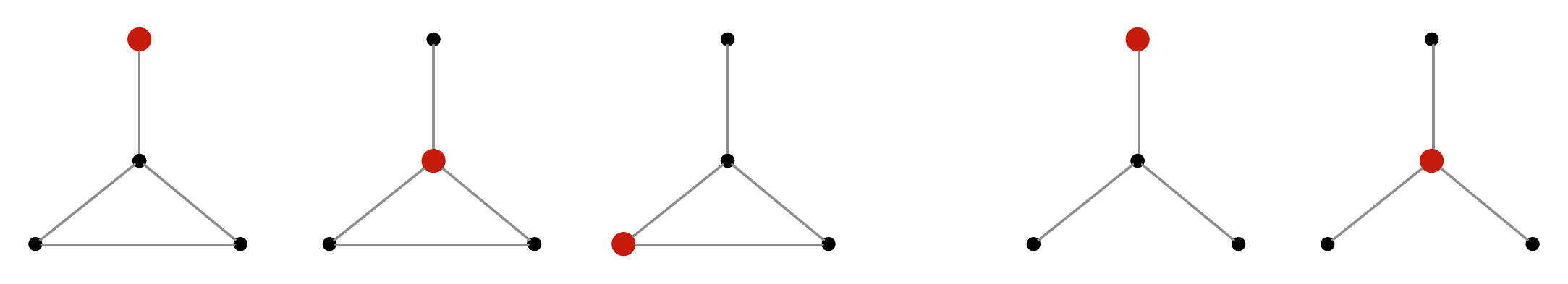}
\caption{Rooted versions of the graphs in Figure~\ref{analysis-graphsexample}.}
\label{analysis-unlabeledexample}
\end{center}
\end{figure}

For example, consider the two graphs of size $4$ in Figure~\ref{analysis-graphsexample}.
If the vertices have distinct labels, then each gives rise to 4 distinct rooted graphs, one for each vertex. However if the vertices are unlabeled, the first graph gives rise to 3 distinct rooted graphs while the second gives rise to only 2, as seen in Figure~\ref{analysis-unlabeledexample}.

Without the $1$-to-$n$ correspondence between the unrooted and rooted classes, the techniques from the previous section cannot be directly applied to derive the enumeration and Boltzmann sampler of the unrooted class from those of the rooted class. In this work we study two techniques for addressing this difficulty. In Section~\ref{dissymmetry} we discuss the dissymmetry theorem, which combines the techniques of pointing at vertices, undirected edges, and directed edges in order to develop enumerations and Boltzmann samplers for unlabeled, unrooted classes of graphs. Then in Section~\ref{cyclepointing} we discuss cycle pointing, a technique which points at certain cycles of vertices of a graph in such a way that each unpointed object of size $n$ gives rise to exactly $n$ pointed objects of size $n$. This establishes a $1$-to-$n$ correspondence for unlabeled graphs that can be employed in much the same way that vertex pointing was used in the labeled case.

\section{The dissymmetry theorem}
\label{dissymmetry}

\subsection{Overview}
\label{dissymmetry-overview}
The dissymmetry theorem, first introduced by Otter~\cite{otter} and popularized by Bergeron~\etal\cite{bergeron} in the context of species theory, relates the enumeration of an unrooted class of trees to the enumerations of three corresponding rooted classes. While it may seem restrictive that this theorem applies only to classes of trees, many non-tree classes of graphs -- some of which we will see in Section~\ref{example} -- can be characterized in terms of structures that are trees. The dissymmetry theorem can then be applied to these classes as well, as mentioned, for instance, by Chapuy~\etal\cite{chapuy}.

\begin{thm}(Dissymmetry theorem)
\label{dissymmetry-overview-thm}
Let $\A$ be an unrooted class of trees, and let $\Av$, $\Au$, and $\Ad$ be the corresponding classes of trees rooted at a vertex, an edge, and a directed edge, respectively. Then there is a bijection $$\A + \Ad\simeq\Av + \Au,$$ in the sense that for a given $n$, there are the same number of objects of size $n$ in both classes. In other words, $$\A_n = \Avn + \Aun - \Adn.$$
\end{thm}
\noindent An elegant proof of this result is given by Drmota~\cite{enumcomb}, which we briefly review here.
\begin{definition}
\label{dissymmetry-overview-centerdef}
Let $T$ be a tree. We define the \emph{center} of $T$ to be the vertex or edge of $T$ that is obtained by the following iterative procedure: at each step, simultaneously delete all leaves of $T$, and repeat until $T$ has size less than $3$. Since every tree has at least one leaf, and deleting a leaf from a tree results in another tree, this process will terminate with either a single vertex or a single edge of $T$.
\end{definition}

\begin{proof}[Proof of Theorem~\ref{dissymmetry-overview-thm}]
Consider each tree $T\in\A$ as being rooted at its center, which is either a vertex or an edge of $T$. Then the class $$\A' = (\Av + \Au)\backslash\A$$ can be thought of as the subclass of $\Av + \Au$ containing all vertex-rooted and edge-rooted trees that are not rooted at their center.

It suffices to show a bijection between this class and $\Ad$. Consider a rooted tree $(T, r)\in \A'$, with root $r$ and center $c = c(T)$. There are four possible cases to consider, which are outlined below -- in each case we define a mapping $\phi((T, r)) = (T, e)$ to a tree $(T, e)\in \Ad$, and afterwards we check that this mapping is indeed a bijection. Also, in each case we denote by $P$ be the unique path from $r$ to $c$ in $T$ (where $P$ contains both endpoints of $r$ (and/or $c$) if $r$ (and/or $c$) is an edge).

\begin{enumerate}
\item $r$ and $c$ are vertices (note that $r\neq c$ by assumption)

Since $r\neq c$, the length of $P$ is $\geq 1$. Let $e$ be the directed edge obtained by directing the first edge of $P$ away from $r$.

\item $r$ is a vertex and $c$ is an edge 

Let $e$ be the directed edge obtained by directing the first edge of $P$ away from $r$.

\item $r$ and $c$ are edges (note that $r\neq c$ by assumption)

Let $e$ be the directed edge obtained by directing the first edge of $P$ (namely $r$) away from $c$ -- since $r\neq c$, this is well-defined.

\item $r$ is an edge and $c$ is a vertex

Let $e$ be the directed edge obtained by directing the first edge of $P$ (namely $r$) away from $c$.

\end{enumerate}
To show that $\phi$ is a bijection, it suffices to show that it has an inverse. Indeed, for $(T, e)\in\Ad$ with center $c$, define $\phi^{-1}((T, e)) = (T, r)$ as follows:

\begin{enumerate}
\item $e = c$, or $e\neq c$ and $e$ is directed towards $c$

Let $r$ be the tail of $e$.

\item $e\neq c$ and $e$ is directed away from $c$

Let $r$ be $e$ with its direction removed.

\end{enumerate}
By inspection we see that $\phi$ and $\phi^{-1}$ are inverses, so $\phi:\A'\ra\Ad$ is a bijection.
\end{proof}

\noindent Following Chauve~\etal\cite{chauvelumbrosofusy}, we note that it is possible to only consider internal nodes when applying the dissymmetry theorem:

\begin{lem}
\label{dissymmetry-overview-noleaveslemma}
The dissymmetry theorem remains true when the three rooted classes are restricted to only contain those trees rooted at internal nodes or edges between two internal nodes.
\end{lem}
\begin{proof}
Let \text{\scriptsize $\square$} and \text{\scriptsize $\blacksquare$} denote a leaf and an internal node, respectively; so, for example, $\A^{\sq\ra\bsq}$ is the class of trees in $\A$ rooted at a directed edge from a leaf to an internal node. Then $$\Av = \A^\sq + \A^\bsq$$ $$\Au = \A^{\sq-\sq} + \A^{\sq-\bsq} + \A^{\bsq-\bsq}$$ $$\Ad = \A^{\sq\ra\sq} + \A^{\sq\ra\bsq} + \A^{\bsq\ra\sq} + \A^{\bsq\ra\bsq},$$ so by the dissymmetry theorem we have $$\A + \A^{\sq\ra\sq} + \A^{\sq\ra\bsq} + \A^{\bsq\ra\sq} + \A^{\bsq\ra\bsq}\simeq \A^\sq + \A^\bsq + \A^{\sq-\sq} + \A^{\sq-\bsq} + \A^{\bsq-\bsq}.$$ Since leaves have degree $1$, we see that $\A^{\sq}\simeq \A^{\sq-\bsq}\simeq \A^{\sq\ra\bsq}\simeq \A^{\bsq\ra\sq}$, so it follows that $$\A + \A^{\sq\ra\sq} + \A^{\bsq\ra\bsq}\simeq \A^\bsq + \A^{\sq-\sq} + \A^{\bsq-\bsq}.$$ Finally, the classes $\A^{\sq-\sq}$ and $\A^{\sq\ra\sq}$ are either both empty or both contain a single graph of size $2$ (depending on whether or not $\A$ contains the tree with two vertices), so $\A^{\sq-\sq}\simeq\A^{\sq\ra\sq}$, and hence $$\A  + \A^{\bsq\ra\bsq}\simeq \A^\bsq + \A^{\bsq-\bsq}.$$
\end{proof}

\subsection{Boltzmann sampler by center-rejection}
\label{dissymmetry-sampler}

In this section, we introduce a new technique for sampling from an unrooted class of trees that is specified by the dissymmetry theorem. As mentioned previously, it is possible to recursively build a Boltzmann sampler for any decomposable combinatorial class. Unfortunately, the equation $$\A\simeq\Av + \Au - \Ad$$ given by the dissymmetry theorem is not a true symbolic specification, because at first glance there is no combinatorial meaning or Boltzmann sampler rule that can be ascribed to the subtraction of the final term.

We overcome this difficulty by describing a Boltzmann sampler rule that accounts for this subtraction. More specifically, we show how to build a Boltzmann sampler for an arbitrary class that is specified by the dissymmetry theorem, assuming that there exist samplers for the corresponding vertex-rooted and edge-rooted classes. The sampler draws repeatedly from the class $\Av + \Au$, each time obtaining a pair $(T, r)$ where $T$ is a tree and $r$ is either a vertex or edge of $T$ that is marked, and stops once this marked vertex/edge happens to be the center $c(T)$ of the tree (which can also be either a vertex or an edge). Thus it utilizes the technique of \emph{rejection} -- sampling from a superclass until the sampled object has a certain property~\cite{devroye}. As we will see, its correctness follows almost immediately from the main idea of the proof of the dissymmetry theorem. 

\begin{thm} 
\label{dissymmetry-sampler-thm}
Let $\A$ be an unrooted class of trees, and suppose that we have Boltzmann samplers $\Gamma\Av(z)$ and $\Gamma\Au(z)$. Then the following procedure is a Boltzmann sampler for $\A$:
\end{thm}
$\Gamma\A(z)$:\nopagebreak

\hspace{2mm} \textbf{do}\nopagebreak

\hspace{5mm} \textbf{if} Bern$\left(\frac{\Av(z)}{\Av(z) + \Au(z)}\right) = 1$ \textbf{then}\nopagebreak

\hspace{8mm} $(T, r)\leftarrow\Gamma\Av(z)$\nopagebreak

\hspace{5mm} \textbf{else}\nopagebreak

\hspace{8mm} $(T, r)\leftarrow\Gamma\Au(z)$\nopagebreak

\hspace{2mm} \textbf{until} $r = c(T)$\nopagebreak

\hspace{2mm} \textbf{return} $T$\nopagebreak

\begin{proof}
Let $T$ be an element of $\A$. To show that the above procedure is a Boltzmann sampler for $\A$, it suffices to show that the probability that it draws $T$ is $$\P_z[T] = \frac{z^{\n{T}}}{\A(z)}.$$

\noindent Let $$\A_v(z) = \sum_{\substack{T'\in \A\\c(T') \text{ is}\\\text{ a vertex}}}z^{\n{T'}}\qquad\text{and}\qquad\A_e(z) = \sum_{\substack{T'\in \A\\c(T') \text{ is} \\\text{ an edge}}}z^{\n{T'}}.$$ Since each element of $\A$ has either a vertex or an edge as its center, we have $$\A(z) = \A_v(z) + \A_e(z).$$  We consider two cases: when $c(T)$ is a vertex, and when it is an edge. Let $P$ denote the procedure inside the loop. In the first case, 
\begin{align*}
\P_z[T] & = \frac{\P[(T, c(T))\text{ is drawn by $P$}]}{\P[T', c(T')) \text{ is drawn by $P$ for some tree } T']}\\
&= \frac{\frac{\Av(z)}{\Av(z) + \Au(z)}\cdot\frac{z^{\n{T}}}{\Av(z)}}{\frac{\Av(z)}{\Av(z) + \Au(z)}\cdot\frac{\A_v(z)}{\Av(z)} + \frac{\Au(z)}{\Av(z) + \Au(z)}\cdot\frac{\A_e(z)}{\Au(z)}}\\
&= \frac{\frac{z^{\n{T}}}{\Av(z) + \Au(z)}}{\frac{\A_v(z) + \A_e(z)}{\Av(z) + \Au(z)}}\\
&= \frac{z^{\n{T}}}{\A(z)}
\end{align*} The second case follows by the same argument, except that the initial expression for the quantity $\P[(T, c(T))\text{ is drawn by $P$}]$ is $$\frac{\Au(z)}{\Av(z) + \Au(z)}\cdot\frac{z^{\n{T}}}{\Au(z)}.$$
\end{proof}

\paragraph{Finding the center of a tree.}
We recall from Definition~\ref{dissymmetry-overview-centerdef} that the center $c(T)$ of a tree $T$ is the vertex or edge of $T$ that is obtained by the iterative process that, at each step, deletes all leaves of $T$, and that halts when $T$ has size less than $3$. This can be computed by the following linear-time algorithm. Initialize a FIFO queue $Q$ that contains the leaves of $T$, and then repeatedly: pop a leaf $\l$ from $Q$, delete $\l$ from $T$, and push the former neighbor of $\l$ into $Q$ if it is now a leaf (\textit{i.e.} if its new degree is $1$). Continue until $Q$ has either one vertex, or two vertices that are connected by an edge. Since any newly-created leaf will not be popped until all already-existent leaves are handled, this algorithm mimics the process of deleting all leaves simultaneously at each step, and since each vertex is popped from $Q$ at most one time, the algorithm runs in linear time in the number of vertices of~$T$.

\begin{lem}
\label{dissymmetry-sampler-rejectionlemma}
The number of iterations made by the do-while loop until a suitable tree is drawn (\textit{i.e.} one that is marked at its center) is, on average, $$E_{\A}(z) = \frac{\Av(z) + \Au(z)}{\A(z)}.$$
\end{lem}
\begin{proof}
The probability of exiting the loop on a given round is the chance of drawing a tree rooted at its center, which is $$\P[T', c(T')) \text{ is drawn by $P$ for some tree } T'].$$ Furthermore, we saw in the proof of Theorem~\ref{dissymmetry-sampler-thm} that this probability is $$p = \frac{\Av(z)}{\Av(z) + \Au(z)}\cdot\frac{\A_v(z)}{\Av(z)} + \frac{\Au(z)}{\Av(z) + \Au(z)}\cdot\frac{\A_e(z)}{\Au(z)} = \frac{\A(z)}{\Av(z) + \Au(z)}.$$ Since the number of rounds of the loop is a geometric random variable with success probability $p$, its expected value is $$\frac{1}{p} = \frac{\Av(z) + \Au(z)}{\A(z)}.$$
\end{proof}

\subsection{Example}
\label{dissymmetry-example}

Let $\T$ be the class of unlabeled, unrooted, non-plane 2-3 trees -- equivalently, the class of trees whose vertices each have degree 1, 3, or 4 (notice that if you hang such a tree from one of its leaves, each internal node has either $2$ or $3$ children). In order to analyze $\T$ with the dissymmetry theorem, we begin by determining combinatorial specifications for the classes $\Tv$, $\Tu$, and $\Td$; respectively, the class of trees in $\T$ where one vertex is marked, the class of trees in $\T$ where one undirected edge is marked, and the class of trees in $\T$ where one directed edge is marked. 

Objects in $\Tv$ are decomposed at their root -- indeed, an object in $\Tv$ is a (marked) root node with 1, 3, or 4 neighbors, where each neighbor is a node with 0, 2, or 3 other neighbors, so we have the decomposition $$\Tv = \Z^{\bullet}\times\Set_{1, 3, 4}(\S),$$$$\S = \Z + \Z\times\Set_{2, 3}(\S).$$ 
\begin{figure}[!htb]
\begin{center}
\includegraphics[width=0.3\linewidth]{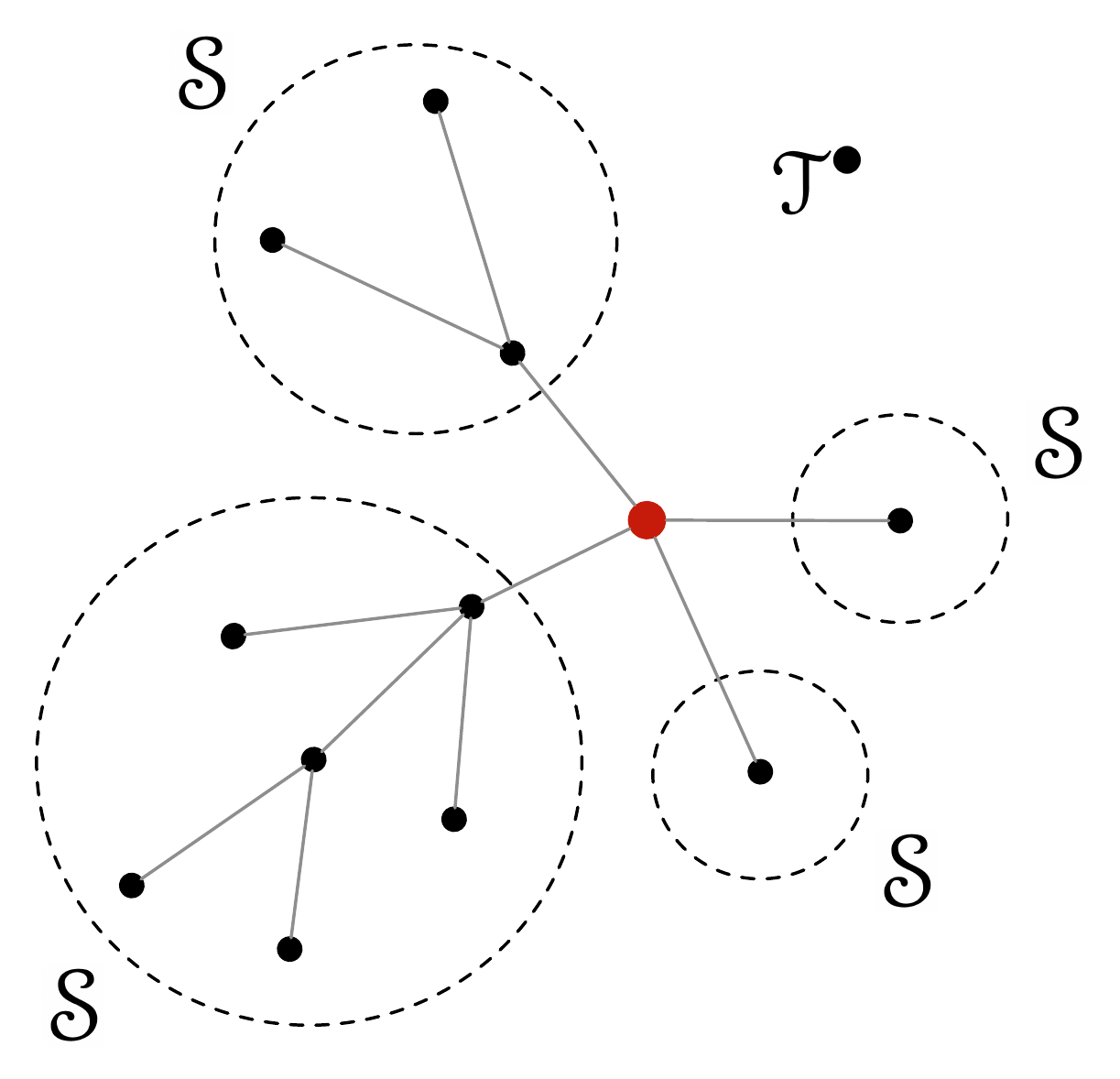}
\caption{Decomposition of vertex-rooted 2-3 trees.}
\label{dissymmetry-twothreetrees}
\end{center}
\end{figure}

\noindent Objects in $\Tu$ and $\Td$ are decomposed at their marked edge -- in the first case, the edge has a set of two neighbors, while in the second case it has a sequence of two neighbors (since the head and tail of the edge are distinguished), so we have $$\Tu = \Set_2(\S),$$$$\Td = \S\times\S.$$ Using the Maple package $\tt{combstruct}$, we compute the ordinary generating functions for these~classes: $$\Tv(z) = z^2 + 2z^4 + 2z^5 + 2z^6 + 4z^7 + 6z^8 + 10z^9 + 17z^{10} + 29z^{11} + 48z^{12} + 85z^{13} + 148z^{14} + 259z^{15} + \ldots$$$$\Tu(z) = z^2 + z^4 + z^5 + 2z^6 + 3z^7 + 5z^8 + 8z^9 + 14z^{10} + 24z^{11} + 42z^{12} + 73z^{13} + 131z^{14} + 230z^{15} + \ldots$$$$\Td(z) = z^2 + 2z^4 + 2z^5 + 3z^6 + 6z^7 + 9z^8 + 16z^9 + 27z^{10} + 48z^{11} + 82z^{12} + 146z^{13} + 259z^{14} + 460z^{15} + \ldots$$ and by Theorem~\ref{dissymmetry-overview-thm} we have 
\begin{align*}
\T(z) &= \Tv(z) + \Tu(z) - \Td(z) \\
&= z^2 + z^4 + z^5 + z^6 + z^7 + 2z^8 + 2z^9 + 4z^{10} + 5z^{11} + 8z^{12} + 12z^{13} + 20z^{14} + 29z^{15} + \ldots.
\end{align*}
In order to build a Boltzmann sampler for $\T$ using Theorem~\ref{dissymmetry-sampler-thm}, we require samplers for $\Tv$ and  $\Tu$. It is known how to construct these samplers \cite{flfupi}, and we will provide more detail on the necessary tools in Section~\ref{cyclepointing-sampler-theory}. However, we can immediately use Lemma~\ref{dissymmetry-sampler-rejectionlemma} to estimate the rejection cost for different values of $z$, as shown in Table~\ref{dissymmetry-rejectioncosttable}.
\begin{table}[!htb]
\begin{center}
\begin{tabular} {cM{4cm}M{4cm}M{4cm}N}
\toprule
$z$ & Expected size & Probability of success on each iteration & Expected number of iterations of the loop & \\[10pt]
\midrule
$0.01$ & $2.000$ & $0.5000$ & $2.000$ & \\[10pt]
$0.1$ & $2.023$ & $0.4972$ & $2.011$ & \\[10pt]
$0.5$ & $3.647$ & $0.3043$ & $3.287$ & \\[10pt]
$\rho\approx0.508256$ & $4.224$ & $0.2455$ & $4.051$ & \\[10pt]
\bottomrule
\end{tabular}
\caption{Rejection cost for the dissymmetry theorem Boltzmann sampler for $\T$.}
\label{dissymmetry-rejectioncosttable}
\end{center}
\end{table}

\begin{figure}[!htb]
\begin{center}
\includegraphics[width=0.425\linewidth]{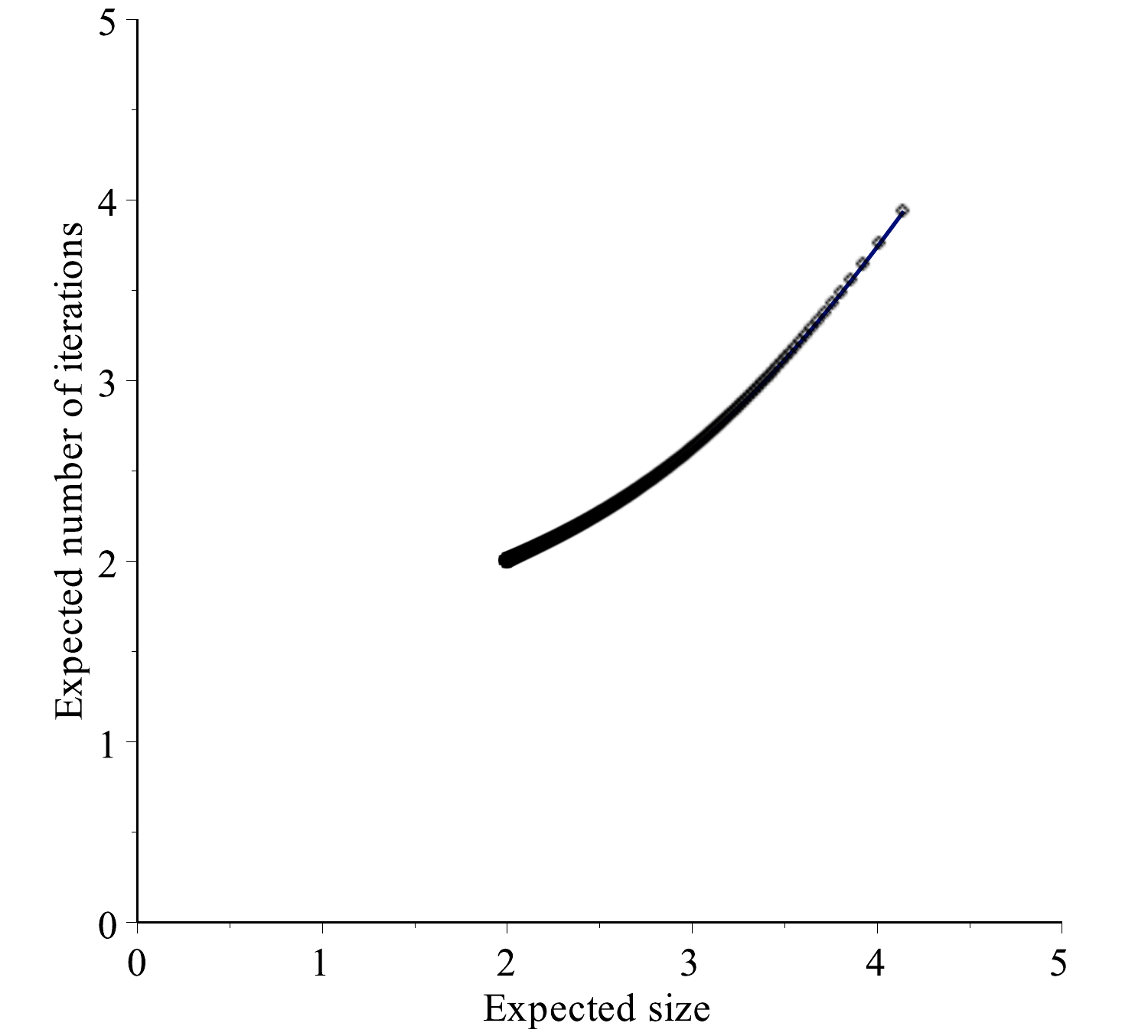}
\caption{Rejection cost vs. expected size for the dissymmetry theorem Boltzmann sampler for $\T$.}
\label{dissymmetry-rejectioncostbysizegraph}
\end{center}
\end{figure}

In Figure~\ref{dissymmetry-rejectioncostbysizegraph} we show a plot of the expected number of iterations as a function of $z\A'(z)/\A(z)$, the expected size of the returned object. It appears that these quantities have an approximately linear relationship with with a slope of about $1$.

\section{Cycle pointing}
\label{cyclepointing}

In Section~\ref{dissymmetry}, we saw that the dissymmetry theorem allows one to enumerate and create a Boltzmann sampler for an unlabeled class of trees by analyzing three corresponding rooted classes.  However, this technique lacks combinatorial strength -- it relies on the very particular notion of the center of a tree to establish a bijection, but fails to provide a symbolic decomposition that carries along with it all the usual trappings such as automatic Boltzmann samplers and asymptotics. Furthermore, our sampler that emerges from the dissymmetry theorem might not be ideal, because it relies on a potentially costly rejection process. Finally, the existence of a method for pointing at labeled objects that establishes a $1$-to-$n$ correspondence between the unpointed and pointed classes (\textit{cf.} Section~\ref{analysis-challenges-labeled}) raises the question of whether or not such a method exists for unlabeled objects as well.

Cycle pointing, a technique introduced by Bodirsky~\etal \cite{cpshort, cplong}, addresses these issues simultaneously. Instead of selecting a certain distinguished vertex of a graph, one selects a \emph{cycle} of vertices satisfying a certain property, and this property is chosen in such a way that there are exactly $n$ pointed objects of size $n$ for each object of size $n$ in the original class. Together with the ability to decompose the pointed class, this immediately provides an unbiased sampler for the unlabeled, unpointed class that does not use rejection. As we will mention again later, cycle pointing in the labeled case exactly reduces to vertex pointing, so cycle pointing can be thought of as a generalization of the method of Section~\ref{analysis-challenges-labeled} to the unlabeled case.

However, using cycle pointing to enumerate and build a sampler for a class is quite a challenging task; two of the main contributions of this work are to help break down this process for future readers, and to apply cycle pointing to enumerate and build the first unbiased samplers for the classes of distance-hereditary and three-leaf power graphs.

\paragraph{Outline.}
We first give a refresher on certain relevant graph theoretic notions, and we then dive into the definitions and main results of cycle pointing in Section~\ref{cyclepointing-introduction}. Section~\ref{cyclepointing-decomposition} shows how to write a combinatorial specification for a cycle-pointed class, Section~\ref{cyclepointing-enumeration} shows how to exploit the decomposition using elements of P\'{o}lya theory to derive an enumeration for the class, and Section~\ref{cyclepointing-sampler} shows how to automatically translate the decomposition into an unbiased sampler for the underlying~class.

\subsection{Introduction}
\label{cyclepointing-introduction}

We begin this section by reviewing some definitions from graph theory, which once in hand allow us to introduce cycle pointing and present the relevant results from Bodirsky~\etal\cite{cpshort, cplong}. All proofs in Section~\ref{cyclepointing} are adapted from these two sources.

\begin{definition}
An \emph{automorphism} of a graph $G$ is a mapping from $G$ to itself that preserves its underlying structure, in particular its adjacencies and nonadjacencies. More formally, an automorphism of $G$ is a bijection $\phi:V(G)\ra V(G)$ such that for any $u, v\in V(G)$, $uv\in E(G)$ iff $\phi(u)\phi(v)\in E(G)$.
\end{definition}
\begin{figure}[!htb]
\begin{center}
\includegraphics[width=0.2\linewidth]{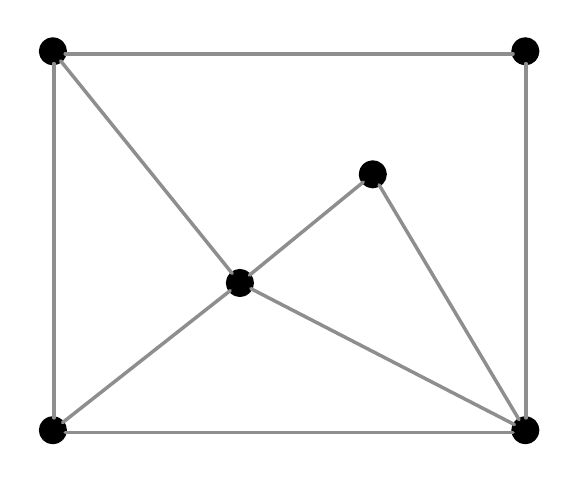}
\caption{An identity graph.}
\label{cyclepointing-identitygraph}
\end{center}
\end{figure}
\noindent Every graph has at least one automorphism -- the identity map -- and graphs with no other automorphisms are called \emph{identity graphs}. All other graphs are said to have \emph{non-trivial} automorphisms. With the operation of composition, the set of automorphisms of a graph forms a group called its \emph{automorphism group}. 

\begin{figure}[!htb]
\begin{center}
\includegraphics[width=0.45\linewidth]{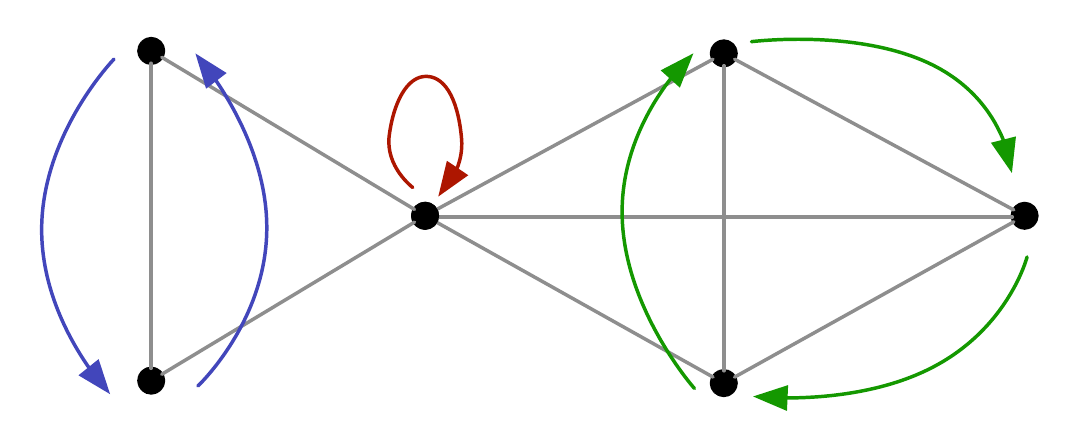}
\caption{An automorphism of a graph, with cycles shown.}
\label{cyclepointing-cyclesexample}
\end{center}
\end{figure}

\begin{figure}[!htb]
\begin{center}
\includegraphics[width=0.45\linewidth]{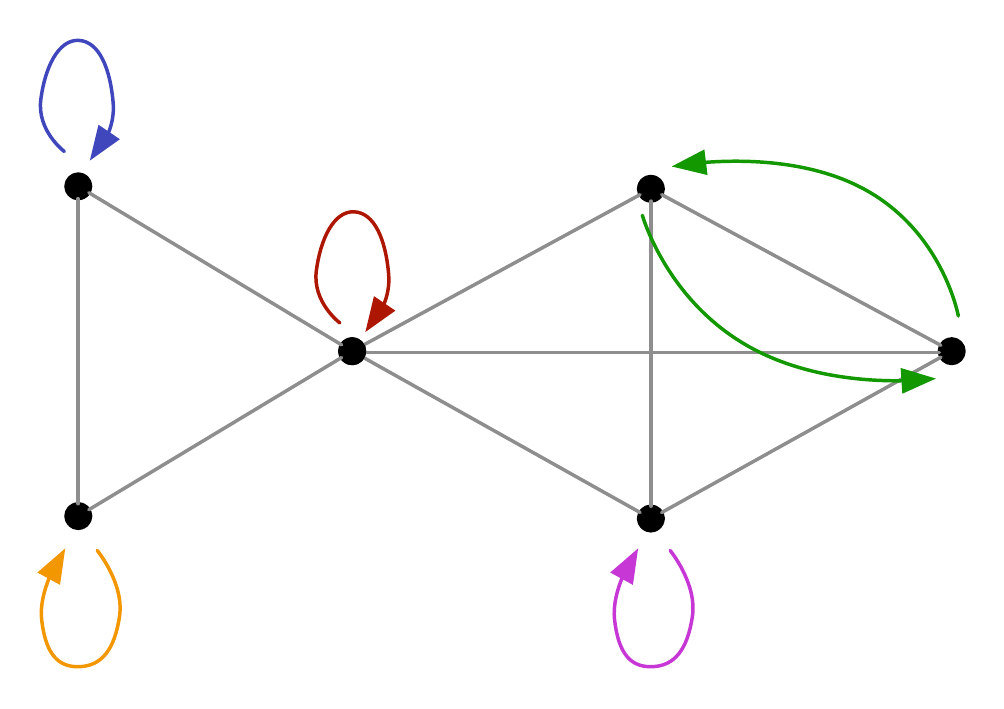}
\caption{A different automorphism on the graph from Figure~\ref{cyclepointing-cyclesexample}.}
\label{cyclepointing-cyclesnewexample}
\end{center}
\end{figure}

Since each automorphism is a permutation of the vertices of $G$, it may be uniquely decomposed as a set of disjoint cycles on the vertices of $G$. For example, the automorphism shown in Figure~\ref{cyclepointing-cyclesexample} has three cycles, shown in different colors. Figure~\ref{cyclepointing-cyclesnewexample} shows a different automorphism on the same graph.

\begin{figure}[!htb]
\begin{minipage}[b]{0.5\textwidth}
\begin{center}
\includegraphics[width=0.5\linewidth]{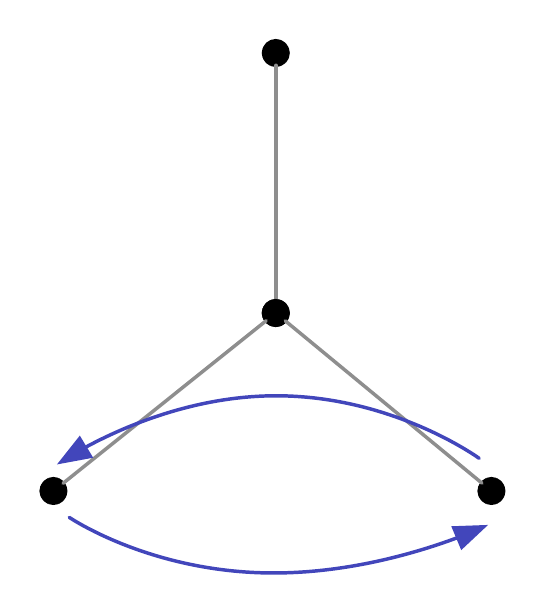}
\caption{Valid cycle-pointed structure.}
\label{cyclepointing-valid}
\end{center}
\end{minipage}
\hfill
\begin{minipage}[b]{0.5\textwidth}
\begin{center}
\includegraphics[width=0.5\linewidth]{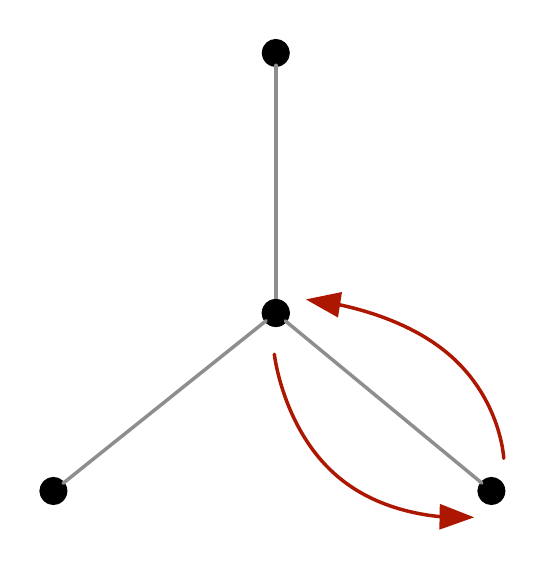}
\caption{Invalid cycle-pointed structure.}
\label{cyclepointing-invalid}
\end{center}
\end{minipage}
\end{figure}

\noindent From this concept of an automorphism and its cycles, we can define cycle-pointed structures and the cycle pointing operator:
\begin{definition}
\label{cyclepointing-introduction-cyclepointeddef}
For a graph $G$, a \emph{cycle-pointed structure} is a pair $P = (G, c)$ such that there exists at least one automorphism of $G$ having $c$ as one of its cycles (such an automorphism is called a $c$-\emph{automorphism} of $G$). Then $c$ is called the \emph{marked cycle} of $P$, and $G$ is called its \emph{underlying structure}. $P$ is called \emph{symmetric} if $c$ has at least two vertices (because in this case $c$ corresponds to a non-trivial automorphism of $G$).
\end{definition}
\noindent For a better intuition of what constitutes a cycle-pointed structure, we note that Figure~\ref{cyclepointing-valid} is a valid cycle-pointed structure, while Figure~\ref{cyclepointing-invalid} is not. One way to see that Figure~\ref{cyclepointing-invalid} is not a cycle-pointed structure is to note that, since automorphisms preserve adjacencies and nonadjacencies, any automorphism must map a vertex of a certain degree to a vertex of the same degree (this condition is necessary but of course not sufficient), so the red marked cycle cannot be part of any automorphism.

\begin{figure}[!htb]
\begin{center}
\includegraphics[width=0.85\linewidth]{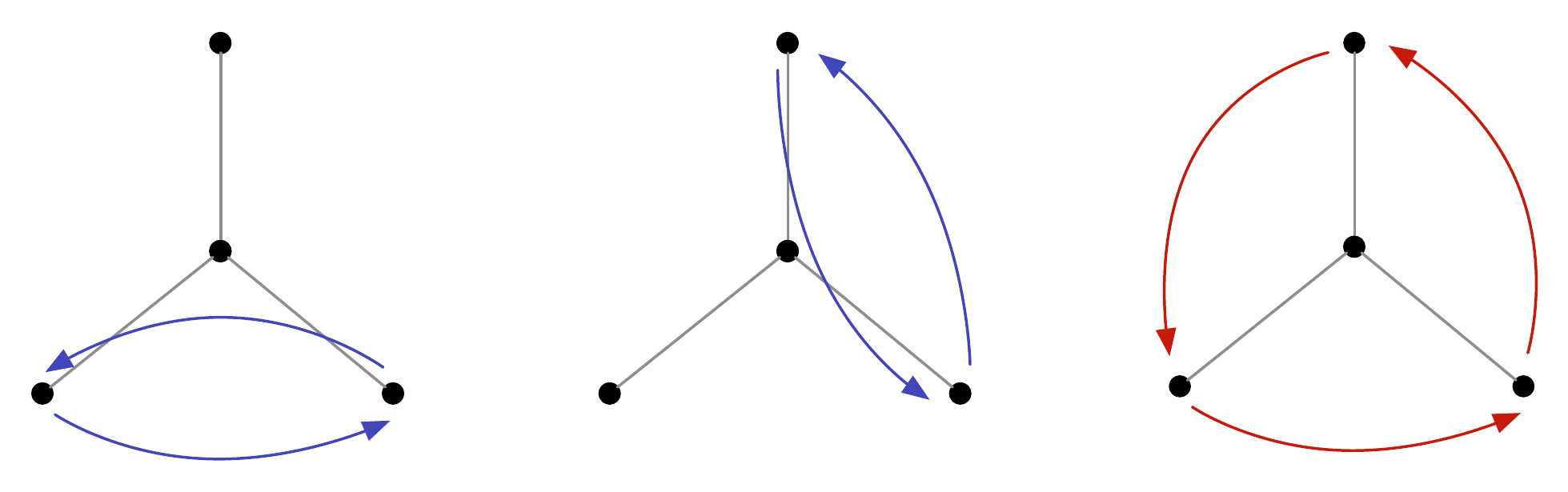}
\caption{Two isomorphic cycle-pointed structures, and one counterexample.}
\label{cyclepointing-isomorphic}
\end{center}
\end{figure}

Two cycle-pointed structures $P = (G, c)$ and $P' = (G', c')$ are considered isomorphic if there exists an isomorphism from $G$ to $G'$ that maps $c$ to $c'$ in a manner that preserves the cyclic order of the cycles. For example, in Figure~\ref{cyclepointing-isomorphic}, the first two cycle-pointed structures are isomorphic, while the third is not isomorphic to either. Beginning in Definition~\ref{cyclepointing-introduction-classdef}, we will consider, unless stated otherwise, two isomorphic cycle-pointed structures to be exactly the same.
\begin{definition}
\label{cyclepointing-introduction-classdef}
For a class $\A$ of graphs, the \emph{cycle-pointed} class corresponding to $\A$ is the class denoted $\Acp$ containing all (non-isomorphic) cycle-pointed structures whose underlying structure is in $\A$, where the size of an element $(G, c)$ in $\Acp$ is defined as the size of $G$ in $\A$. We refer to the map $\A\ra\Acp$ as the \emph{cycle-pointing operator}. Also more generally, we refer to any class of cycle-pointed structures as a cycle-pointed class. 
\end{definition}

\begin{remark}
The cycle-pointing operator, owing to its differential nature (\textit{cf.} Theorem~\ref{cyclepointing-introduction-correspondencethm}), satisfies the following compatibility properties:
\begin{itemize}
\item $\cp{(\A + \B)} = \Acp + \Bcp$
\item $\cp{(\A\times\B)} = \Acp\times\B + \A\times\Bcp$
\end{itemize}
\end{remark}

For a labeled graph, the only automorphism is the identity (due to the distinct labels on each vertex), so the only valid cycle-pointed structures have singleton cycles as their marked cycles. Thus in the labeled case, cycle pointing reduces to vertex pointing, so there are $n$ cycle-pointed objects of size $n$ for each unpointed object of size $n$. 

\begin{figure}[!htb]
\begin{center}
\includegraphics[width=0.55\linewidth]{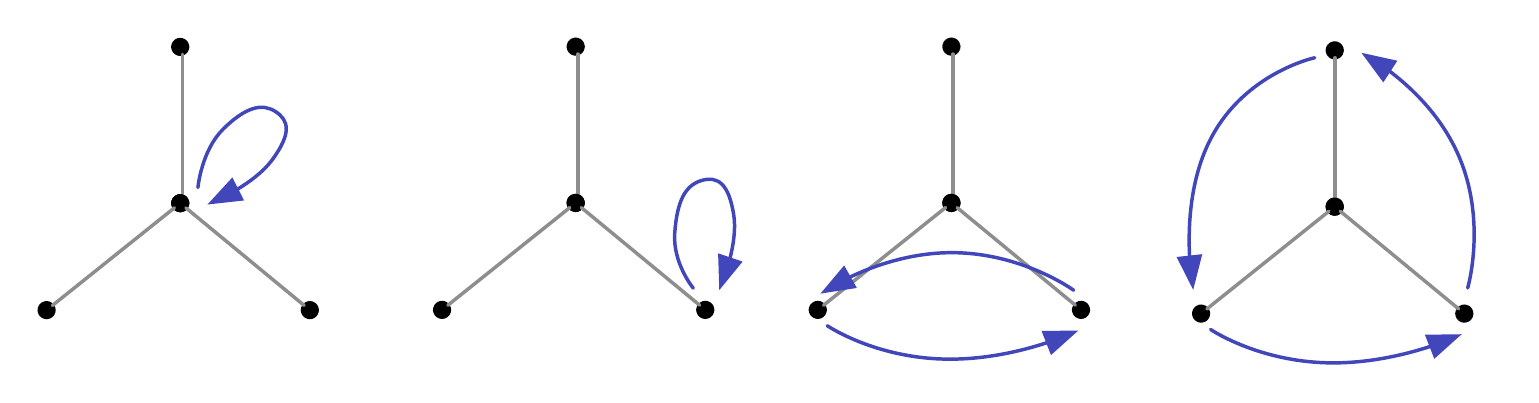}
\caption{Cycle-pointed structures of the claw graph.}
\label{cyclepointing-claw}
\end{center}
\end{figure}

\begin{figure}[!htb]
\begin{center}
\includegraphics[width=0.55\linewidth]{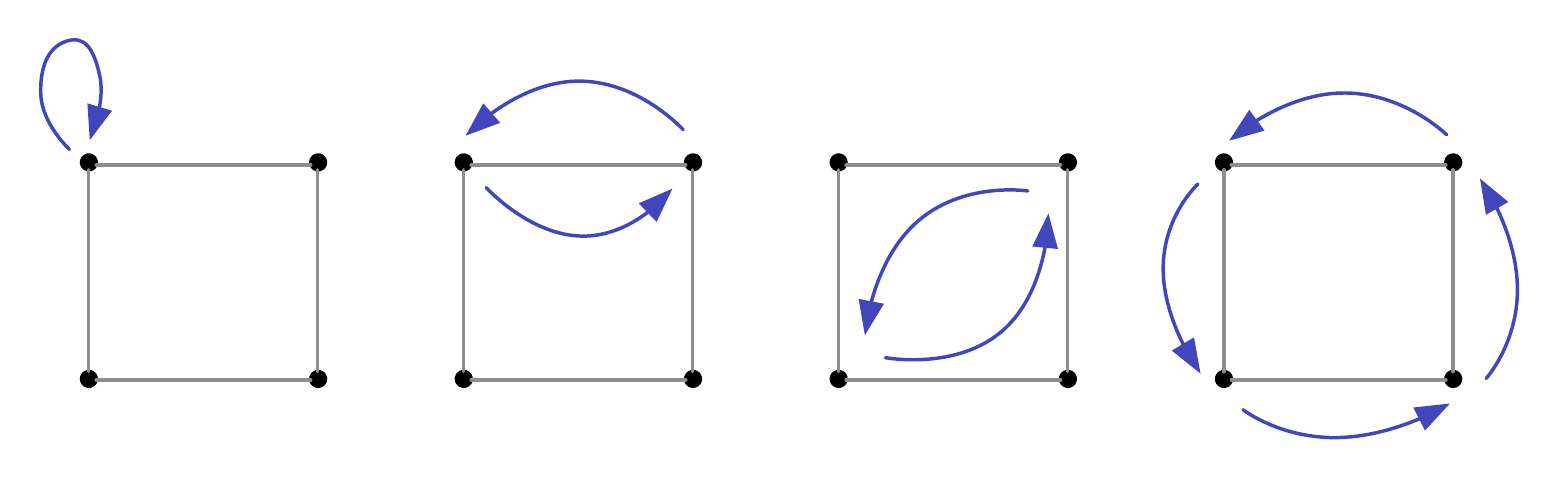}
\caption{Cycle-pointed structures of the square graph.}
\label{cyclepointing-square}
\end{center}
\end{figure}

\begin{figure}[!htb]
\begin{center}
\includegraphics[width=0.7\linewidth]{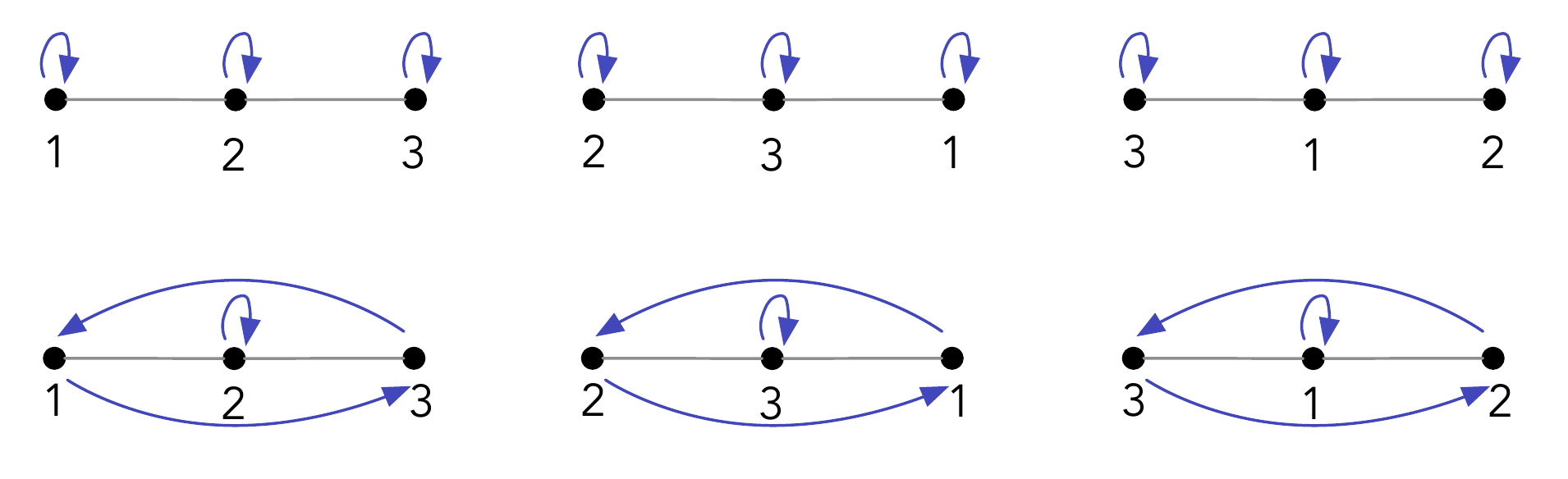}
\caption{Symmetries of the path on three vertices.}
\label{cyclepointing-pathsymmetries}
\end{center}
\end{figure}

As alluded to earlier, this $1$-to-$n$ correspondence carries over to the unlabeled case as well -- for each unlabeled graph of size $n$, there are exactly $n$ cycle-pointed structures of size $n$ whose underlying structure is that graph. For example, Figure~\ref{cyclepointing-claw} shows the four cycle-pointed structures of the claw graph, and Figure~\ref{cyclepointing-square} shows the four cycle-pointed structures of the square graph.
\begin{thm}
\label{cyclepointing-introduction-correspondencethm}
Let $\A$ be an unlabeled class of graphs. Then for each graph $G\in\A$ of size $n$, there are exactly $n$ objects of size $n$ in $\Acp$ whose underlying structure is $G$. Thus, the OGF for $\Acp$ satisfies $$\Acp(z) = z\A'(z).$$
\end{thm}
\noindent In order to prove this theorem, we must first introduce a few new concepts.
\begin{definition}
For an unlabeled graph $G$, a \emph{symmetry} of $G$ is a pair $(G^\ell, \sigma)$ where $G^\ell$ is a labeled graph whose unlabeled structure is $G$ and $\sigma$ is an automorphism of $G$. The set of symmetries of $G$ is denoted $\Sym(G)$.
\end{definition}
\noindent For example, Figure~\ref{cyclepointing-pathsymmetries} shows the symmetries of the path graph on three vertices.
\begin{lem}
\label{cyclepointing-introduction-symmetrylemma}
An unlabeled graph $G$ of size $n$ has $n!$ symmetries.
\end{lem}
\begin{proof}
Let $\G$ be the automorphism group of $G$, and let $X$ be the set of $n!$ labeled graphs whose unlabeled structure is $G$, considered \textbf{not} up to isomorphism (so for example, the path 1-2-3 and the path 3-2-1 would be considered distinct in $X$, even though they are not distinct labeled graphs). Then we may define a group action of $\G$ on $X$ -- essentially, a rule by which every element of $\G$ maps each element of $X$ to another element of $X$ -- by, for $\sigma\in\G$ and $x\in X$, defining $\sigma\cdot x$ to be the element of $X$ obtained by applying the automorphism $\sigma$ to the labeled graph $x$. 

We say that two elements $x, y$ are in the same orbit of this group action if there exists some $\sigma\in\G$ such that $\sigma\cdot x = y$; in other words, if there is an automorphism mapping $x$ to $y$. Since $\G$ is the automorphism group of $G$, the orbits $X/\G$ of this action are in bijection with the labeled graphs whose unlabeled structure is $G$, considered up to isomorphism.

Since there is exactly one symmetry of $G$ for each choice of one orbit of the action and one automorphism of $G$, the number of symmetries of $G$ is $$\n{\Sym(G)} = \n{\G}\cdot\n{X/\G}.$$ By Burnside's lemma, this is equal to $$\sum_{g\in\G}\n{\Fix(g)},$$ where $\Fix(g)$ is the set of elements $x\in X$ such that $g\cdot x = x$~\cite{artin}. However the only element of $\G$ that fixes any elements of $X$ is the identity automorphism, and this fixes all $n!$ elements of $X$, so it follows that $$\n{\Sym(G)} = \n{\G}\cdot\n{X/\G} = n!.$$
\end{proof}
\noindent Similarly, we define a symmetry of a cycle-pointed structure $P = (G, c)$ as a tuple $(G^\ell, c^\ell, \sigma)$ such that $(G^\ell, \sigma)$ is a symmetry of $G$, $c^\ell$ is a labeled cycle whose unlabeled structure is $c$, and $\sigma$ preserves the cycle $c^\ell$. By a parallel argument to the one given above, every unlabeled cycle-pointed object of size $n$ has $n!$ symmetries.

\begin{figure}[!htb]
\begin{center}
\includegraphics[width=0.75\linewidth]{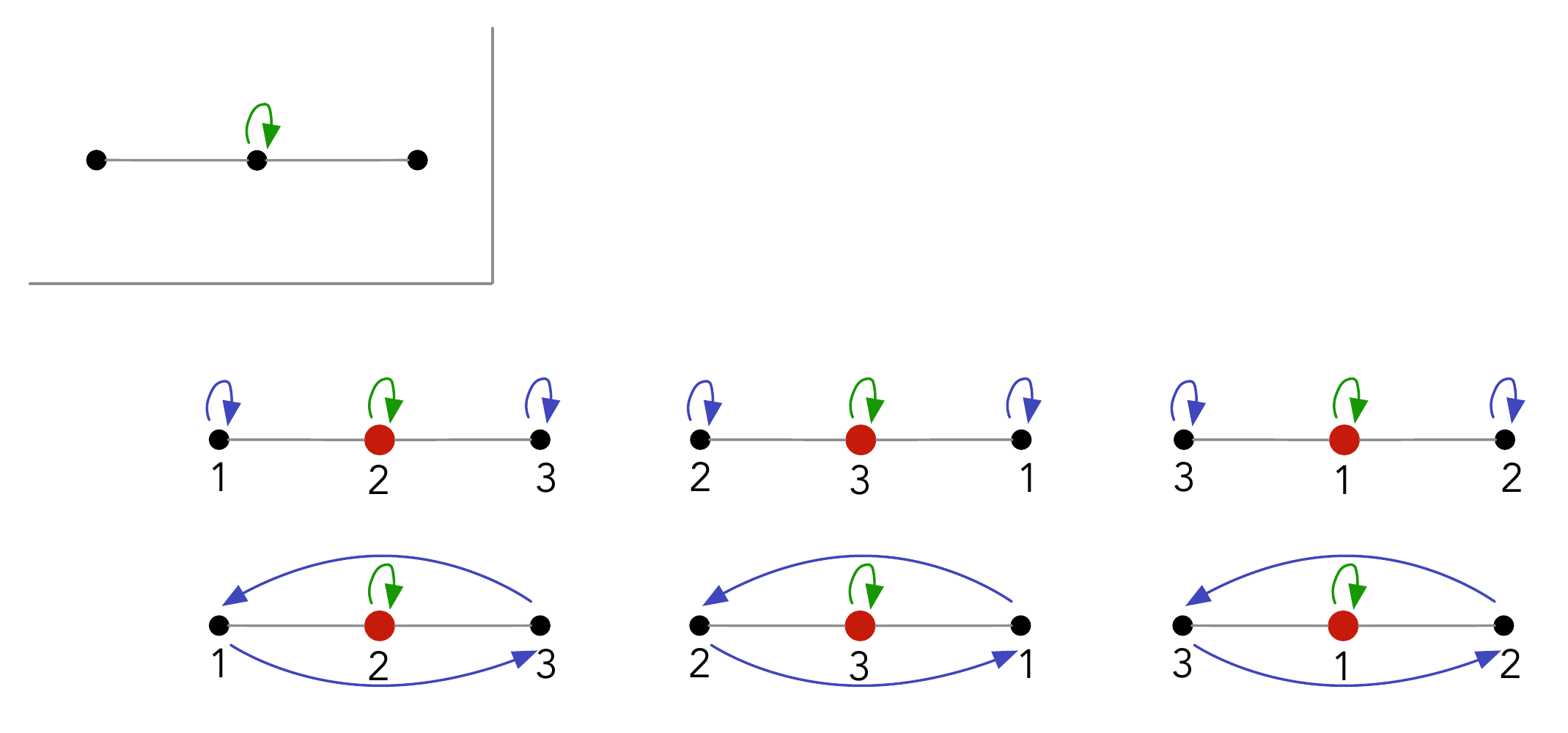}
\caption{A cycle-pointed structure and its rooted $c$-symmetries.}
\label{cyclepointing-pathrootedcsymmetries}
\end{center}
\end{figure}

\begin{definition}
For an unlabeled cycle-pointed structure $P = (G, c)$, a \emph{$c$-symmetry} of $P$ is a tuple $(G^\ell, c^\ell, \sigma)$ where $G^\ell$ is a labeled graph whose unlabeled structure is $G$, $c^\ell$ is a labeled cycle whose unlabeled structure is $c$, and $\sigma$ is a $c$-automorphism of $P$ (note that since the set of automorphisms having $c$ as a cycle is a subset of the set of automorphisms that respect $c$, every $c$-symmetry of $P$ is also a symmetry of $P$, but the reverse is not necessarily true). Furthermore, a \emph{rooted $c$-symmetry} of $P$ is a tuple $(G^\ell, c^\ell, \sigma, v)$ where $(G^\ell, c^\ell, \sigma)$ is a $c$-symmetry of $P$ and $v$ is a vertex on $c^\ell$. The set of rooted $c$-symmetries of $P$ is denoted $\Rcsym(P)$.
\end{definition}
\noindent For example, Figure~\ref{cyclepointing-pathrootedcsymmetries} shows the rooted $c$-symmetries of a cycle-pointed structure $(G, c)$, where the cycle $c$ is shown in green, and for each rooted $c$-symmetry $(G^\ell, c^\ell, \sigma, v)$, the cycles of $\sigma$ that are not $c^\ell$ are shown are blue and the root $v$ is shown in red.
\begin{lem}
\label{cyclepointing-introduction-rootedcsymmetrylemma}
An unlabeled cycle-pointed structure $P = (G, c)$ of size $n$ has $n!$ rooted $c$-symmetries.
\end{lem}
\begin{proof}
Since $P$ has $n!$ symmetries, it suffices to establish a bijection between the symmetries of $P$ and the rooted $c$-symmetries of $P$. We will accomplish this by choosing some fixed $c$-symmetry of $P$, and then for each symmetry of $P$, using its ``distance'' from this fixed $c$-symmetry to map it to a rooted $c$-symmetry of $P$.

Fix a $c$-symmetry $(G^\ell_0, c^\ell_0, \sigma_0)$ of $P$ (so $c^\ell_0$ is a cycle of $\sigma_0$), and consider an arbitrary symmetry $(G^\ell, c^\ell, \sigma)$ of $P$ (so $\sigma$ preserves $c^\ell$, but $c^\ell$ is not necessarily one of its cycles). Since $\sigma$ preserves the cycle $c^\ell$, it shifts the atoms $(v_1, \ldots, v_{\n{c^\ell}})$ of $c^\ell$ by some value $1\leq r\leq \n{c^\ell}$ modulo $\n{c^\ell}$, mapping $v_i$ to $v_{i + r\text{ (mod }{\n{c^\ell}})}$. Also, since $(G^\ell_0, c^\ell_0)$ and $(G^\ell, c^\ell)$ are labeled cycle-pointed structures with the same unlabeled structure, there is an isomorphism $\sigma_1$ from $G^\ell$ to $G^\ell_0$ that maps $c^\ell$ to $c^\ell_0$.

Let $\tau = \sigma_1^{-1}\sigma_0\sigma_1$, and let $v$ be the atom on $c^\ell$ having the $r^{th}$ smallest label.  Then we claim that $(G^\ell, c^\ell, \tau^{-r + 1}\sigma, v)$ is a rooted $c$-symmetry of $P$. To show this, it suffices to show that $\tau^{-r + 1}\sigma$ has $c^\ell$ as one of its cycles. Indeed, since $\sigma_1$ maps $c^\ell$ to $c^\ell_0$, $\sigma_0$ has $c^\ell_0$ as one of its cycles, and $\sigma_1^{-1}$ maps $c^\ell_0$ to $c^\ell$, it follows that $\tau$ has $c^\ell$ as one of its cycles. Thus $\tau^{-1}$ shifts $c^\ell$ backwards by one step, and since $\sigma$ shifts $c^\ell$ forward by $r$ steps it follows that $(\tau^{-1})^{r-1}\sigma = \tau^{-r + 1}\sigma$ shifts $c^\ell$ forward by one step and hence has $c^\ell$ as one of its cycles.

In the reverse direction, for a given rooted $c$-symmetry $(G^\ell, c^\ell, \sigma', v)$, let $r\geq 1$ be such that $v$ has the $r^{th}$ smallest label on $c^\ell$. Then defining $\sigma_1$ as the first paragraph, we recover $\sigma = (\sigma_1^{-1}\sigma_0\sigma_1)^{r-1}\sigma'$, and by the reverse argument to the one above it follows that $(G^\ell, c^\ell, \sigma)$ is a symmetry of $P$.
\end{proof}

\begin{lem}
\label{cyclepointing-introduction-correspondencelemma}
For a given symmetry $(G^\ell, \sigma)$ of an unlabeled graph $G$ of size $n$, there are $n$ rooted $c$-symmetries that have $\sigma$ as their automorphism and $G^\ell$ as their graph.
\end{lem}
\begin{proof}
Since $G^\ell$ has $n$ vertices and each vertex is on only one cycle of $\sigma$, there can be at most $n$ suitable rooted $c$-symmetries. On the other hand, there is indeed one for each vertex $v$ of $G^\ell$, obtained by choosing $c^\ell$ as the cycle of $\sigma$ containing $v$.
\end{proof}
\noindent With these results in hand, we return to the proof of Theorem~\ref{cyclepointing-introduction-correspondencethm}.
\begin{thm*}
Let $\A$ be an unlabeled class of graphs. Then for each graph $G\in\A$ of size $n$, there are exactly $n$ objects of size $n$ in $\Acp$ whose underlying structure is $G$. Thus, the OGF for $\Acp$ satisfies $$\Acp(z) = z\A'(z).$$
\end{thm*}
\begin{proof}
Let $G\in\A$ be a graph of size $n$, and let $S$ be the set of objects in $\Acp$ whose underlying structure is $G$. From Lemma~\ref{cyclepointing-introduction-symmetrylemma} we know that $\n{\Sym(G)} = n!$, and from Lemma~\ref{cyclepointing-introduction-rootedcsymmetrylemma} we know that each element of $S$ has $n!$ rooted $c$-symmetries. From Lemma~\ref{cyclepointing-introduction-correspondencelemma}, each symmetry of $G$ gives rise to exactly $n$ rooted $c$-symmetries, and since the cycle-pointed structure of each of these rooted $c$-symmetries is in $S$ it follows that $n\cdot\n{\Sym(G)} = \n{S}\cdot n!.$ Since $\n{\Sym(G)} = n!$, $\n{S} = n$.
\end{proof}
 
\begin{cor}
\label{cyclepointing-introduction-correspondencecor}
For a class $\A$ of unlabeled graphs, $\A_n = \frac{1}{n}\Acpn$. Furthermore, if $\Gamma\Acp(z)$ is a Boltzmann sampler for $\Acp$, then $$\tilde{\Gamma}\A(z) = ((G, c)\leftarrow\Gamma\Acp(z);\textbf{ return }G);$$ is an unbiased sampler for $\A$, in the sense that $$\P_z[\,G\,\vb\n{G} = n] = \frac{1}{\A_n}.$$
\end{cor}
\begin{proof}
The first claim is a restatement of Theorem~\ref{cyclepointing-introduction-correspondencethm}, and the second claim follows by a similar argument to the one given in Lemma~\ref{analysis-challenges-correspondencelemma}.
\end{proof}

\subsection{Decomposition of cycle-pointed classes}
\label{cyclepointing-decomposition}

In order to analyze an unlabeled class $\A$ with Corollary~\ref{cyclepointing-introduction-correspondencecor}, it is first necessary to enumerate and build a Boltzmann sampler for the cycle-pointed class $\Acp$, and the first step in accomplishing this is to determine a symbolic specification for this class. While objects in $\A$ lack a distinguished feature at which they can be decomposed, objects in $\Acp$ have such a feature -- the marked cycle -- and it is this cycle which is used to decompose the graphs and develop a specification. 

The details of the decomposition technique vary depending on the nature of the class $\A$; as we are primarily interested in classes of trees, we will defer consideration of the decomposition techniques used for non-tree cycle-pointed graph classes, and instead focus on the techniques used for cycle-pointed classes of trees.

\subsubsection{Theory}
\label{cyclepointing-decomposition-theory}

Let $\A$ be an unlabeled class of trees, and let $\Acp$ be its cycle-pointed class. To decompose $\Acp$, we begin by partitioning it into two classes $$\Acp = \Av + \Ascp$$ where the first contains all elements of $\Acp$ whose marked cycle has length $1$, and the second, called the \emph{symmetric cycle-pointed} class of $\A$, contains all elements whose marked cycle has length at least $2$ (\textit{i.e.} all symmetric elements of $\Acp$). By considering a marked singleton cycle on a vertex as equivalent to marking the vertex itself, we may consider the class $\Av$ to be the vertex-rooted class of $\A$. 

Elements in the first class, $\Av$, are decomposed at their pointed vertex (similarly to how Cayley trees were decomposed in Section~\ref{analysis}), which is generally a straightforward process. Decomposing elements in the second class is more challenging, and for this we introduce two new concepts: the \emph{cycle-pointed substitution operator}, and the \emph{center of symmetry} of a cycle-pointed tree. This center of symmetry provides the distinguished location at which a symmetric cycle-pointed tree can be decomposed, and the cycle-pointed substitution operator allows us to decompose a cycle-pointed tree at its center of symmetry.

\paragraph{A note on notation.}
The concepts of cycle-pointed and symmetric cycle-pointed classes were introduced by Bodirsky~\etal\cite{cplong}, who use the notations $\A^{\circ}$ and $\A^{\circledast}$ (in \cite{cpshort}, $\Av$ is used in place of $\A^{\circ}$). We reintroduce them here with slightly different notations -- $\Acp$ and $\Ascp$ -- for the sake of clarity. We have chosen these both to provide a visual representation that we are marking \emph{cycles}, and to minimize the chance of confusion with other common uses of $\A^{\circ}$ and $\Av$ (e.g. vertex pointing).

\begin{definition}
Let $c$ be a cycle of length $k$, and let $c_1 = (v_{11}, \ldots, v_{1k}), \ldots, c_\ell = (v_{\ell1}, \ldots, v_{\ell k})$ be $\ell$ isomorphic copies of $c$ (on disjoint sets of vertices). Then the composition $c_1\circ \cdots \circ c_\ell$ of these cycles is defined as the cycle $$(v_{11}, v_{21}, \ldots, v_{\ell1}, v_{12}, v_{22}, \ldots, v_{\ell2}, \ldots, v_{1k}, v_{2k}, \ldots, v_{\ell k}).$$
\end{definition}

\begin{definition}
Let $\Bcp$ (or $\Bscp$, etc.) be a cycle-pointed class, and let $\C$ be a \emph{non} cycle-pointed class. Then the \emph{cycle-pointed substitution} $\Bcp\sub\C$ is defined as the cycle-pointed class containing all structures obtained as follows:
\begin{enumerate}
\item Let $P = (G, c)$ be an element of $\Bcp$, and let $c = (v_1, \ldots, v_k)$.
\item Replace the vertices of $G$ that are on $c$ with elements of $\Ccp$, and those that are not on $c$ with elements of $\C$, in a manner that respects at least one $c$-automorphism of $P$ (we say that a replacement \emph{respects} an automorphism $\sigma$ if for any vertex $v$ of $G$, the structures that replace $v$ and $\sigma(v)$ are isomorphic). Note that the vertices $v_1, \ldots, v_k$ of $c$ must be replaced with isomorphic copies $Q_1 = (H_1, c'_1), \ldots, Q_k = (H_k, c'_k)$ of the same cycle-pointed structure $Q = (H, c')$.
\item Let the marked cycle of the composed structure be $c'_1\circ \cdots\circ c'_k$.
\end{enumerate}
\end{definition}

\begin{figure}[!htb]
\begin{center}
\includegraphics[width=0.8\linewidth]{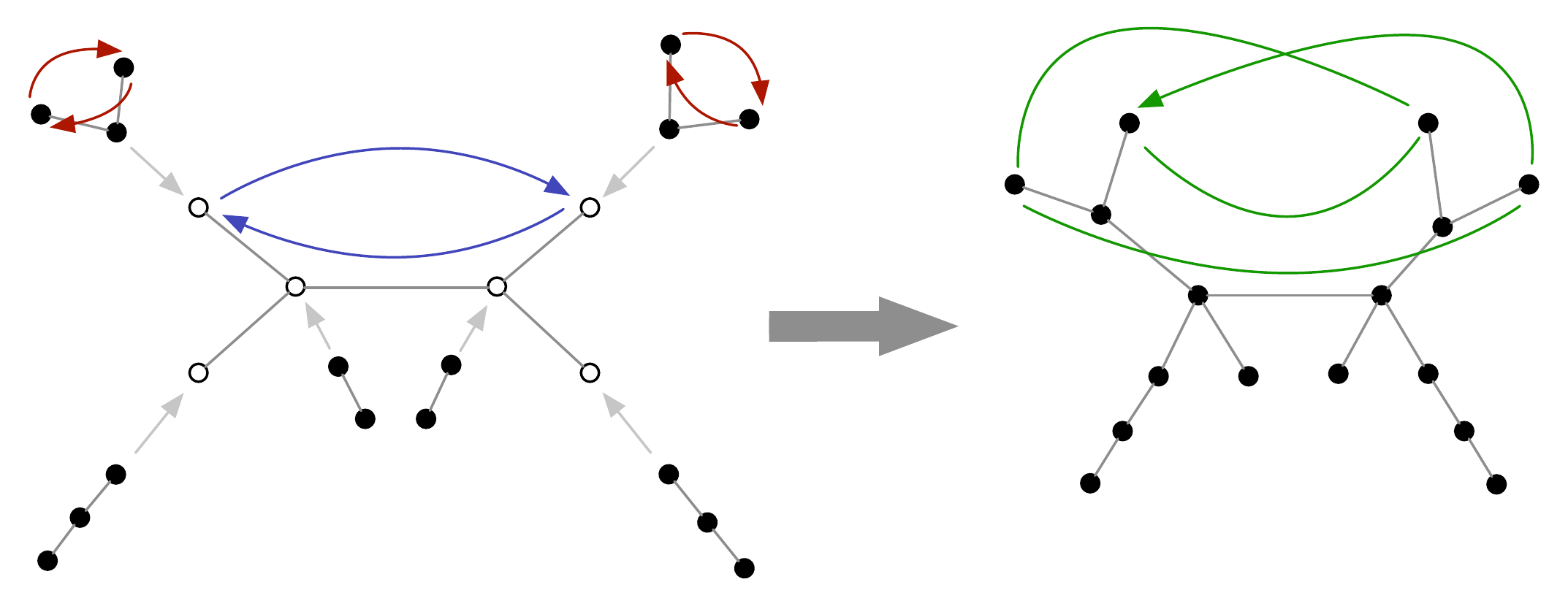}
\caption{A cycle-pointed substitution.}
\label{cyclepointing-substitution}
\end{center}
\end{figure}

\noindent For example, letting $\B$ be the class of trees, Figure~\ref{cyclepointing-substitution} illustrates the construction of an element of $\Bcp\sub\B$, where the element $P = (T, c)$ from $\Bcp$ is depicted with hollow vertices and a blue cycle, the cycles $c_1'$ and $c_2'$ are shown in red, and the composed cycle is shown in green. Note that in order for the substitution to respect a (in fact, the only) $c$-automorphism of $P$, the graphs substituted at the two bottom vertices of $G$ must be isomorphic, and the graphs substituted at the two middle vertices of $G$ must be as well.

\begin{lem}
Let $P = (T, c)$ be a symmetric cycle-pointed tree. For each consecutive pair of vertices $(v_i, v_{i+1})$ on $c$, call the path $p_i$ through $T$ that connects these vertices a \emph{connecting path} of $P$ (note: all indices are taken modulo $\n{c}$). Then all connecting paths of $P$ share the same middle element, called the \emph{center of symmetry} of $P$, which is either a vertex (if the paths all have even length) or an edge (if they all have odd length).
\end{lem}
\begin{proof}
Let $U$ be the subgraph of $T$ formed by the union of all the paths $p_i$, and let $\sigma$ be a $c$-automorphism of $T$ (such a $\sigma$ must exist by Definition~\ref{cyclepointing-introduction-cyclepointeddef}). Since $p_i$ is connected to $p_{i+1}$ at $v_{i+1}$ for each $i$, it follows that $p_i$ is connected to $p_j$ for any two indices $i$ and $j$, so $U$ is a connected subgraph of $T$. Hence, $U$ is a tree. Furthermore, since $c$ is a cycle of $\sigma$, $\sigma$ maps each connecting path to another connecting path (indeed it maps $p_i$ to $p_{i+1}$), so $U$ is fixed by $\sigma$ and hence $\sigma\vert_U$ ($\sigma$ \emph{restricted to} $U$) is an automorphism of $U$.

Let $x = c(U)$ be the center of $U$ (\textit{cf.} Definition~\ref{dissymmetry-overview-centerdef}). We claim that $x$ is the middle element of all connecting paths of $P$. First, since the iterative procedure which defines the center of a tree is invariant to automorphism, it follows that $x$ is fixed by any automorphism of $U$. Since $\sigma\vert_U$ is a $c$-automorphism of $U$, the cyclic group $\G = \left\langle\sigma\vert_U\right\rangle$ acts transitively on the set of vertices of $c$, and since $x$ is fixed by all elements of $\G$ it follows that $x$ is equidistant from all vertices of $c$. Furthermore, since $\sigma\vert_U$ is a $c$-automorphism of $U$, $\G$ acts transitively on the set of paths $p_i$, and since $x$ is fixed by all elements of $\G$ and is on at least one $p_i$ (after all, it is in $U$), it follows that $x$ is on all the $p_i$. Combining these two results, it follows that $x$ is the middle of all connecting paths of $P$.
\end{proof}

\noindent Informally, the center of symmetry of a cycle-pointed tree is obtained by first deleting all vertices and edges that are not on some path connecting two vertices of the marked cycle, and then taking the traditional center of the resulting tree. Thus it can be thought of as the center of the marked cycle, and it is fixed by any automorphism of the tree that respects the marked cycle. 

With these tools in hand, we use the following procedure to decompose a cycle-pointed class $\Acp$ of trees:

\begin{enumerate}
\item Write $\Acp$ as $\Acp = \Av + \Ascp$.
\item Decompose an object in $\Av$ at its pointed vertex.
\item Write $\Ascp = \Ascpv + \Ascpe$, where $\Ascpv$ ($\Ascpe$, respectively) contains the elements of $\Ascp$ whose center of symmetry is a vertex (edge, respectively).
\item Decompose a structure in $\Ascpv$ at its center of symmetry, $v$. Since $v$ is the center of symmetry and the marked cycle has length at least 2, $v$ is attached to at least two isomorphic copies of the same tree across which the marked cycle can be decomposed into isomorphic cycles, as well as possibly other trees with no marked cycle. Thus the neighbors of $v$ can be accounted for by substituting into $\Setscp$/$\scp{\Cyc}$ if $\A$ is a non-plane/plane class (with restrictions on the size of the set/cycle if $\A$ has restrictions on the degrees of its vertices). 
\item Decompose a structure in $\Ascpe$ at its center of symmetry, $e$. Since $e$ is the center of symmetry and the marked cycle of the structure has length at least 2, the trees attached to the two endpoints of $e$ must be two isomorphic copies of the same tree, and the marked cycle must be a composition of two isomorphic cycles on these trees. Thus, these can be accounted for by substituting into $\Setscp_2$.
\end{enumerate}
\begin{figure}[!htb]
\begin{center}
\includegraphics[width=0.7\linewidth]{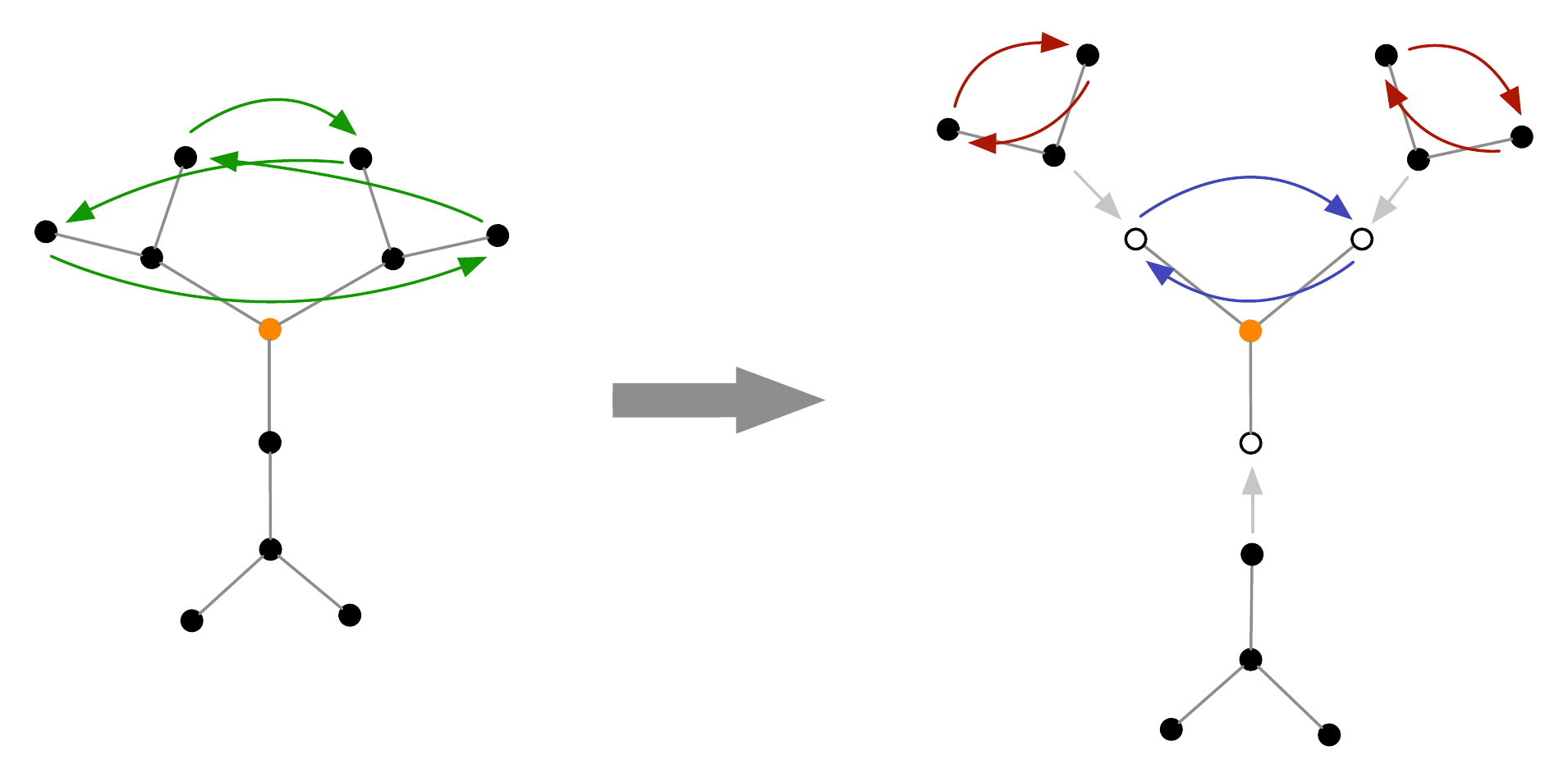}
\caption{A cycle-pointed tree decomposed at its center of symmetry, which is the orange vertex.}
\label{cyclepointing-vertexdecomposition}
\end{center}
\end{figure}
\begin{figure}[!htb]
\begin{center}
\includegraphics[width=0.8\linewidth]{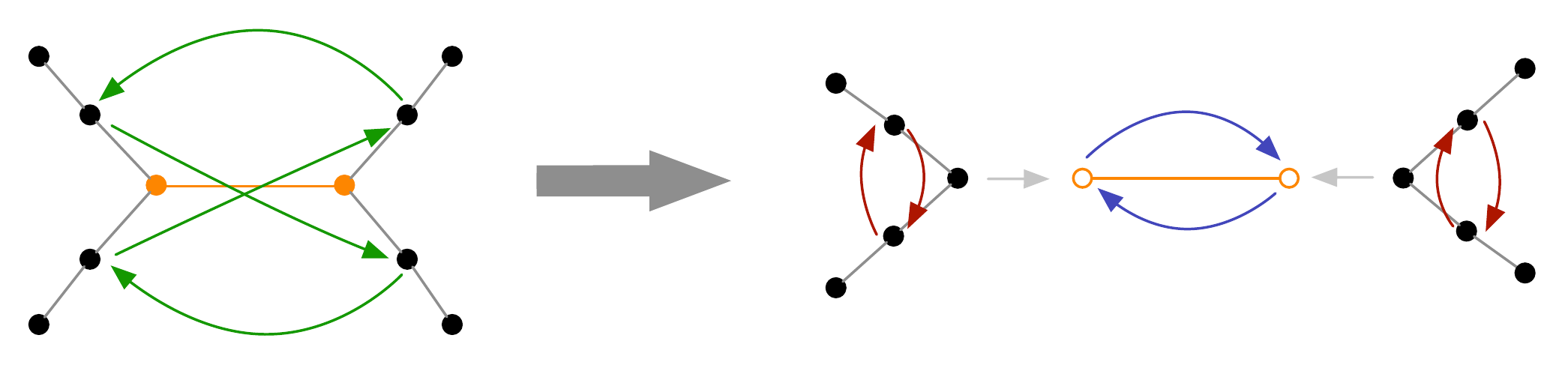}
\caption{A cycle-pointed tree decomposed at its center of symmetry, which is the orange edge.}
\label{cyclepointing-edgedecomposition}
\end{center}
\end{figure}

\subsubsection{Example}
\label{cyclepointing-decomposition-example}

Recall the class $\T$ of unlabeled, unrooted, non-plane 2-3 trees, which we analyzed using the dissymmetry theorem in Section~\ref{dissymmetry-example}. In order to decompose $\Tcp$, we begin by partitioning the class into its symmetric and non-symmetric parts: $$\Tcp = \Tv + \Tscp.$$ Objects in $\Tv$ are decomposed at their root (the vertex of the marked singleton cycle) -- indeed, an object in $\Tv$ is a root node with 1, 3, or 4 neighbors, where each neighbor is a node with 0, 2, or 3 other neighbors, so we have the decomposition $$\Tv = \Zcp\times\Set_{1, 3, 4}(\S)$$$$\S = \Z + \Z\times\Set_{2, 3}(\S).$$ Next, $\Tscp$ is partitioned as $$\Tscp = \Tscpv + \Tscpe.$$ Objects in $\Tscpv$ are decomposed at their center of symmetry, which is a vertex -- indeed, an object in $\Tscpv$ is a node (the center of symmetry) with a symmetric cycle-pointed set of either 3 or 4 elements of $\S$ (the center of symmetry cannot be a leaf since the marked cycle has length at least $2$), so we have the decomposition $$\Tscpv = \Z\times(\Setscp_{3, 4}\sub\S).$$ Finally, objects in $\Tscpe$ are decomposed at their center of symmetry, which is an edge -- indeed, an object in $\Tscpe$ is an edge (the center of symmetry) whose extremities form a symmetric cycle-pointed set of 2 elements of $\S$ (it must be symmetric since the marked cycle of objects in $\Tscpe$ has length at least 2), so we have the decomposition $$\Tscpe = \Setscp_2\sub\S.$$ Combining these results, we have the system of equations $$\Tcp = \Zcp\times\Set_{1, 3, 4}(\S) + \Z\times(\Setscp_{3, 4}\sub\S) + \Setscp_2\sub\S$$$$\S=\Z+\Z\times\Set_{2, 3}(\S).$$

\subsection{Enumeration}
\label{cyclepointing-enumeration}

\subsubsection{Theory}
\label{cyclepointing-enumeration-theory}

After obtaining a combinatorial specification for $\Acp$, the next step is to use this specification to derive a system of generating function equations for the OGF $\Acp(z)$, and from there to compute the coefficients $\Acpn$ and $\A_n = \frac{1}{n}\Acpn$. In order to translate such a specification to a generating function equation, we use \emph{cycle index sums}, first introduced by P\'{o}lya~\cite{polya}. These are generalizations of ordinary generating functions that capture all symmetries of the objects in a combinatorial class within a single power series over infinitely many variables.

\begin{definition}
\label{cyclepointing-enumeraton-theory-cycleindexsumdef}
If $\B$ is a non cycle-pointed class, the \emph{cycle index sum} of $\B$ is defined as the following formal power series over the variables $s_1, s_2, \ldots$: $$Z_{\B}(s_1, s_2, \ldots) = \sum_{G\in\B}\sum_{(G^\ell, \sigma)\in\Sym(G)}\frac{1}{\n{G}!}\prod_{i = 1}^{\n{G}}s_i^{c_i(\sigma)}$$ where $c_i(\sigma)$ is defined as the number of cycles in $\sigma$ of length $i$. If $\B$ is a cycle-pointed class, the \emph{cycle index sum} of $\B$ is defined as the following formal power series over the variables $s_1, s_2, \ldots, t_1, t_2, \ldots$: $$Z_{\B}(s_1, s_2, \ldots; t_1, t_2, \ldots) = \sum_{(G, c)\in\B}\sum_{(G^\ell, c^\ell, \sigma, v)\in\Rcsym((G, c))}\frac{1}{\n{G}!}t_{\n{c^\ell}}\prod_{i = 1}^{\n{G}}s_i^{c'_i(\sigma)}$$ where $c'_i(\sigma)$ is defined as the number of cycles in $\sigma$ of length $i$ that are not equal to $c^\ell$ (in a sense, the marked cycle $c^\ell$ is moved outside of the inner product and accounted for with a $t_i$ variable instead of an $s_i$ variable).
\end{definition}
\noindent Cycle index sums commute with sums and products of combinatorial classes, in the sense that $$Z_{\B + \C} = Z_{\B} + Z_{\C}\qquad\text{and}\qquad Z_{\B\times\C} = Z_{\B}\cdot Z_{\C}.$$ Furthermore, the cycle index sum is a generalization of its ordinary generating function, in the sense that replacing $s_i$ with $z^i$ and $t_i$ with $z^i$ in a cycle index sum results in the OGF of the class.

\begin{lem}
\label{cyclepointing-enumeration-theory-reductionlemma}
If $\B$ is a non cycle-pointed class, $Z_{\B}(z, z^2, \ldots) = \B(z)$. Similarly, if $\B$ is a cycle-pointed class, $Z_{\B}(z, z^2, \ldots; z, z^2, \ldots) = \B(z)$.
\end{lem}
\begin{proof}
First consider the case when $\B$ is non cycle-pointed. Then
\begin{align*}
Z_{\B}(z, z^2, \ldots) &= \sum_{G\in\B}\sum_{(G^\ell, \sigma)\in\Sym(G)}\frac{1}{\n{G}!}\prod_{i=1}^{\n{G}}z^{ic_i(\sigma)}\\
&=\sum_{G\in\B}\sum_{(G^\ell, \sigma)\in\Sym(G)}\frac{1}{\n{G}!}z^{\sum_{i=1}^{\n{G}}ic_i(\sigma)}\\
&=\sum_{G\in\B}\sum_{(G^\ell, \sigma)\in\Sym(G)}\frac{1}{\n{G}!}z^{\n{G}}.
\end{align*}
By Lemma~\ref{cyclepointing-introduction-symmetrylemma}, $G$ has $\n{G}!$ symmetries, so this expression reduces to $$\sum_{G\in\B}z^{\n{G}} = \B(z).$$ In the case when $\B$ is cycle-pointed, we have
\begin{align*}
Z_{\B}(z, z^2, \ldots; z, z^2, \ldots) &=  \sum_{(G, c)\in\B}\sum_{(G^\ell, c^\ell, \sigma, v)\in\Rcsym((G, c))}\frac{1}{\n{G}!}z^{\n{c^\ell}}\prod_{i = 1}^{\n{G}}z^{ic'_i(\sigma)}\\
&=  \sum_{(G, c)\in\B}\sum_{(G^\ell, c^\ell, \sigma, v)\in\Rcsym((G, c))}\frac{1}{\n{G}!}z^{\n{c^\ell} + \sum_{i=1}^{\n{G}}ic'_i(\sigma)}\\
&= \sum_{(G, c)\in\B}\sum_{(G^\ell, c^\ell, \sigma, v)\in\Rcsym((G, c))}\frac{1}{\n{G}!}z^{\n{G}}.
\end{align*}
By Lemma~\ref{cyclepointing-introduction-rootedcsymmetrylemma}, $(G, c)$ has $\n{G}!$ rooted $c$-symmetries, so this expression reduces to $$\sum_{(G, c)\in\B}z^{\n{G}} = \B(z).$$
\end{proof}
\noindent The cycle index sums for some common classes are shown in Table~\ref{cyclepointing-cycleindexsumstable} (we refer to Bodirsky~\etal\cite{cplong} for proof).
\begin{table}[!htb]
\begin{center}
\begin{tabular} {clN}
\toprule
Class & Cycle index sum & \\[10pt]
\midrule
$\eps$ & $Z_\eps = 0$ & \\[10pt]
$\Z$ & $Z_{\Z} = s_1$ & \\[10pt]
$\Set$ & $Z_{\Set} = \lgexp{\sum_{i = 1}^{\infty}\frac{s_i}{i}}$ & \\[12pt]
$\Seq$ & $Z_{\Seq} = \frac{1}{1-s_1}$ & \\[15pt]
$\Cyc$ & $Z_{\Cyc} = 1 + \sum_{i = 1}^{\infty}\frac{\phi(i)}{i}\log{\left(\frac{1}{1 - s_i}\right)}$ & \\[20pt]
$\Setcp$ & $Z_{\Setcp} = \left(\sum_{\l\geq1}t_\l\right)\cdot \lgexp{\sum_{i = 1}^{\infty}\frac{s_i}{i}}$ & \\[30pt]
$\Setscp$ & $Z_{\Setscp} = \left(\sum_{\l\geq2}t_\l\right)\cdot \lgexp{\sum_{i = 1}^{\infty}\frac{s_i}{i}}$ & \\[30pt]
\bottomrule
\end{tabular}
\caption{Some common cycle index sums.}
\label{cyclepointing-cycleindexsumstable}
\end{center}
\end{table}
In addition, the cycle index sums for size-restricted $\Set$, $\Seq$, and $\Cyc$ classes and their cycle-pointed versions can often be easily computed from Definition~\ref{cyclepointing-enumeraton-theory-cycleindexsumdef}. For example, $\Set_2$ has one element $\{\bullet, \bullet\}$ with two symmetries, the identity $\sigma_{\mathrm{id}}$ and the permutation $\sigma_{\mathrm{sw}}$ that swaps the two atoms, so $$Z_{\Set_2} = \frac{1}{2!}\prod_{i = 1}^2s_i^{c_i(\sigma_{\mathrm{id}})} + \frac{1}{2!}\prod_{i=1}^2s_i^{c_i(\sigma_{\mathrm{sw}})} = \frac{1}{2}s_1^2 + \frac{1}{2}s_2.$$ Similarly, $\Setcp_2$ has two elements each with two rooted $c$-symmetries, where the first element has a marked cycle of length $1$ and the second has a marked cycle of length $2$, so its cycle index sum is $$Z_{\Setcp_2} = 2\cdot\frac{1}{2!}t_1\prod_{i=1}^2s_i^{c'_i(\sigma_{\mathrm{id}})} + 2\cdot\frac{1}{2!}t_2\prod_{i=1}^2s_i^{c'_i(\sigma_{\mathrm{sw}})} = s_1t_1 + t_2.$$ Since $\Setscp_2$ excludes the element of $\Setcp_2$ whose marked cycle has length $1$, its cycle index sum is simply $Z_{\Setscp_2} = t_2.$

The power of cycle index sums is that they provide us with transfer theorems for the substitution and cycle-pointed substitution operators, as shown in Table~\ref{cyclepointing-transfertheoremstable}~\cite{cpshort, cplong}.
\begin{table}[!htb]
\begin{center}
\begin{tabular} {clN}
\toprule
Class & OGF & \\[10pt]
\midrule
$\B + \C$ & $\B(z) + \C(z)$ & \\[10pt]
$\B\times\C$ & $\B(z)\cdot\C(z)$ & \\[10pt]
$\B(\C)$ & $Z_{\B}(\C(z), \C(z^2), \C(z^3), \ldots)$ & \\[20pt]
$\B\sub\C$ & $Z_{\B}(\C(z), \C(z^2), \ldots; z\C'(z), z^2\C'(z^2), \ldots)$ & \\[16pt]
\bottomrule
\end{tabular}
\caption{Operator transfer theorems for unlabeled classes.}
\label{cyclepointing-transfertheoremstable}
\end{center}
\end{table}
With these rules, we can use the following procedure to develop a system of equations for $\Acp(z)$ and thereby determine the enumeration of $\A$:
\begin{enumerate}
\item Develop a combinatorial specification for $\Acp$ (\textit{cf.} Section~\ref{cyclepointing-decomposition-theory}).
\item For each class that appears as the first argument of a substitution or cycle-pointed substitution, determine its cycle index sum.
\item For all other classes that appear in the specification, determine their ordinary generating function.
\item Apply the rules in Table~\ref{cyclepointing-transfertheoremstable} to translate the specification to a system of generating function equations.
\item Compute the coefficients $\Acpn$ and $\A_n = \frac{1}{n}\Acpn$ from this system of equations. A tool such as $\tt{combstruct}$ is often helpful at this step in order to determine the coefficients of recursively specified generating functions.
\end{enumerate}

\subsubsection{Example}
\label{cyclepointing-enumeration-example}
Recall the decomposition of the class $\Tcp$ from Section~\ref{cyclepointing-decomposition-example}:$$\Tcp = \Zcp\times\Set_{1, 3, 4}(\S) + \Z\times(\Setscp_{3, 4}\sub\S) + \Setscp_2\sub\S$$$$\S=\Z+\Z\times\Set_{2, 3}(\S).$$ The cycle index sums for $\Set_{1, 3, 4}, \Set_{2, 3}, \Setscp_2, $ and $\Setscp_{3, 4}$ are\footnote{For example, $Z_{\Set_{2, 3}}$ can be derived as follows. The single element of $\Set_3$ has six symmetries: the identity, which contributes a $s_1^3$ term (one $s_1$ for each $1$-cycle) to the cycle index sum; three symmetries that swap a pair of vertices, each of which contributes an $s_1s_2$ term (for the $1$-cycle and the $2$-cycle); and two 3-cycle symmetries, each of which contributes an $s_3$ term. Summing these and dividing by $3!$ gives the expression $Z_{\Set_3} = \frac{1}{6}s_1^3 + \frac{1}{2}s_1s_2 + \frac{1}{3}s_3$, and adding this to the expression for $Z_{\Set_2}$ from Section~\ref{cyclepointing-enumeration-theory} gives the stated expression for $Z_{\Set_{2, 3}}$.\\\indent As a second example, $Z_{\Setscp_{3, 4}}$ can be derived as follows. $\Setscp_3$ has two elements; one with a marked $2$-cycle, and the other with a marked $3$-cycle. The first has six rooted $c$-symmetries, corresponding to the three possible labelings and two possible rootings on the marked cycle, and each contributes a $t_2s_1$ term to the cycle index sum (for the marked $2$-cycle and the unmarked $1$-cycle). The second also has six rooted $c$-symmetries, corresponding to the two possible labelings and three possible rootings on the marked cycle, and each contributes a $t_3$ term. Summing these and dividing by $3!$ gives the expression $Z_{\Setscp_3} = t_2s_1 + t_3$. \\\indent In addition, $\Setscp_4$ has three elements; one with a marked $2$-cycle, one with a marked $3$-cycle, and one with a marked $4$-cycle. The first has $24$ rooted $c$-symmetries, $12$ of which have an unmarked $2$-cycle and contribute a $t_2s_2$ term to the cycle index sum, and the other $12$ of which have two unmarked $1$-cycles and contribute a $t_2s_1^2$ term. The second also has $24$ rooted $c$-symmetries, corresponding to the eight possible labelings and three possible rootings on the marked cycle, and each contributes a $t_3s_1$ term (for the marked $3$-cycle and the unmarked $1$-cycle). The third also has $24$ rooted $c$-symmetries, corresponding to the six possible labelings and four possible rootings on the marked cycle, and each contributes a $t_4$ term. Summing these and dividing by $4!$ gives the expression $Z_{\Setscp_4} = \frac{1}{2}t_2s_1^2 + \frac{1}{2}t_2s_2 + t_3s_1 + t_4$. Finally, summing the expressions for $Z_{\Setscp_3}$ and $Z_{\Setscp_4}$ results in the stated expression for $Z_{\Setscp_{3, 4}}$} $$Z_{\Set_{1, 3, 4}} = s_1 + \frac{1}{6}s_1^3 + \frac{1}{2}s_1s_2 + \frac{1}{3}s_3 + \frac{1}{24}s_1^4 + \frac{1}{4}s_1^2s_2 + \frac{1}{8}s_2^2 + \frac{1}{3}s_1s_3 + \frac{1}{4}s_4$$$$Z_{\Set_{2, 3}} = \frac{1}{2}s_1^2 + \frac{1}{2}s_2 + \frac{1}{6}s_1^3 + \frac{1}{2}s_1s_2 + \frac{1}{3}s_3$$$$Z_{\Setscp_2} = t_2$$$$Z_{\Setscp_{3, 4}} = t_2s_1 + t_3 + \frac{1}{2}t_2s_1^2 + \frac{1}{2}t_2s_2 + t_3s_1 + t_4$$ so by Table~\ref{cyclepointing-transfertheoremstable} we have the following system of ordinary generating function equations:\begin{align*}
\Tcp(z) &= z\cdot\left[\S(z) + \frac{1}{6}\S(z)^3 + \frac{1}{2}\S(z)\S(z^2) + \frac{1}{3}\S(z^3) + \frac{1}{24}\S(z)^4 + \frac{1}{4}\S(z)^2\S(z^2) + \frac{1}{8}\S(z^2)^2 + \frac{1}{3}\S(z)\S(z^3) + \frac{1}{4}\S(z^4)\right] \\
&+ z\cdot\left[z^2\S'(z^2)\S(z) + z^3\S'(z^3) + \frac{1}{2}z^2\S'(z^2)\S(z)^2 + \frac{1}{2}z^2\S'(z^2)\S(z^2) + z^3\S'(z^3)\S(z) + z^4\S'(z^4)\right] \\
&+ z^2\S'(z^2)
\end{align*}
$$\S(z) = z + z\cdot\left[\frac{1}{2}\S(z)^2 + \frac{1}{2}\S(z^2) + \frac{1}{6}\S(z)^3 + \frac{1}{2}\S(z)\S(z^2) + \frac{1}{3}\S(z^3)\right]$$
Using $\tt{combstruct}$, we can compute $\S(z)$ to arbitrary accuracy: $$\S(z) = z + z^3 + z^4 + z^5 + 2z^6 + 3z^7 + 5z^8 + 8z^9 + 14z^{10} + 23z^{11} + 40z^{12} + 70z^{13} + 122z^{14} + 217z^{15} + \ldots,$$ and from the expression for $\Tcp(z)$, by differentiating $\S(z)$ and substituting $z\ra z^i$ as appropriate, we can compute $\Tcp(z)$ to arbitrary accuracy: $$\Tcp(z) = 2z^2 + 4z^4 + 5z^5 + 6z^6 + 7z^7 + 16z^8 + 18z^9 + 40z^{10} + 55z^{11} + 96z^{12} + 156z^{13} + 280z^{14} + 435z^{15} + \ldots.$$ Finally, from Corollary~\ref{cyclepointing-introduction-correspondencecor} we have $$\T(z) = \int \frac{\Tcp(z)}{z}dz = z^2 + z^4 + z^5 + z^6 + z^7 + 2z^8 + 2z^9 + 4z^{10} + 5z^{11} + 8z^{12} + 12z^{13} + 20z^{14} + 29z^{15} + \ldots.$$
We note that this agrees with the enumeration of $\A$ computed in Section~\ref{dissymmetry-example}.

\subsection{Boltzmann samplers}
\label{cyclepointing-sampler}

\subsubsection{Theory}
\label{cyclepointing-sampler-theory}

In this section we describe how to build a Boltzmann sampler for $\Acp$ from its combinatorial specification, which (by Corollary~\ref{cyclepointing-introduction-correspondencecor}) then provides an unbiased sampler for $\A$. To accomplish this we use \emph{P\'{o}lya-Boltzmann samplers}, which generalize Boltzmann samplers in the same manner that cycle index sums generalize ordinary generating functions.

\begin{definition}
\label{cyclepointing-sampler-theory-polyaboltzmanndef}
Suppose that $\B$ is a non cycle-pointed class. For fixed parameters $(s_1, s_2, \ldots)$ at which $Z_{\B}(s_1, s_2, \ldots)$ converges, a \emph{P\'{o}lya-Boltzmann sampler} for $\B$ is a random generator $\Gamma Z_{\B}(s_1, s_2, \ldots)$ that draws an object $(G^\ell, \sigma)$ from $\bigcup_{G\in\B}\Sym(G)$ with probability $$\P_{(s_1, s_2, \ldots)}[G^\ell, \sigma] = \frac{1}{Z_{\B}(s_1, s_2, \ldots)}\frac{1}{\n{G^\ell}!}\prod_{i = 1}^{\n{G^\ell}}s_i^{c_i(\sigma)}.$$ If $\B$ is a cycle-pointed class, then for fixed parameters $(s_1, s_2, \ldots; t_1, t_2, \ldots)$ at which $Z_{\B}(s_1, s_2, \ldots; t_1, t_2, \ldots)$ converges, a \emph{P\'{o}lya-Boltzmann sampler} for $\B$ is a random generator $\Gamma Z_{\B}(s_1, s_2, \ldots; t_1, t_2, \ldots)$ that draws an object $(G^\ell, c^\ell, \sigma, v)$ from $\bigcup_{(G, c)\in\B}\Rcsym((G, c))$ with probability $$\P_{(s_1, s_2, \ldots; t_1, t_2, \ldots)}[G^\ell, c^\ell, \sigma, v] = \frac{1}{Z_{\B}(s_1, s_2, \ldots; t_1, t_2, \ldots)}\frac{1}{\n{G^\ell}!}t_{\n{c^\ell}}\prod_{i = 1}^{\n{G^\ell}}s_i^{c'_i(\sigma)}.$$
\end{definition} 
\noindent A P\'{o}lya-Boltzmann sampler for a class $\B$ is a generalization of the ordinary Boltzmann sampler, in the sense that sampling at the parameters $s_i = z^i$ and $t_i = z^i$ and returning the underlying structure of the symmetry/rooted $c$-symmetry results in a Boltzmann sampler for $\B$.

\begin{figure}[!htb]
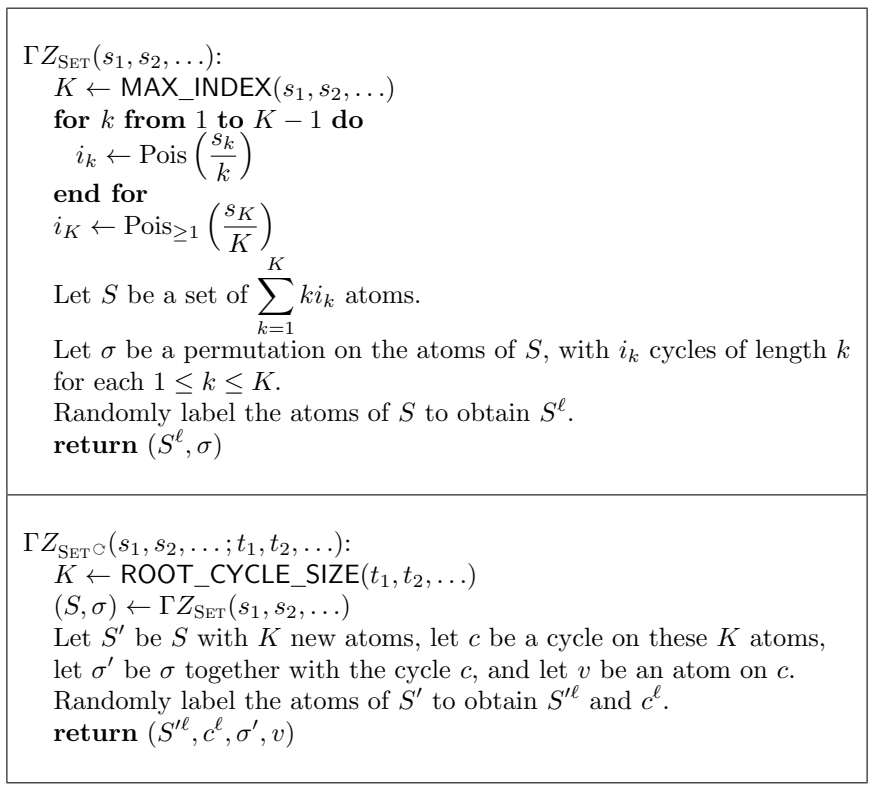

\begin{center}
\begin{tabular} {|p{11cm}|N}
\hline
$\Gamma Z_{\Set}(s_1, s_2, \ldots)$:\newline
\hspace*{4mm}$K\leftarrow$ \textsf{MAX\TextUnderscore{}INDEX}$(s_1, s_2, \ldots)$\newline
\hspace*{4mm}\textbf{for }$k$ \textbf{from} $1$ \textbf{to} $K-1$ \textbf{do}\newline
\hspace*{7mm}$i_k\leftarrow\text{Pois}\left(\frac{s_k}{k}\right)$\newline
\hspace*{4mm}\textbf{end for}\newline
\hspace*{4mm}$i_K\leftarrow\text{Pois}_{\geq1}\left(\frac{s_K}{K}\right)$\newline
\hspace*{4mm}Let $S$ be a set of $\sum_{k = 1}^Kki_k$ atoms.\newline
\hspace*{4mm}Let $\sigma$ be a permutation on the atoms of $S$, with $i_k$ cycles of length $k$\newline
\hspace*{4mm}for each $1\leq k\leq K.$\newline
\hspace*{4mm}Randomly label the atoms of $S$ to obtain $S^\ell$.\newline
\hspace*{4mm}\textbf{return} $(S^\ell, \sigma)$\newline
& \\[25pt]
\hline
$\Gamma Z_{\Setcp}(s_1, s_2, \ldots; t_1, t_2, \ldots)$:\newline
\hspace*{4mm}$K\leftarrow$ \textsf{ROOT\TextUnderscore{}CYCLE\TextUnderscore{}SIZE}$(t_1, t_2, \ldots)$\newline
\hspace*{4mm}$(S, \sigma)\leftarrow\Gamma Z_{\Set}(s_1, s_2, \ldots)$\newline
\hspace*{4mm}Let $S'$ be $S$ with $K$ new atoms, let $c$ be a cycle on these $K$ atoms,\newline
\hspace*{4mm}let $\sigma'$ be $\sigma$ together with the cycle $c$, and let $v$ be an atom on $c$.\newline
\hspace*{4mm}Randomly label the atoms of $S'$ to obtain $S'^{\ell}$ and $c^\ell$.\newline
\hspace*{4mm}\textbf{return} $(S'^{\ell}, c^\ell, \sigma', v)$\newline 
& \\[25pt]
\hline
\end{tabular}
\caption{P\'{o}lya-Boltzmann samplers for $\protect\Set$ and $\protect\Setcp$.\protect\footnotemark}
\label{cyclepointing-setpolyaboltzmann}
\end{center}
\end{figure}

\begin{lem}
\label{cyclepointing-sampler-theory-reductionlemma}
If $\B$ is a non cycle-pointed class, $$\Gamma\B(z) = ((G^{\ell}, \sigma)\leftarrow\Gamma Z_{\B}(z, z^2, \ldots);\textbf{ return }G)$$ is a Boltzmann sampler for $\B$. Similarly, if $\B$ is a cycle-pointed class, $$\Gamma\B(z) = ((G^{\ell}, c^{\ell}, \sigma, v)\leftarrow\Gamma Z_{\B}(z, z^2, \ldots; z, z^2, \ldots);\textbf{ return }(G, c))$$ is a Boltzmann sampler for $\B$.
\end{lem}
\begin{proof}
If $\B$ is not cycle-pointed, then by Lemma~\ref{cyclepointing-enumeration-theory-reductionlemma} the probability of returning $(G^\ell, \sigma)\in\Sym(G)$ from $\Gamma\B(z)$ is $$\frac{1}{\B(z)}\frac{1}{\n{G}!}z^{\sum_{i = 1}^{\n{G}}ic_i(\sigma)} = \frac{1}{\B(z)}\frac{1}{\n{G}!}z^{\n{G}}.$$ Since $G$ has $\n{G}!$ symmetries (\textit{cf.} Lemma~\ref{cyclepointing-introduction-symmetrylemma}), the probability of returning one such symmetry from $\Gamma Z_{\B}(z, z^2, \ldots)$, and hence of returning $G$ from $\Gamma\B(z)$, is $$\frac{z^{\n{G}}}{\B(z)}.$$ A parallel argument holds in the case when $\B$ is cycle-pointed, by using Lemma~\ref{cyclepointing-introduction-rootedcsymmetrylemma} in place of Lemma~\ref{cyclepointing-introduction-symmetrylemma}.
\end{proof}
\noindent P\'{o}lya-Boltzmann samplers for various common classes are described by Bodirsky~\etal\cite{cplong}. We do not list them here for the sake of brevity, but for the purpose of example we include the P\'{o}lya-Boltzmann samplers for $\Set$ and $\Setcp$ in Figure~\ref{cyclepointing-setpolyaboltzmann}.

\footnotetext{\textsf{MAX\TextUnderscore{}INDEX}$(s_1, s_2, \ldots)$ is a random generator over the integers $k\geq 1$ for the distribution $$\P[K\leq k] = \frac{1}{Z_{\Set}(s_1, s_2, \ldots)}\prod_{i\leq k}\lgexp{\frac{s_i}{i}},$$ and \textsf{ROOT\TextUnderscore{}CYCLE\TextUnderscore{}SIZE}$(t_1, t_2, \ldots)$ is a random generator over the integers $k\geq 1$ for the distribution $$\P[K =  k] = \frac{t_k}{\sum_{i = 1}^{\infty}t_i}.$$ 

In practice, these distributions are drawn from by inversion sampling. The infinite sums present in the probability expressions (recall that $Z_{\Set}$ has an infinite sum) can be computed to arbitrary precision using a tool such as Maple, assuming that the values of $s_i$ and $t_i$ are explicit functions of $i$.}

Just as cycle index sums provide transfer theorems for the substitution and cycle-pointed substitution operators, P\'{o}lya-Boltzmann samplers allow us to build Boltzmann samplers for classes specified with these operators. To compute a Boltzmann sampler for a class specified by a substitution or cycle-pointed substitution, it suffices to have a P\'{o}lya-Boltzmann sampler for the class that appears as the first argument of the substitution and an ordinary Boltzmann sampler for the class that appears as the second argument. The corresponding algorithmic rules are shown in Table~\ref{cyclepointing-polyaboltzmanntransfertable}.
\begin{table}[!htb]
\begin{center}
\begin{tabular} {cp{10.5cm}N}
\toprule
Class & Boltzmann sampler & \\[10pt]
\midrule
$\B + \C$ & \textbf{if} Bern$\left(\frac{\B(z)}{\B(z) + \C(z)}\right)$ \textbf{then} $\Gamma\B(z)$ \textbf{else} $\Gamma\C(z)$ & \\[20pt]
$\B\times\C$ & $(\Gamma\B(z), \Gamma\C(z))$ & \\[10pt]
$\B(\C)$ & $(B, \sigma)\leftarrow\Gamma Z_{\B}(\C(z), \C(z^2), \ldots)$ \newline 
\textbf{for each} cycle $c$ of $\sigma$ \textbf{do}\newline
\hspace*{2mm} $C\leftarrow\Gamma\C(z^{\n{c}})$\newline
\hspace*{2mm} Replace each atom of $c$ by a copy of $C$\newline
\textbf{end for}\newline
\textbf{return} the resulting structure\newline
& \\[20pt]
$\B\sub\C$ & $((B, c), \sigma, v)\leftarrow\Gamma Z_{\B}(\C(z), \C(z^2), \ldots; z\C'(z), z^2\C'(z^2), \ldots)$ \newline 
\textbf{for each} unmarked cycle $c'$ of $\sigma$ \textbf{do}\newline
\hspace*{2mm} $C\leftarrow\Gamma\C(z^{\n{c'}})$\newline
\hspace*{2mm} Replace each atom of $c'$ by a copy of $C$\newline
\textbf{end for}\newline
$(C, \tilde{c})\leftarrow\Gamma\Ccp(z^{\n{c}})$\newline
Replace each atom of $c$ by a copy of $(C, \tilde{c})$\newline
Mark the cycle that is the composition of all $\n{c}$ copies of $\tilde{c}$\newline
\textbf{return} the resulting cycle-pointed structure\newline
& \\[20pt]
\bottomrule
\end{tabular}
\caption{Boltzmann sampler constructions for operators on unlabeled classes.}
\label{cyclepointing-polyaboltzmanntransfertable}
\end{center}
\end{table}

With these rules, we can use the following procedure to develop a Boltzmann sampler for $\Acp(z)$ and thereby obtain an unbiased sampler for $\A$:\begin{enumerate}
\item Develop a combinatorial specification for $\Acp$ (\textit{cf.} Section~\ref{cyclepointing-decomposition-theory}).
\item For each class that appears as the first argument of a substitution or cycle-pointed substitution, compute a P\'{o}lya-Boltzmann sampler for it.
\item For all other classes that appear in the specification, compute an ordinary Boltzmann sampler.
\item Apply the rules in Table~\ref{cyclepointing-polyaboltzmanntransfertable} to translate the specification to a Boltzmann sampler for $\Acp$.
\item Apply Corollary~\ref{cyclepointing-introduction-correspondencecor} to obtain an unbiased sampler for $\A$.
\end{enumerate}

\subsubsection{Example}
\label{cyclepointing-sampler-example}
Again, recall the decomposition of the class $\Tcp$ from Section~\ref{cyclepointing-decomposition-example}:$$\Tcp = \Zcp\times\Set_{1, 3, 4}(\S) + \Z\times(\Setscp_{3, 4}\sub\S) + \Setscp_2\sub\S$$$$\S=\Z+\Z\times\Set_{2, 3}(\S).$$ From the P\'{o}lya-Boltzmann samplers for $\Set_k$ and $\Setscp_k$ given by Bodirsky~\etal\cite{cplong}, we may apply the substitution rules in Table~\ref{cyclepointing-polyaboltzmanntransfertable} to derive Boltzmann samplers for $\Set_k(\A)$, $\Setcp_k\sub\A$, and $\Setscp_k\sub\A$ (for arbitrary $k$ and $\A$). By combining these with the sum and product rules in Table~\ref{cyclepointing-polyaboltzmanntransfertable}, we may compute Boltzmann samplers for $\S$, $\Scp$, and $\Tcp$. By Corollary~\ref{cyclepointing-introduction-correspondencecor}, running $\Gamma\Tcp(z)$ and forgetting the marked cycle provides an unbiased sampler for the class $\T$. The expected size of an object drawn from this sampler when the parameter is taken at the singularity is $39.710$, compared to the corresponding value of $4.224$ for a Boltzmann sampler of $\T$ (\textit{cf.} Section~\ref{dissymmetry-example}). We provide pseudocode for these samplers in Appendix~\ref{appendices-twothreetrees}.

\section{Two interesting classes of graphs}
\label{example}

In this section, we apply the techniques of the previous sections to analyze two interesting classes of unlabeled graphs -- distance-hereditary graphs, and three-leaf power graphs. These were previously studied from an analytic combinatorics point of view by Chauve~\etal\cite{chauvelumbrosofusy}, where the authors develop grammars for these classes of graphs using the dissymmetry theorem and then derive exact enumerations and asymptotics from these grammars (we review this work in Section~\ref{example-dissymmetry}). We aim to extend this result by analyzing the same graph classes using cycle pointing, with which we obtain enumerations that match the ones provided by Chauve~\etal and build unbiased samplers for these two classes of graphs.

\subsection{Distance-hereditary graphs}
\label{example-dh}

\begin{definition} 
For a graph $G$ and vertices $u, v\in V(G)$, the \emph{distance} between $u$ and $v$ -- denoted $d_G(u, v)$ -- is defined as the length of the shortest path between $u$ and $v$ (or $\infty$ if $u$ and $v$ are in different connected components of $G$).
\end{definition}

\begin{definition} 
A subgraph $H\leq G$ is said to be \emph{induced} if, for any $u, v\in V(H)$ such that $uv\in E(G)$, $uv\in E(H)$ as well.
\end{definition}

\begin{definition}\cite{graphclassesdh} 
A connected graph $G$ is said to be \emph{distance-hereditary} if, for any connected induced subgraph $H\leq G$ and any vertices $u, v\in V(H)$, $d_H(u, v) = d_G(u, v)$.
\end{definition}
\noindent Distance-hereditary graphs were first introduced in 1977 by Howorka~\cite{howorka}, and many properties of these graphs have been discovered since then. For example, various optimization problems that are NP-hard in the general case, such as finding a Hamiltonian cycle or a maximal clique, can be solved in polynomial time for distance-hereditary graphs~\cite{muller, hsieh, cogis}. One alternate characterization that is particularly useful for recognizing a distance-hereditary graph is the following:

\begin{lem}
A connected graph $G$ is distance-hereditary iff every cycle of length five or higher has at least one pair of crossing diagonals.
\end{lem}

\subsection{Three-leaf power graphs}
\label{example-3lp}
\begin{definition} \cite{graphclassesklp}
For an integer $k > 0$, a connected graph $G$ is said to be a \emph{$k$-leaf power} if there exists a tree $T$ with $V(T) = V(G)$ such that for any vertices $u, v\in V(G)$, $uv\in E(G)$ iff $d_T(u, v)\leq k$.
\end{definition}

\noindent $k$-leaf power graphs were first introduced in 2002 by Nishimura~\etal\cite{nishimura}, who were interested in building \emph{phylogenetic trees} that reconstruct the evolutionary history of a set of species or genes. We study the particular case $k = 3$, called three-leaf power graphs.

\subsection{Split decomposition}
Recently, various techniques have emerged for analyzing classes of graphs by using bijective representations of these graphs in terms of trees -- see, for example, the correspondence between binary trees and edge-rooted 3-connected planar graphs used by Fusy~\cite{fusy}, or the correspondence between Apollonian networks and $k$-trees used by Darrasse~\cite{darrassecorr}. In order to study distance-hereditary and three-leaf power graphs, we employ the representation of these classes developed by Gioan and Paul~\cite{gioanpaul}, who use the technique of \emph{split decomposition} to characterize distance-hereditary and three-leaf power graphs in terms of \emph{graph-labeled trees}. We begin by giving an overview of this characterization, which we use in the following sections to enumerate and generate graphs from these classes.

\begin{figure}[!htb]
\begin{center}
\includegraphics[width=0.57\linewidth]{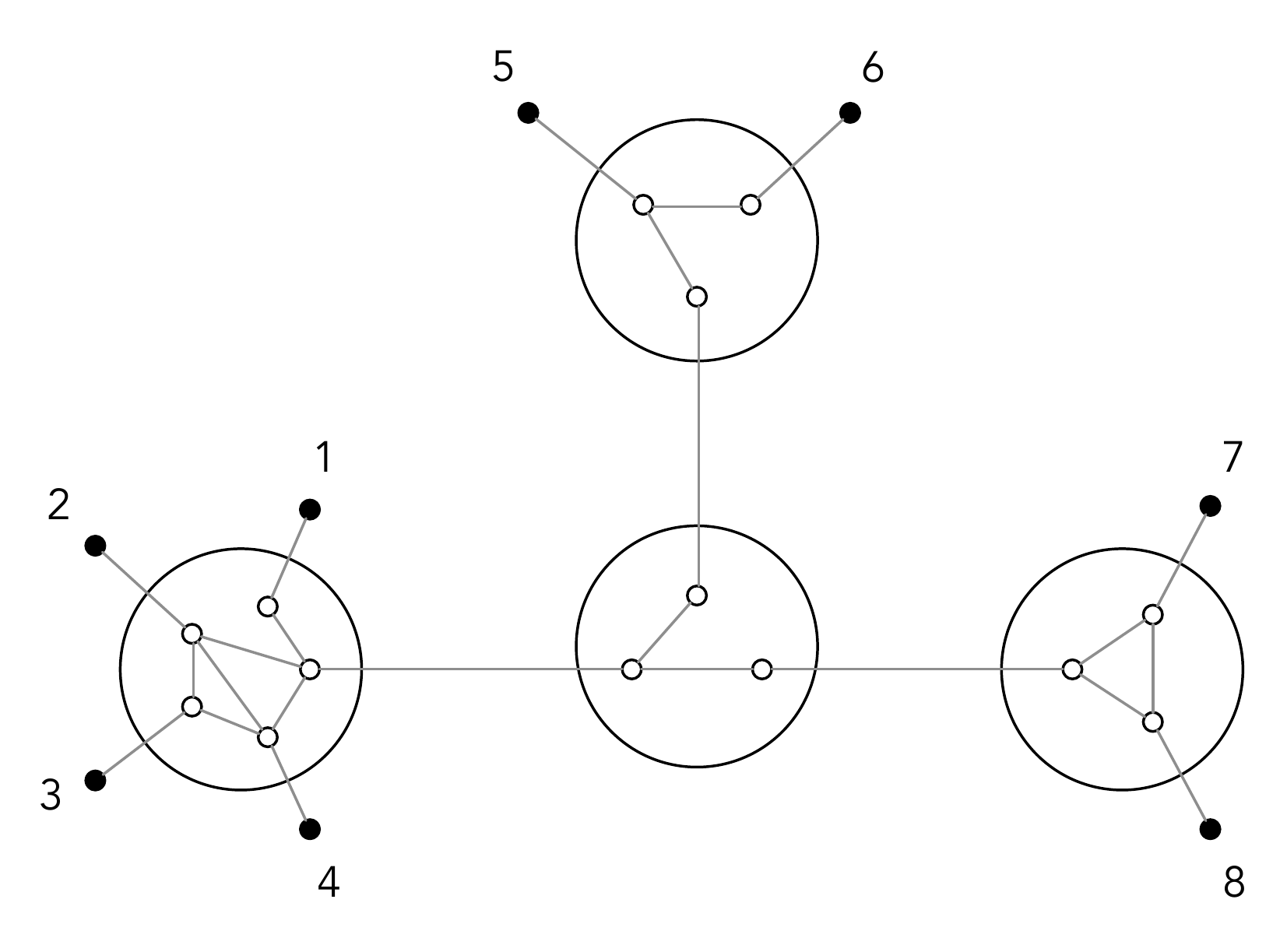}
\caption{A graph-labeled tree.}
\label{example-graphlabeledtree}
\end{center}
\end{figure}

\begin{definition}
A \emph{graph-labeled tree} $(T, \F)$ is a tree $T$ in which every internal node $v$ of degree $k$ is labeled by a graph $G_v\in\F$ on $k$ vertices, called \emph{marker vertices}, such that there is a bijection $\rho_v$ from the tree-edges of $T$ incident to $v$ to the marker vertices of $G_v$.
\end{definition}
\noindent For example, in Figure~\ref{example-graphlabeledtree} the internal nodes of $T$ are denoted with large circles, the marker vertices are denoted with small hollow circles, the leaves of $T$ are denoted with small solid circles, and the bijection $\rho_v$ is denoted by each edge that crosses the boundary of an internal node and ends at a marker vertex. (Note: the node labels are only for convenience in discussing the vertices - the tree itself is unlabeled.)

\begin{definition}
Let $(T, \F)$ be a graph-labeled tree and let $l, l'\in V(T)$ be leaves of $T$. We say that $\l$ and $\l'$ are \emph{accessible} (or equivalently, $\l$ is \emph{accessible} from $\l'$) if there exists a path from $\l$ to $\l'$ in $T$ such that for any adjacent edges $e = uv$ and $e' = vw$ on the path, $\rho_v(e)\rho_v(e')\in E(G_v)$.
\end{definition}
\noindent Informally, $\l$ and $\l'$ are accessible if it is possible to draw a path through the graph-labeled tree from $\l$ to $\l'$ that uses at most one interior edge from each graph label $G_v$. For example, in Figure~\ref{example-graphlabeledtree}, leaf $1$ is accessible from leaves $5$, $7$, and $8$, leaf $2$ is accessible from leaves $3$, $4$, $5$, $7$, and $8$, and leaf $3$ is only accessible from leaves $2$ and $4$.

\begin{figure}[!htb]
\begin{center}
\includegraphics[width=0.57\linewidth]{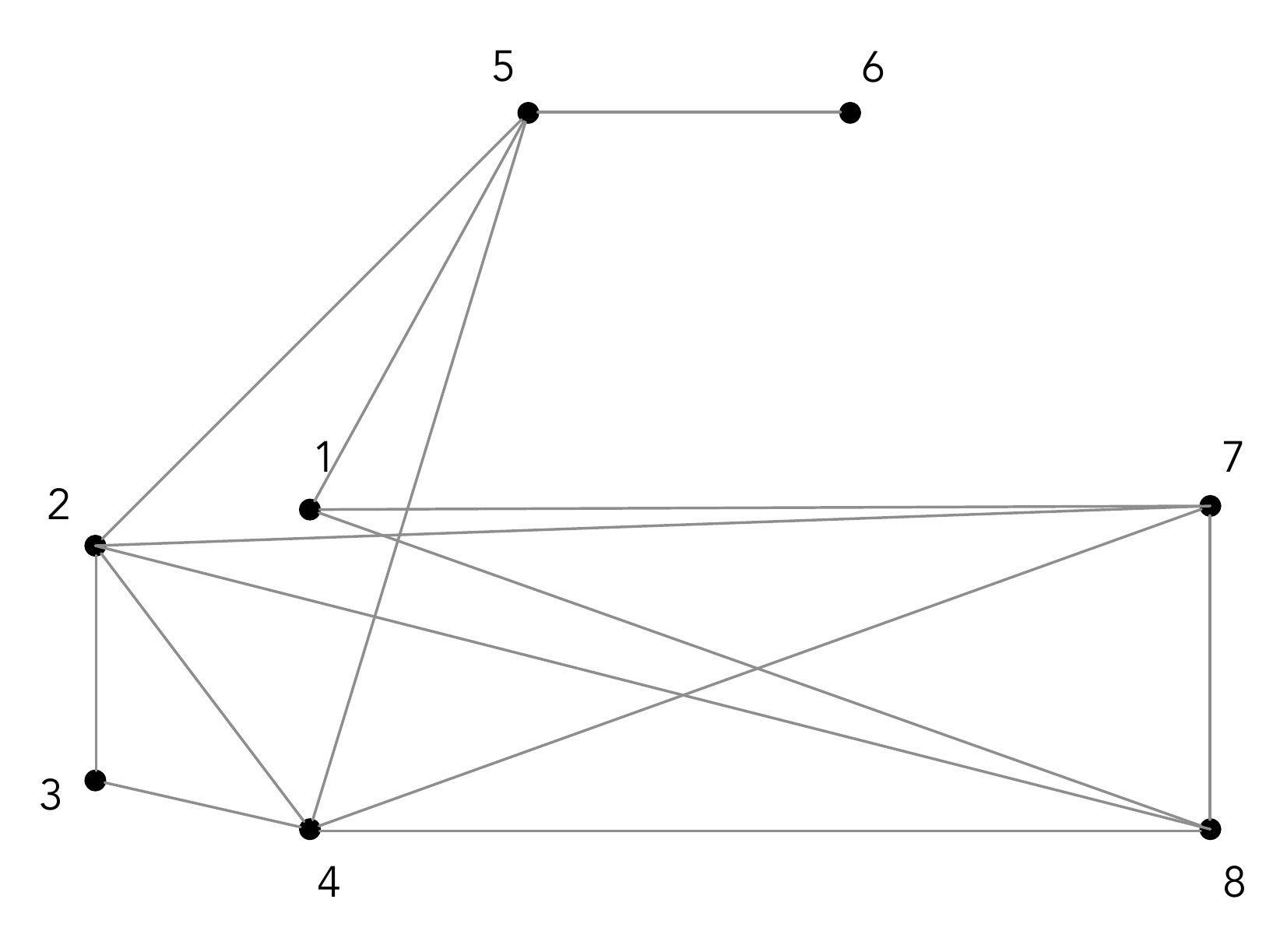}
\caption{Original graph for the graph-labeled tree in Figure~\ref{example-graphlabeledtree}.}
\label{example-originalgraph}
\end{center}
\end{figure}

\begin{definition}
\label{example-split-originalgraph}
The \emph{original graph} (called the \emph{accessibility graph} by Gioan and Paul~\cite{gioanpaul}) of a graph-labeled tree $(T, \F)$ is the graph $G = Gr(T, \F)$ where $V(G)$ is the leaf set of $T$ and, for $x, y\in V(G)$, $xy\in E(G)$ iff $x$ and $y$ are accessible in $(T, \F)$.
\end{definition}
\noindent For example, Figure~\ref{example-originalgraph} shows the original graph for the graph-labeled tree in Figure~\ref{example-graphlabeledtree}. We now define the \emph{split tree} of a connected graph $G$, which is a particular graph-labeled tree whose original graph is $G$.

\begin{definition}
For $V\subseteq V(G)$, the \emph{neighborhood} of $V$, denoted $N(V)$, is defined as the set of vertices in $V(G)\backslash V$ that are adjacent to at least one vertex in $V$.
\end{definition}

\begin{figure}[!htb]
\begin{minipage}[b]{0.5\textwidth}
\begin{center}
\includegraphics[width=0.5\linewidth]{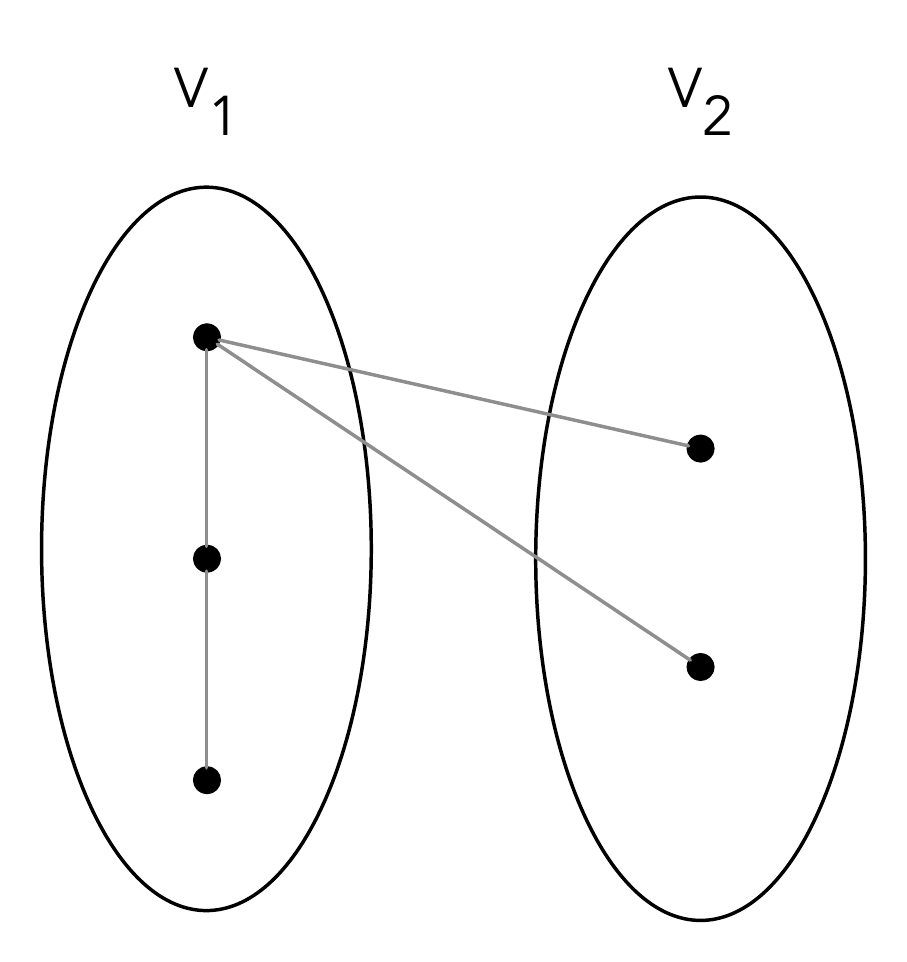}
\caption{Valid split.}
\label{example-validsplit}
\end{center}
\end{minipage}
\hfill
\begin{minipage}[b]{0.5\textwidth}
\begin{center}
\includegraphics[width=0.5\linewidth]{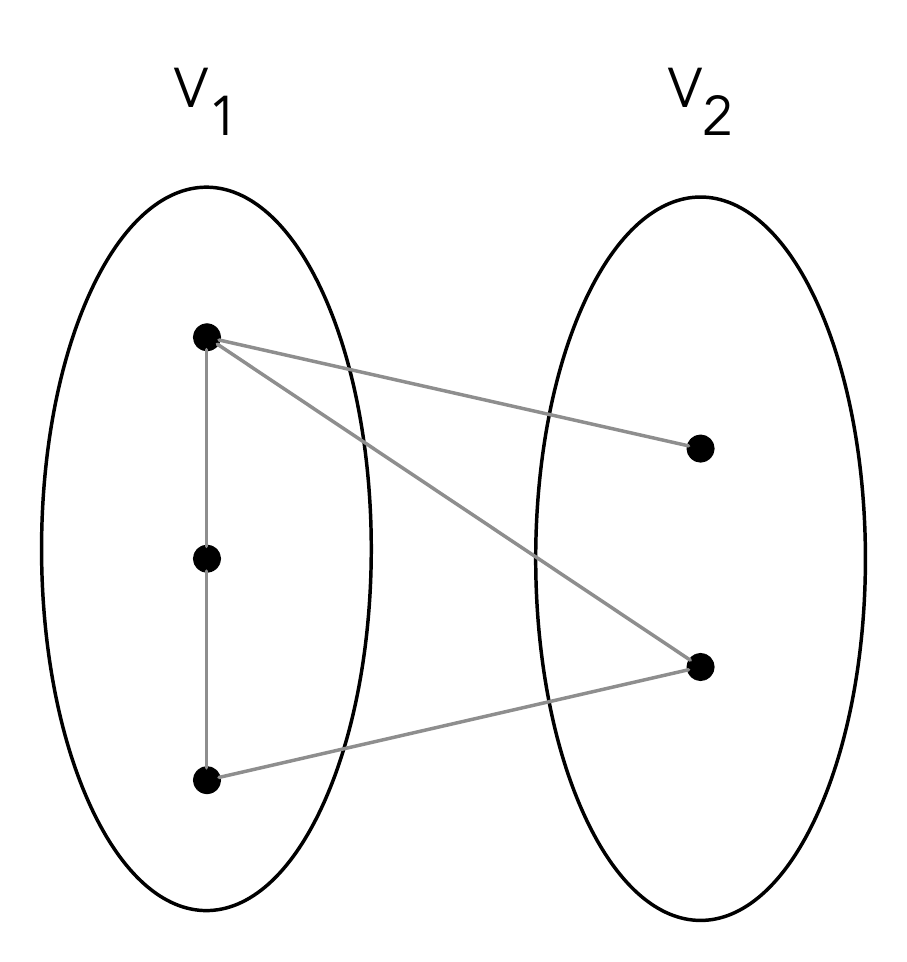}
\caption{Invalid split.}
\label{example-invalidsplit}
\end{center}
\end{minipage}
\end{figure}

\begin{definition}
A \emph{split} of a graph $G$ is a bipartition $(V_1, V_2)$ of $V(G)$ such that
\begin{enumerate}
\item $\n{V_1}\geq 2$ and $\n{V_2}\geq 2$, and
\item every vertex of $N(V_1)$ is adjacent to every vertex of $N(V_2)$.
\end{enumerate}
\end{definition}
\noindent For example, Figure~\ref{example-validsplit} is a split, while Figure~\ref{example-invalidsplit} is not a split because the bottom vertex of $V_1$ is in $N(V_2)$ and the top vertex of $V_2$ is in $N(V_1)$, but these vertices are not adjacent.
 
\begin{definition}
A graph $G$ is called \emph{prime} if it has no split, and \emph{degenerate} if every partition of $V(G)$ into two sets of size $\geq2$ is a split. It is known that the only degenerate graphs are the cliques $K_n$ and the stars $K_{1, n}$ for $n\geq 0$.
\end{definition}
\begin{figure}[!htb]
\begin{minipage}[b]{0.5\textwidth}
\begin{center}
\includegraphics[width=0.5\linewidth]{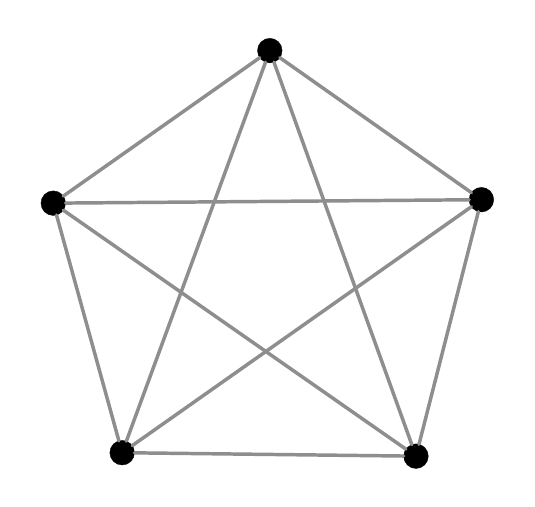}
\caption{Clique $K_5$.}
\label{example-clique}
\end{center}
\end{minipage}
\hfill
\begin{minipage}[b]{0.5\textwidth}
\begin{center}
\includegraphics[width=0.5\linewidth]{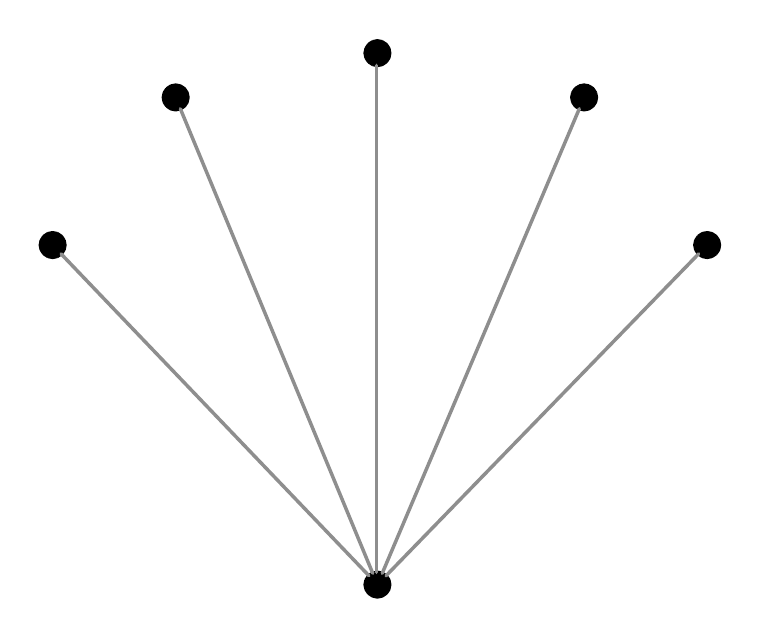}
\caption{Star $K_{1, 5}$.}
\label{example-star}
\end{center}
\end{minipage}
\end{figure}

\begin{definition}
A graph-labeled tree $(T, \F)$ is called \emph{reduced} if 
\begin{enumerate}
\item every $v\in V(T)$ has degree $\geq 3$,
\item there does not exist $uv\in E(T)$ such that $G_u$ and $G_v$ are both cliques, and
\item there does not exist $e = uv\in E(T)$ such that $G_u$ and $G_v$ are both stars, $\rho_u(e)$ is the center of $G_u$, and $\rho_v(e)$ is an extremity of $G_v$.
\end{enumerate}
\end{definition}

\begin{figure}[!htb]
\begin{minipage}[b]{0.475\textwidth}
\begin{center}
\includegraphics[width=\linewidth]{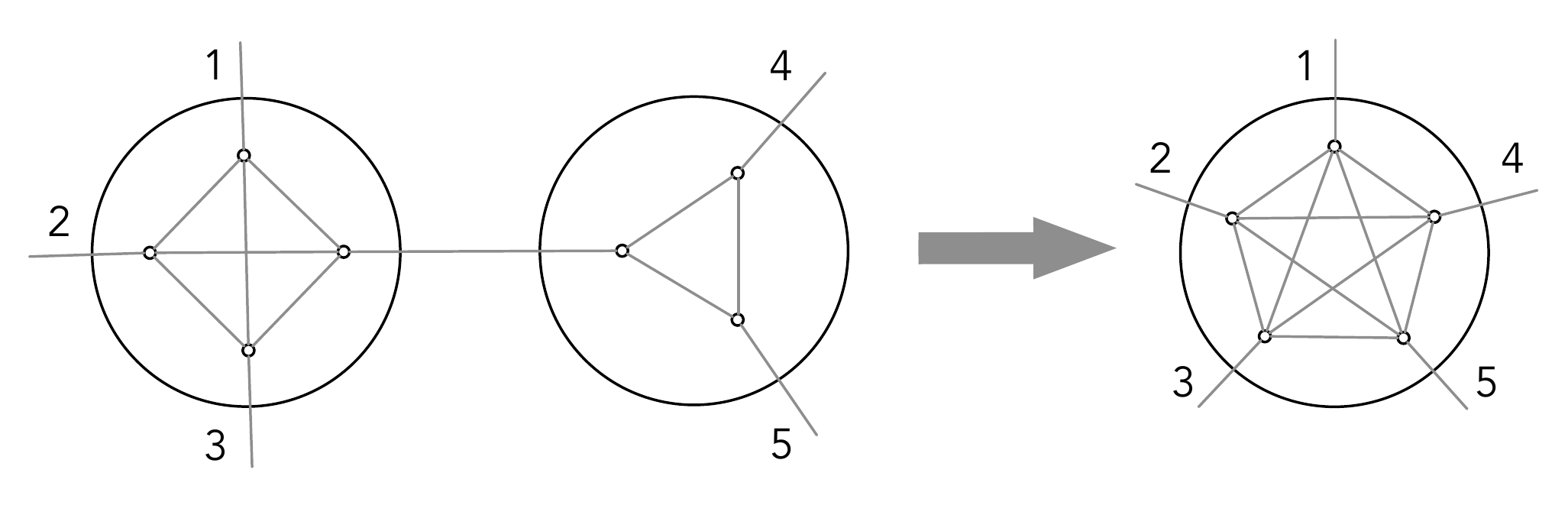}
\caption{Reduction of two adjacent clique nodes in a split tree to a single clique (recreated from \protect\cite{gioanpaul}).}
\label{example-cliquereduction}
\end{center}
\end{minipage}
\hfill
\begin{minipage}[b]{0.475\textwidth}
\begin{center}
\includegraphics[width=\linewidth]{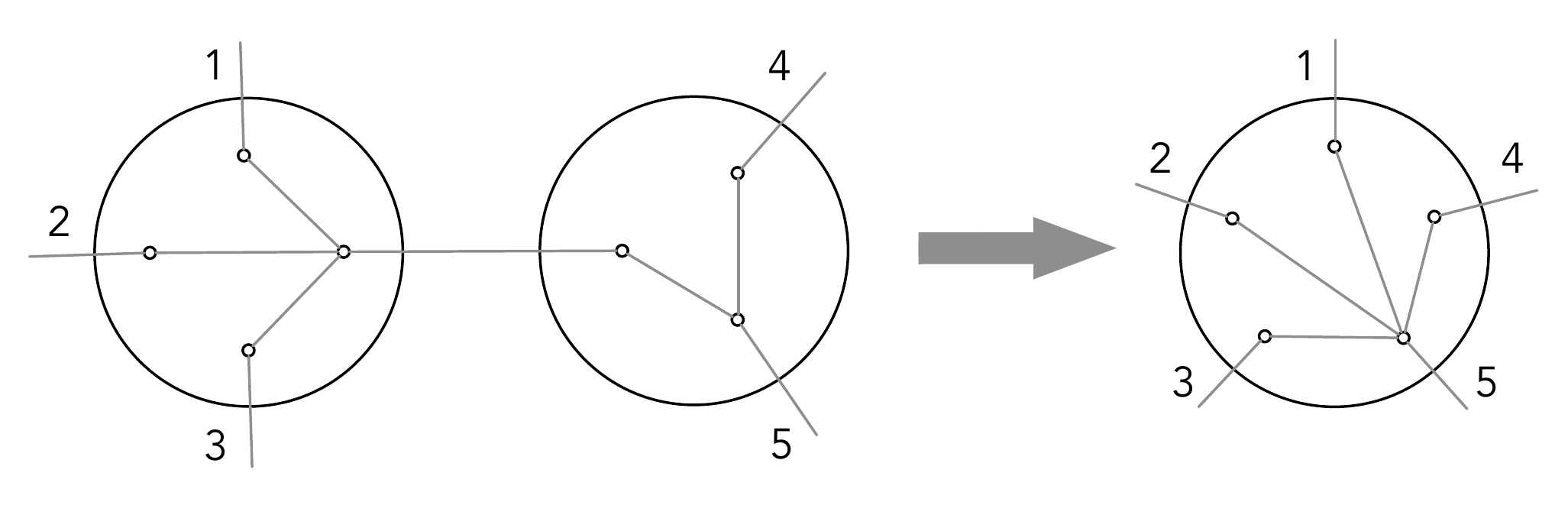}
\caption{Reduction of two adjacent star nodes in a split tree to a single star (recreated from \protect\cite{gioanpaul}).}
\label{example-starreduction}
\end{center}
\end{minipage}
\end{figure}

\noindent The intuition behind this definition is the following: if two clique nodes of sizes $m$ and $n$ are adjacent in $(T, \F)$, then they can be replaced with a single clique node of size $m+n-2$ in such a way that the original graph of $(T, \F)$ does not change (see Figure~\ref{example-cliquereduction}). A similar reduction can be performed when the center of one star node is adjacent to an extremity of another star node (see Figure~\ref{example-starreduction}).

\noindent The following characterization of connected graphs is shown by Gioan and Paul~\cite{gioanpaul}:

\begin{thm}
For any connected graph $G$, there exists a unique reduced graph-labeled tree $(T, \F)$ such that $G = Gr(T, \F)$ and every node label $G_v\in\F$ is either prime or degenerate. We call this graph-labeled tree the \emph{split tree} of $G$, and denote it by $ST(G)$.
\end{thm}

\begin{figure}[!htb]
\begin{center}
\includegraphics[width=0.45\linewidth]{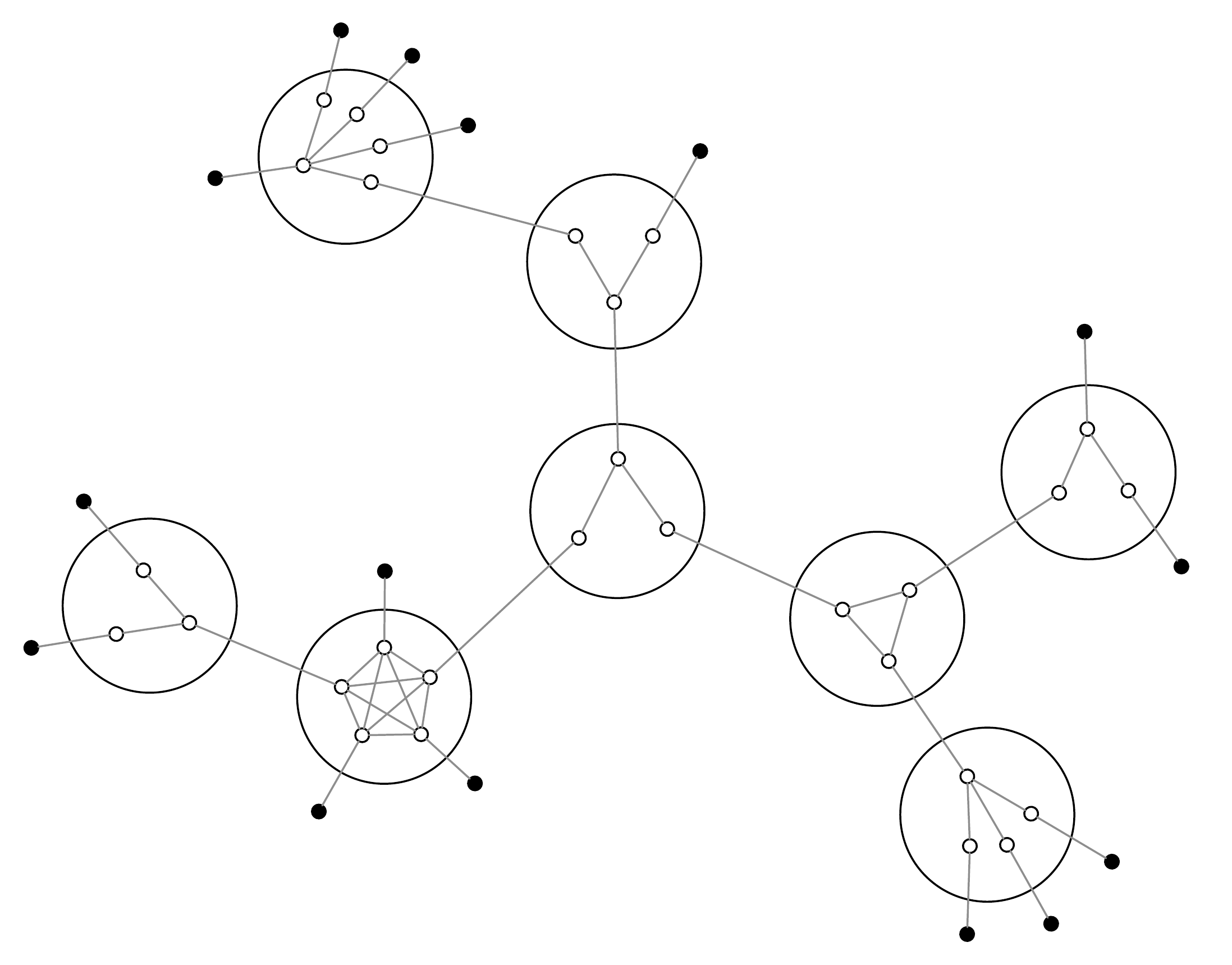}
\caption{Split tree of a distance-hereditary graph.}
\label{example-dhsplittree}
\end{center}
\end{figure}

\noindent Gioan and Paul further show that distance-hereditary and three-leaf power graphs can be characterized by the following conditions on their split trees:

\begin{thm}
\label{example-split-dhthm}
A graph $G$ is distance-hereditary iff its split tree has only clique and star nodes (\textit{i.e.} has only degenerate nodes). For this reason, distance-hereditary graphs are called \emph{totally decomposable} with respect to the split decomposition.
\end{thm}
\noindent Figure~\ref{example-dhsplittree} shows a split tree of a distance-hereditary graph.

\begin{figure}[!htb]
\begin{center}
\includegraphics[width=0.45\linewidth]{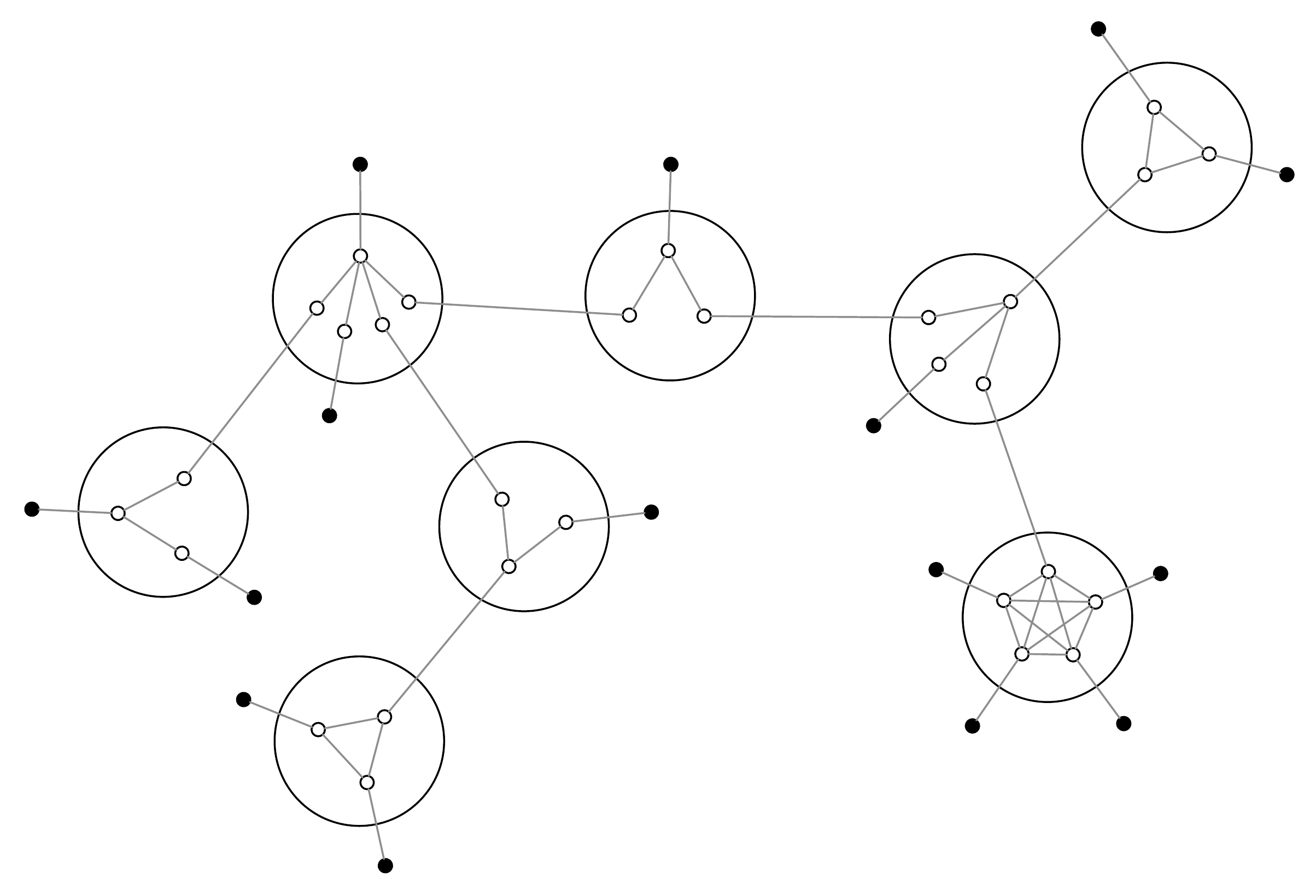}
\caption{Split tree of a three-leaf power graph.}
\label{example-3lpsplittree}
\end{center}
\end{figure}

\begin{thm}
\label{example-split-3lpthm}
A graph $G$ is a three-leaf power iff its split tree $ST(G) = (T, \F)$
\begin{enumerate}
\item has only clique and star nodes,
\item the set of star nodes forms a subtree of $T$, and
\item the center of every star node is adjacent to either a clique node or a leaf.
\end{enumerate}
\end{thm}
\noindent We note that by condition 1, all three-leaf power graphs are distance-hereditary. Figure~\ref{example-3lpsplittree} shows a split tree of a three-leaf power graph.

\subsection{Enumeration with the dissymmetry theorem}
\label{example-dissymmetry}

Chauve~\etal\cite{chauvelumbrosofusy} use the dissymmetry theorem and the split decomposition characterization of distance-hereditary and three-leaf power graphs described in Theorems~\ref{example-split-dhthm} and \ref{example-split-3lpthm} to construct combinatorial grammars and enumerations for these classes. We give an overview of these results here, and make use of the same grammars in our cycle pointing analysis in Section~\ref{example-cyclepointing}.

\subsubsection{Distance-hereditary graphs}
\label{example-dissymmetry-dh}

From Theorem~\ref{example-split-dhthm}, Theorem~\ref{dissymmetry-overview-thm}, and Lemma~\ref{dissymmetry-overview-noleaveslemma}, Chauve~\etal\cite{chauvelumbrosofusy} derive the following grammar for the class $\DH$ of split trees of distance-hereditary graphs:
\begin{align*}
\mathcal{DH + T_{K-S} + T_{S\ra S}}&\mathcal{\simeq T_K + T_S + T_{S-S}}\\
\mathcal{T_K} &= \Set_{\geq 3}(\mathcal{Z + S_X + S_C})\\
\mathcal{T_S} &= \mathcal{S_C \times(Z + K + S_C)}\\
\mathcal{T_{K-S}} &= \mathcal{K\times(S_C + S_X)}\\
\mathcal{T_{S-S}} &= \Set_2(\mathcal{S_C}) + \Set_2(\mathcal{S_X})\\
\mathcal{T_{S\ra S}} &= \mathcal{S_C\times S_C + S_X\times S_X}\\
\mathcal{K} &= \Set_{\geq 2}(\mathcal{Z + S_C + S_X})\\
\mathcal{S_C} &= \Set_{\geq 2}(\mathcal{Z + K + S_X})\\
\mathcal{S_X} &= (\mathcal{Z + K + S_C})\times\Set_{\geq 1}(\mathcal{Z + K + S_X})\\
\end{align*}

\noindent\textbf{Explanation of the classes.}
\begin{itemize}
\item $\K$: A clique node with one of its incident subtrees having been removed
\item $\SC$: A star node with the subtree incident to its center having been removed
\item $\SX$: A star node with the subtree incident to one of its extremities having been removed.
\item $\mathcal{T_K}$: A distance-hereditary split tree rooted at a clique node 
\item $\mathcal{T_S}$: A distance-hereditary split tree rooted at a star node
\item $\mathcal{T_{K-S}}$: A distance-hereditary split tree rooted at an undirected edge connecting a clique node and a star node
\item $\mathcal{T_{S-S}}$: A distance-hereditary split tree rooted at an undirected edge connecting two star nodes
\item $\mathcal{T_{S\ra S}}$: A distance-hereditary split tree rooted at a directed edge connecting two star nodes
\end{itemize}

\paragraph{Sketch of proof.}
The three mutually-recursive expressions for $\K$, $\SC$, and $\SX$ encode the requirements that a split tree must have all nodes of degree at least $3$, no adjacent clique nodes, and no extremity of a star node adjacent to the center of another star node.

By Theorem~\ref{dissymmetry-overview-thm} and Lemma~\ref{dissymmetry-overview-noleaveslemma}, $$\mathcal{DH + T_{S\ra S} + T_{K\ra K} + T_{S\ra K} + T_{K\ra S}} $$ $$\mathcal{\simeq T_K + T_S + T_{S-S} + T_{K-S} + T_{K-K}}.$$ Noting that $\mathcal{T_{K\ra K}}$ and $\mathcal{T_{K-K}}$ are empty (since clique nodes cannot be adjacent in a split tree) and $\mathcal{T_{S\ra K}\simeq T_{K\ra S}\simeq T_{K-S}}$, it follows that $$\mathcal{DH + T_{K-S} + T_{S\ra S}\simeq T_K + T_S + T_{S-S}}.$$

\begin{flushright}$\square$\end{flushright}

\noindent Using the Maple package $\tt{combstruct}$, Chauve~\etal compute the enumeration of each of the above classes, and hence of the class $\DH$.

\begin{cor}
The first few terms of the OGF of the class of distance-hereditary graphs are $$\DH(z) = z + z^2 + 2z^3 + 6z^4 + 18z^5 + 73z^6 + 308z^7 + 1484z^8 + 7492z^9 + 40010z^{10} + \ldots$$
\end{cor}

\subsubsection{Three-leaf power graphs}
\label{example-dissymmetry-3lp}

From Theorem~\ref{example-split-3lpthm}, Theorem~\ref{dissymmetry-overview-thm}, and Lemma~\ref{dissymmetry-overview-noleaveslemma}, Chauve~\etal\cite{chauvelumbrosofusy} derive the following grammar for the class $\TLP$ of (split trees of) three-leaf power graphs:
\begin{align*}
3\mathcal{LP + T_{S\ra S}} &\simeq\mathcal{K + T_S + T_{S-S}}\\
\mathcal{T_S} &= \mathcal{A\times S_C}\\
\mathcal{T_{S-S}} &= \Set_2{(\mathcal{S_X})}\\
\mathcal{T_{S\ra S}} &= \mathcal{S_X\times S_X}\\
\mathcal{S_C} &= \Set_{\geq 2}{(\mathcal{A + S_X})}\\
\mathcal{S_X} &= \mathcal{A\times} \Set_{\geq 1}{(\mathcal{A + S_X})}\\
\A &= \Z + \Set_{\geq 2}{(\Z)} = \Set_{\geq 1}(\Z)\\
\mathcal{K} &= \Set_{\geq 3}{(\Z)}\\
\end{align*}

\noindent\textbf{Explanation of the classes.}
\begin{itemize}
\item $\A$: The disjoint union of a leaf and a clique having been entered through one of its edges
\item $\SC$: A star node with the subtree incident to its center having been removed
\item $\SX$: A star node with the subtree incident to one of its extremities having been removed
\item $\mathcal{T_S}$: A three-leaf power split tree rooted at a star node
\item $\K$: A three-leaf power split tree consisting of a single clique node
\item $\mathcal{T_{S-S}}$: A three-leaf power split tree rooted at an undirected edge connecting two star nodes
\item $\mathcal{T_{S\ra S}}$: A three-leaf power split tree rooted at a directed edge connecting two star nodes
\end{itemize}

\paragraph{Sketch of proof.}
The expressions for $S_C$ and $S_X$ encode the requirements that the split tree of a three-leaf power graph consists of either a single clique node or a subtree of star nodes joined at their extremities together with either a leaf or a clique node pending from each center and remaining extremity of these star nodes. Due to this structure, the leaves/clique nodes represented by $\A$ can be thought of as ``meta-leaves'' and excluded from the dissymmetry theorem by a similar argument as the one in Lemma~\ref{dissymmetry-overview-noleaveslemma}. Then $$3\mathcal{LP + T_{S\ra S}\simeq K + T_S + T_{S-S}}$$ follows immediately from Theorem~\ref{dissymmetry-overview-thm}.

\begin{flushright}$\square$\end{flushright}

\noindent Using $\tt{combstruct}$, Chauve~\etal compute the enumeration of each of the above classes, and hence of the class $3\mathcal{LP}$.

\begin{cor}
The first few terms of the OGF of the class of three-leaf power graphs are $$\TLP(z) = z + z^2 + 2z^3 + 5z^4 + 12z^5 + 32z^6 + 82z^7 + 227z^8 + 629z^9 + 1840z^{10} + \ldots$$
\end{cor}

\subsection{Enumeration and sampling with cycle pointing}
\label{example-cyclepointing}

\subsubsection{Distance-hereditary graphs}
\label{example-cyclepointing-dh}

In this section we apply the steps outlined in Sections~\ref{cyclepointing-decomposition}, \ref{cyclepointing-enumeration}, and \ref{cyclepointing-sampler} to enumerate and build an unbiased sampler for the class $\DH$ of distance-hereditary graphs. Instead of working directly with the graphs, we will work with their split tree decompositions.

Let $\SX$, $\SC$, and $\K$ be defined as in Section~\ref{example-dissymmetry-dh}. Then we have $$\SX = (\Z + \K + \SC)\times \Set_{\geq 1}(\Z + \K + \SX)$$ $$\SC = \Set_{\geq2}(\Z + \K + \SX)$$ $$\K = \Set_{\geq2}(\Z + \SC + \SX)$$ Since the cycle index sums of $\Set, \Set_0$, and $\Set_1$ are $$Z_{\Set}(s_1, s_2, \dots) = \lgexp{\sum_{i = 1}^{\infty}\frac{s_i}{i}}\qquad Z_{\Set_0}(s_1, s_2, \dots) = 1\qquad Z_{\Set_1}(s_1, s_2, \dots) = s_1$$ it follows that $$Z_{\Set_{\geq1}}(s_1, s_2, \dots) = \lgexp{\sum_{i = 1}^{\infty}\frac{s_i}{i}} - 1$$ and $$Z_{\Set_{\geq2}}(s_1, s_2, \dots) = \lgexp{\sum_{i = 1}^{\infty}\frac{s_i}{i}} - 1 - s_1.$$ From Table~\ref{cyclepointing-transfertheoremstable}, we obtain the equations $$\SX(z) = (z + \K(z) + \SC(z))\cdot\left[\lgexp{\sum_{i=1}^{\infty}\frac{1}{i}\left[z^i+\K(z^i)+\SX(z^i)\right]}-1\right]$$ $$\SC(z) = \lgexp{\sum_{i=1}^{\infty}\frac{1}{i}\left[z^i+\K(z^i)+\SX(z^i)\right]} - 1 - z - \K(z) - \SX(z)$$ $$\K(z) = \lgexp{\sum_{i=1}^{\infty}\frac{1}{i}\left[z^i+\SC(z^i)+\SX(z^i)\right]} - 1 - z - \SC(z) - \SX(z),$$ and using $\tt{combstruct}$ we can compute the coefficients of these generating functions:
$$\SX(z) = z^2 + 5z^3 + 23z^4 + 119z^5 + 639z^6 + 3629z^7 + 21257z^8 + 127995z^9 + 786481z^{10} + \ldots$$ $$\SC(z) = z^2 + 3z^3 + 14z^4 + 67z^5 + 367z^6 + 2065z^7 + 12150z^8 + 73177z^9 + 450322z^{10} + \ldots$$ $$\K(z) = z^2 + 3z^3 + 14z^4 + 67z^5 + 367z^6 + 2065z^7 + 12150z^8 + 73177z^9 + 450322z^{10} + \ldots.$$ 
We now build a specification for $\DHcp$ in terms of $\SX, \SC, $ and $\K$: 
\begin{align*}
\DHcp &= \cp{\Z}\\
&+ \Setcp_2\sub\Z \\
&+ \cp{\Z}\times(\SX + \SC + \K)\\
&+ \Setscp_2\sub\,\SX \\
&+ \Setscp_2\sub\,\SC \\
&+ \Setscp_{\geq 3}\sub(\Z + \SX + \SC) \\
&+ (\Z + \K + \SC)\times\Setscp\sub(\Z + \K + \SX)
\end{align*}
Term $i$ in this specification corresponds to case $i$ of the following cases:

\begin{enumerate}
\item The tree has one leaf.
\item The tree has two leaves.
\item The marked cycle has length 1 and the tree has $>2$ leaves.
\item The marked cycle has length $\geq 2$ and has as its center an edge connecting two star nodes at extremities.
\item The marked cycle has length $\geq 2$ and has as its center an edge connecting two star nodes at their centers.
\item The marked cycle has length $\geq 2$ and has as its center a clique node.
\item The marked cycle has length $\geq 2$ and has as its center a star node.
\end{enumerate}
In a split tree, there can be no edge connecting two clique nodes; furthermore, while there can be an edge connecting a clique node and a star node, such an edge cannot be the center of a cycle of an automorphism of the split tree. Thus these seven cases cover all possibilities for a cycle-pointed split tree of a distance-hereditary graph.

Next, we translate this specification into a generating function equation. Recall that $$Z_{\Setscp_2}(s_1, s_2, \ldots; t_1, t_2, \ldots) = t_2$$ $$Z_{\Setscp}(s_1, s_2, \ldots; t_1, t_2, \ldots) = \left(\sum_{l = 2}^{\infty}t_l\right)\cdot\lgexp{\sum_{i = 1}^{\infty}\frac{s_i}{i}}$$ $$Z_{\Setscp_{\geq3}}(s_1, s_2, \ldots; t_1, t_2, \ldots) = \left(\sum_{l = 2}^{\infty}t_l\right)\cdot\lgexp{\sum_{i = 1}^{\infty}\frac{s_i}{i}} - t_2$$ By Theorem~\ref{cyclepointing-introduction-correspondencethm}, we know that the OGF for $\DHcp$ is $z\cdot\DH'(z)$, so by the transfer theorems from Table~\ref{cyclepointing-transfertheoremstable} it follows that
\begin{align*}
z\cdot\DH'(z) &= z + 2z^2 + z[\SX(z) + \SC(z) + \K(z)] + z^2\SX'(z^2) + z^2\SC'(z^2) \\
& + \left(\sum_{l = 2}^{\infty}z^l\left[1 + \SX'(z^l) + \SC'(z^l)\right]\right)\cdot\lgexp{\sum_{i = 1}^{\infty}\frac{1}{i}\left[z^i+\SX(z^i)+\SC(z^i)\right]} - z^2(1 + \SX'(z^2) + \SC'(z^2)) \\
&+ (z + \K(z) + \SC(z))\left(\sum_{l = 2}^{\infty}z^l\left[1 + \K'(z^l) + \SX'(z^l)\right]\right)\cdot\lgexp{\sum_{i = 1}^{\infty}\frac{1}{i}\left[z^i+\K(z^i)+\SX(z^i)\right]}
\end{align*}
By differentiating the earlier expressions for $\SX(z), \SC(z)$, and $\K(z)$ and using these to simplify the above formula, we reduce it to
\begin{align*}
z\cdot\DH'(z) &= z + 2z^2 + z[\SX(z) + \SC(z) + \K(z)] + z^2\SX'(z^2) + z^2\SC'(z^2) + z[1+\K'(z)+\SC'(z)+\SX'(z)] \\
& - z[1 + \SC'(z)+\SX'(z)][1 + z + \K(z) + \SC(z) + \SX(z)] - z^2[1 + \SX'(z^2) + \SC'(z^2)] \\
&+ [z + \K(z) + \SC(z)][z(1 + \K'(z) + \SC'(z) + \SX'(z)) \\
&- z(1 + \K'(z) + \SX'(z))(1 + z + \K(z) + \SC(z) + \SX(z))].
\end{align*}
Evaluating this in Maple, we find that $$z\cdot\DH'(z) = z + 2z^2 + 6z^3 + 24z^4 + 90z^5 + 438z^6 + 2156z^7 + 11872z^8 + 67428z^9 + 400100z^{10} + \ldots,$$ so $$\DH(z) = z + z^2 + 2z^3 + 6z^4 + 18z^5 + 73z^6 + 308z^7 + 1484z^8 + 7492z^9 + 40010z^{10} + \ldots.$$ 
In order to build a Boltzmann sampler for $\DHcp$ and hence (by Corollary~\ref{cyclepointing-introduction-correspondencecor}) an unbiased sampler for $\DH$, we apply the rules in Table~\ref{cyclepointing-polyaboltzmanntransfertable} and the P\'{o}lya-Boltzmann samplers described by Bodirsky~\etal\cite{cplong} to translate our symbolic specification for $\DHcp$ into a Boltzmann sampler. We discuss some aspects of the implementation in Section~\ref{implementation}, and refer to the accompanying Maple code for full details.

\subsubsection{Three-leaf power graphs}
\label{example-cyclepointing-3lp}

In this section we apply the steps outlined in Sections~\ref{cyclepointing-decomposition}, \ref{cyclepointing-enumeration}, and \ref{cyclepointing-sampler} to enumerate and build an unbiased sampler for the class $\TLP$ of three-leaf power graphs. Instead of working directly with the graphs, we will again work with their split tree decompositions.

Let $\SX$, $\SC$, and $\A$ be as in Section~\ref{example-dissymmetry-3lp}. Then we have $$\SX = \A\times \Set_{\geq 1}{(\A+\SX)}$$ $$\SC = \Set_{\geq2}(\A + \SX)$$ $$\A = \Set_{\geq1}(\Z)$$ Recalling the cycle index sums for $\Set, \Set_0$, and $\Set_1$ from Section~\ref{example-cyclepointing-dh}, we apply the transfer theorems from Table~\ref{cyclepointing-transfertheoremstable} to obtain the following equations: $$\SX(z) = \A(z)\cdot\left[\lgexp{\sum_{i=1}^{\infty}\frac{1}{i}(\A(z^i)+\SX(z^i))}-1\right]$$ $$\SC(z) = \lgexp{\sum_{i=1}^{\infty}\frac{1}{i}(\A(z^i)+\SX(z^i))} - 1 - \A(z) - \SX(z)$$ $$\A(z) = \lgexp{\sum_{i=1}^{\infty}\frac{1}{i}z^i} - 1 = \frac{z}{1-z}$$ Using $\tt{combstruct}$, we can compute the coefficients of these generating functions:
$$\SX(z) =  z^2 + 4z^3 + 12z^4 + 36z^5 + 107z^6 + 331z^7 + 1041z^8 + 3359z^9 + 11018z^{10} + \ldots$$ $$\SC(z) = z^2 + 3z^3 + 11z^4 + 34z^5 + 116z^6 + 378z^7 + 1276z^8 + 4299z^9 + 14684z^{10} + \ldots$$ $$\A(z) = z + z^2 + z^3 + z^4 + z^5 + z^6 + z^7 + z^8 + z^9 + z^{10} + \ldots.$$
We now build a specification for $\TLPcp$ in terms of $\SX, \SC,$ and $\A$: 
\begin{align*}
\TLPcp &= \cp{\Z} \\
&+ \Setcp_2\sub\Z \\
&+ \cp{\Z}\times(\SX + \SC + \A\times(\SX + \SC) + \Set_{\geq2}(\Z)) \\
&+ \Setscp_2\sub\,\SX \\
&+ \A\times\Setscp\sub(\SX + \A) \\
&+ \Setscp_{\geq3}\sub\Z \\
&+ (\SX + \SC)\times\Setscp\sub\Z
\end{align*}
Term $i$ in this specification corresponds to case $i$ of the following cases:

\begin{enumerate}
\item The tree has one leaf.
\item The tree has two leaves.
\item The marked cycle has length 1.
\item The marked cycle has length $\geq 2$ and has as its center an edge connecting two star nodes at extremities.
\item The marked cycle has length $\geq 2$ and has as its center a star node.
\item The marked cycle has length $\geq 2$ and has as its center an isolated clique node (\textit{i.e.} a clique node that is the entire graph).
\item The marked cycle has length $\geq 2$ and has as its center a clique node connected to a star node.
\end{enumerate}
Recall that in addition to the restrictions from being a distance-hereditary split tree, the split tree of a three-leaf power graph also cannot have an edge connecting the centers of two star nodes, and can only have cliques as meta-leaves connected to star nodes. Thus these seven cases cover all possibilities for a cycle-pointed split tree of a three-leaf power graph.

Next, we translate this specification into a generating function equation. Recalling the cycle index sums for $Z_{\Setscp_2}$, $Z_{\Setscp}$, and $Z_{\Setscp_{\geq3}}$ from Section~\ref{example-cyclepointing-dh}, we apply the transfer theorems from Table~\ref{cyclepointing-transfertheoremstable} to obtain an equation for the OGF of $\TLPcp$:
\begin{align*}
z\cdot\TLP'(z) &= z + 2z^2 + z[\SX(z) + \SC(z) + \A(z)(\SX(z) + \SC(z)) + \A(z) - z] + z^2\SX'(z^2) \\
&+ \A(z)\left(\sum_{l = 2}^{\infty}z^l(\A'(z^l) + \SX'(z^l))\right)\cdot\lgexp{\sum_{i = 1}^{\infty}\frac{1}{i}[\A(z^i) + \SX(z^i)]}\\
&+ \left(\sum_{l=2}^{\infty}z^l\right)\cdot\lgexp{\sum_{i=1}^{\infty}\frac{1}{i}z^i} - z^2 + [\SX(z) + \SC(z)]\left(\sum_{l=2}^{\infty}z^l\right)\cdot\lgexp{\sum_{i=1}^{\infty}\frac{1}{i}z^i}
\end{align*}
By differentiating the earlier expressions for $\SX(z), \SC(z)$, and $\A(z)$ and using these to simplify the above formula, it follows that
\begin{align*}
z\cdot\TLP'(z) &= z + 2z^2 + z[\SX(z) + \SC(z) + \A(z)(\SX(z) + \SC(z)) + \A(z) - z] + z^2\SX'(z^2) \\
&+ \A(z)[z(\A'(z) + \SX'(z) + \SC'(z)) - z(\A'(z) + \SX'(z))(1 + \A(z) + \SX(z) + \SC(z))] \\
&+ \frac{z^2}{(1-z)^2} - z^2 + (\SX(z) + \SC(z))\cdot\frac{z^2}{(1-z)^2}.
\end{align*}
Evaluating this in Maple, we find that $$z\cdot\TLP'(z) = z + 2z^2 + 6z^3 + 20z^4 + 60z^5 + 192z^6 + 574z^7 + 1816z^8 + 5661z^9 + 18400z^{10} + \ldots,$$ so $$\TLP(z) = z + z^2 + 2z^3 + 5z^4 + 12z^5 + 32z^6 + 82z^7 + 227z^8 + 629z^9 + 1840z^{10} + \ldots.$$ 
In order to build a Boltzmann sampler for $\TLPcp$ and hence (by Corollary~\ref{cyclepointing-introduction-correspondencecor}) an unbiased sampler for $\TLP$, we apply the rules in Table~\ref{cyclepointing-polyaboltzmanntransfertable} and the P\'{o}lya-Boltzmann samplers described by Bodirsky~\etal\cite{cplong} to translate our symbolic specification for $\TLPcp$ into a Boltzmann sampler. We discuss some aspects of the implementation in Section~\ref{implementation}, and refer to the accompanying Maple code for full details.

\section{Implementation and empirical study of samplers}
\label{implementation}

\subsection{Overview}
\label{implementation-overview}

Using the computer algebra system Maple, we have implemented unbiased samplers for the class $\DH$ of split trees of distance-hereditary graphs and the class $\TLP$ of split trees of three-leaf power graphs. These samplers take a real parameter $z$ that is between $0$ and the singularity of the generating function of the class (we discuss estimation of the singularity in Section~\ref{implementation-details-roc}), and output a string representation of the split tree that obeys the semantics in Table~\ref{implementation-semanticstable}. Some examples of split tree strings and their descriptions are provided in Table~\ref{implementation-splittreedescriptions}.
\begin{table}[!htb]
\begin{center}
\begin{tabular} {cp{12cm}N}
\toprule
Symbol & Meaning & \\[10pt]
\midrule
\textsf{Z} & a leaf node & \\[10pt]
\textsf{KR} & a clique root (can only appear as the root of the tree) & \\[10pt]
\textsf{SR} & a star root (can only appear as the root of the tree) & \\[10pt]
\textsf{K} & a clique that has been entered from another node & \\[10pt]
\textsf{SX} & a star that has been entered from another node at one of its extremities & \\[10pt]
\textsf{SC} & a star that has been entered from another node at its center & \\[10pt]
\textsf{e(A, B)} & an edge that connects nodes \textsf{A} and \textsf{B} & \\[10pt]
\textsf{A(}$\textsf{B}_1$, \ldots, $\textsf{B}_{\textsf{k}}$\textsf{)} & $\textsf{B}_1$, \ldots, $\textsf{B}_{\textsf{k}}$ are neighbors of \textsf{A} (if \textsf{A} is \textsf{SR} or \textsf{SX}, then $\textsf{B}_1$ is connected to the center of \textsf{A}) & \\[15pt]
\bottomrule
\end{tabular}
\caption{Semantics for the strings returned by our Maple implementations of the $\DH$ and $\TLP$ samplers.}
\label{implementation-semanticstable}
\end{center}
\end{table}

\begin{table}[!htb]
\begin{center}
\begin{tabular} {cp{11cm}N}
\toprule
Split tree string & Description & \\[10pt]
\midrule
\textsf{Z(K(Z, Z))} & a leaf connected to a clique that has two other leaves as neighbors & \\[10pt]
\textsf{KR(Z, Z, Z)} & a clique with three leaves as neighbors (same as previous) & \\[10pt]
\textsf{Z(SC(Z, Z))} & a leaf connected to the center of a star that has two leaves as its extremities & \\[20pt]
\textsf{Z(SX(Z, Z))} & a leaf connected to an extremity of a star that has a leaf as its center and a leaf as its other extremity (same as previous) & \\[20pt]
\textsf{e(SC(Z, Z), SC(Z, Z))} & an edge joining two star nodes at their centers, each of which has two leaves as its extremities & \\[20pt]
\textsf{SR(K(Z, Z), Z, Z, Z)} & a star with three leaves as its extremities and whose center is connected to a clique with two other leaves as neighbors & \\[20pt]
\bottomrule
\end{tabular}
\caption{Some examples of split tree strings and their descriptions.}
\label{implementation-splittreedescriptions}
\end{center}
\end{table}

In order to visualize the graphs that are being generated, we recall that the original graph of a split tree (\textit{cf.} Definition~\ref{example-split-originalgraph}) has one node for each leaf of the split tree, and has an edge between two nodes iff there exists a path connecting the corresponding leaves in the split tree that uses at most one internal edge of each node label. Note that, since the number of nodes of the graph is equal to the number of leaves of the split tree, when we refer to the ``size'' of a split tree we mean its number of leaves.

We have built a Python package called $\tt{split\TextUnderscore{}tree}$, which computes the original graph corresponding to a given distance-hereditary or three-leaf power split tree string and visualizes the graph. This package has functions to translate a split tree string into an object-oriented representation of that split tree, to compute the original graph of a split tree object with the quadratic-time algorithm that checks whether each pair of vertices is or is not accessible, and to draw and export the generated graphs using $\tt{NetworkX}$. An original graph of size $n$ is provided as a list of adjacencies over a canonical set $\{1, \ldots, n\}$ of vertices, so the drawing functionality can easily be replaced with another package such as $\tt{Graphviz}$.

We use this combination of Maple and Python to take advantages of the strengths of each of the languages -- Maple has the powerful $\tt{combstruct}$ package, which allows for the automatic computation of generating function coefficients from non cycle-pointed combinatorial specifications, while Python is better suited to the object-oriented computation used in $\tt{split\TextUnderscore{}tree}$.

\begin{figure}[!htb]
\begin{center}
\includegraphics[width=0.6\linewidth]{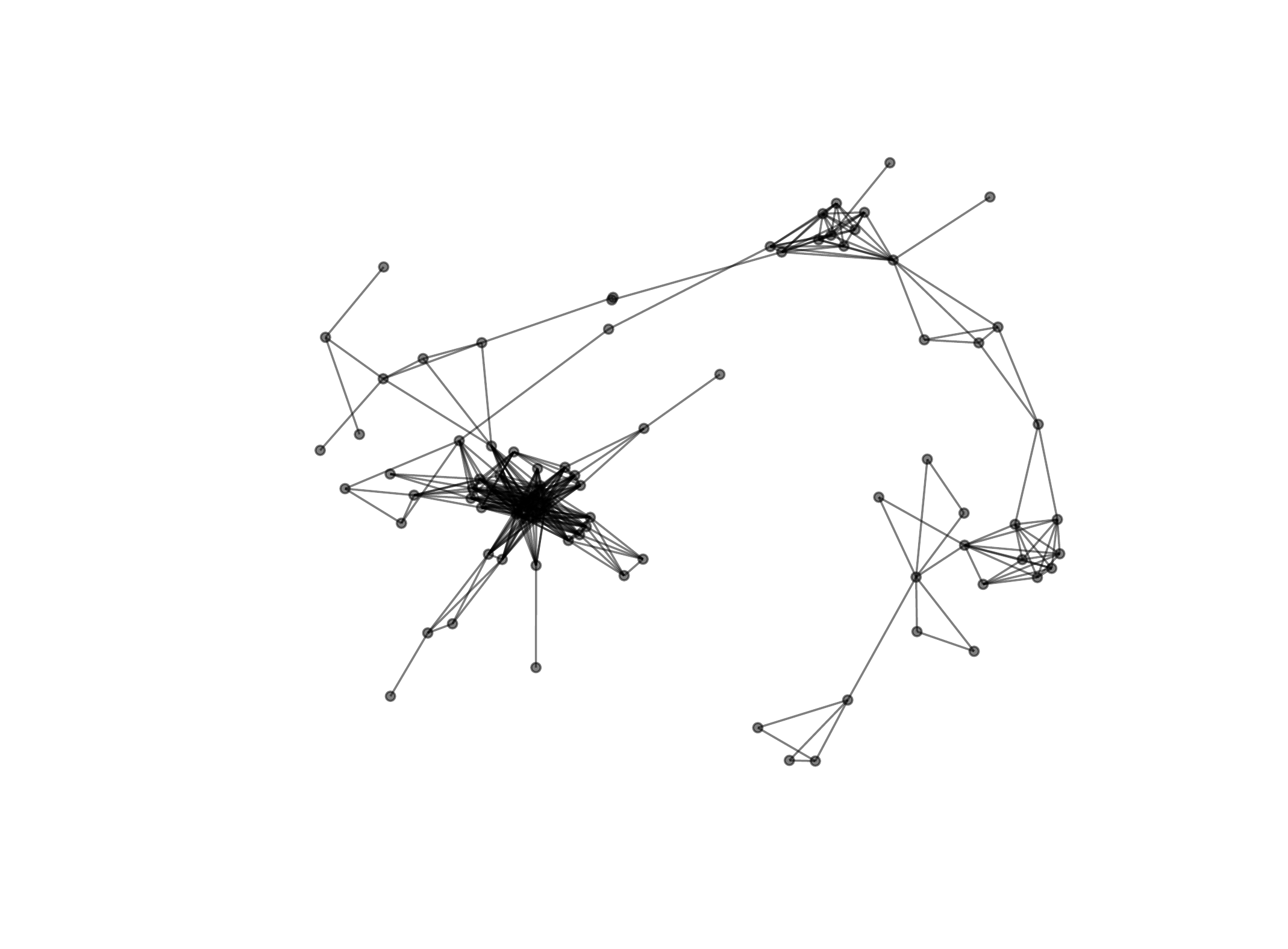}
\caption{A randomly generated distance-hereditary graph with $86$ vertices.}
\label{implementation-dhgraph}
\end{center}
\end{figure}

\begin{figure}[!htb]
\begin{center}
\includegraphics[width=0.6\linewidth]{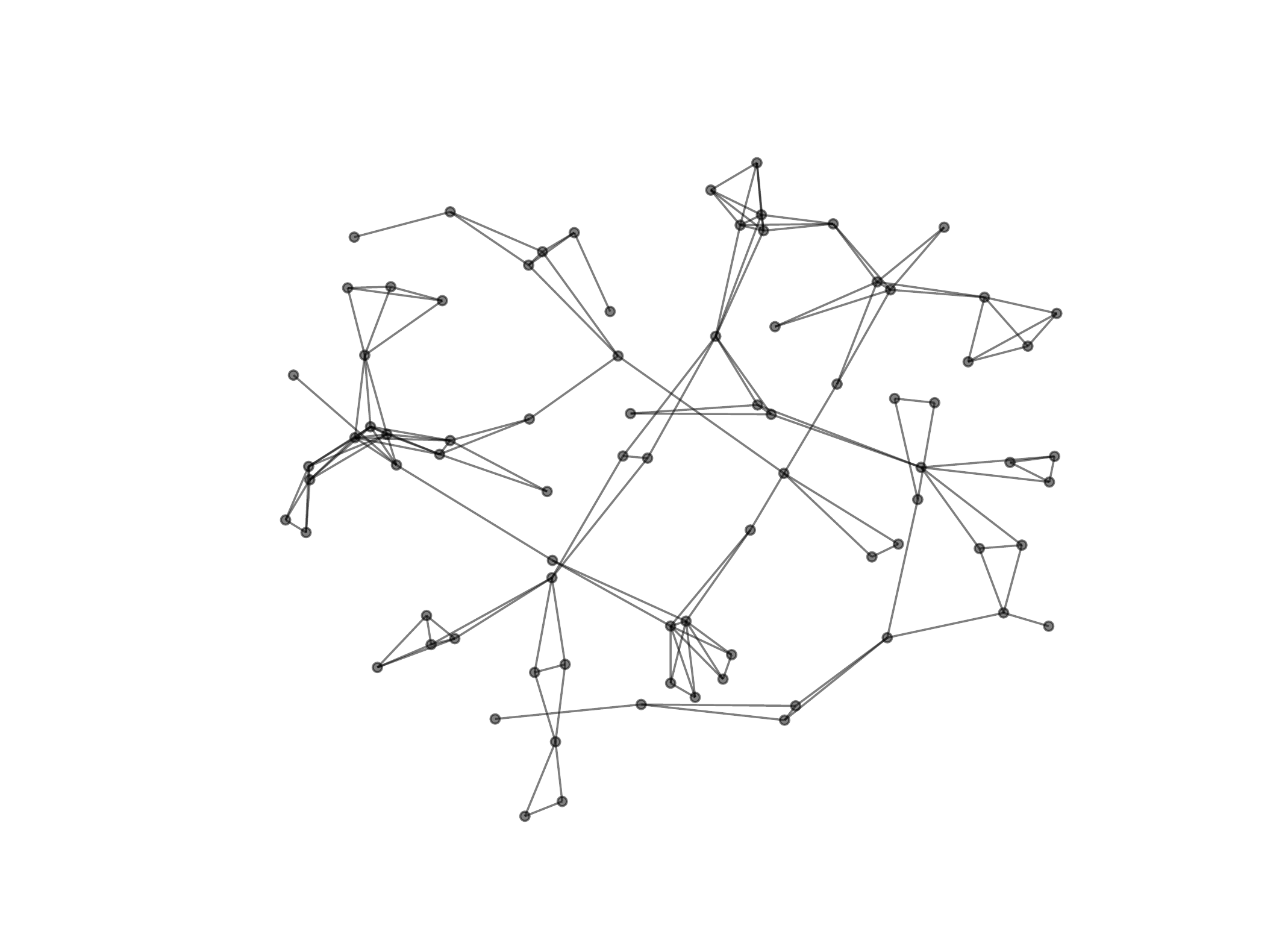}
\caption{A randomly generated three-leaf power graph with $82$ vertices.}
\label{implementation-3lpgraph}
\end{center}
\end{figure}

\begin{figure}[!htb]
\begin{center}
\includegraphics[width=0.6\linewidth]{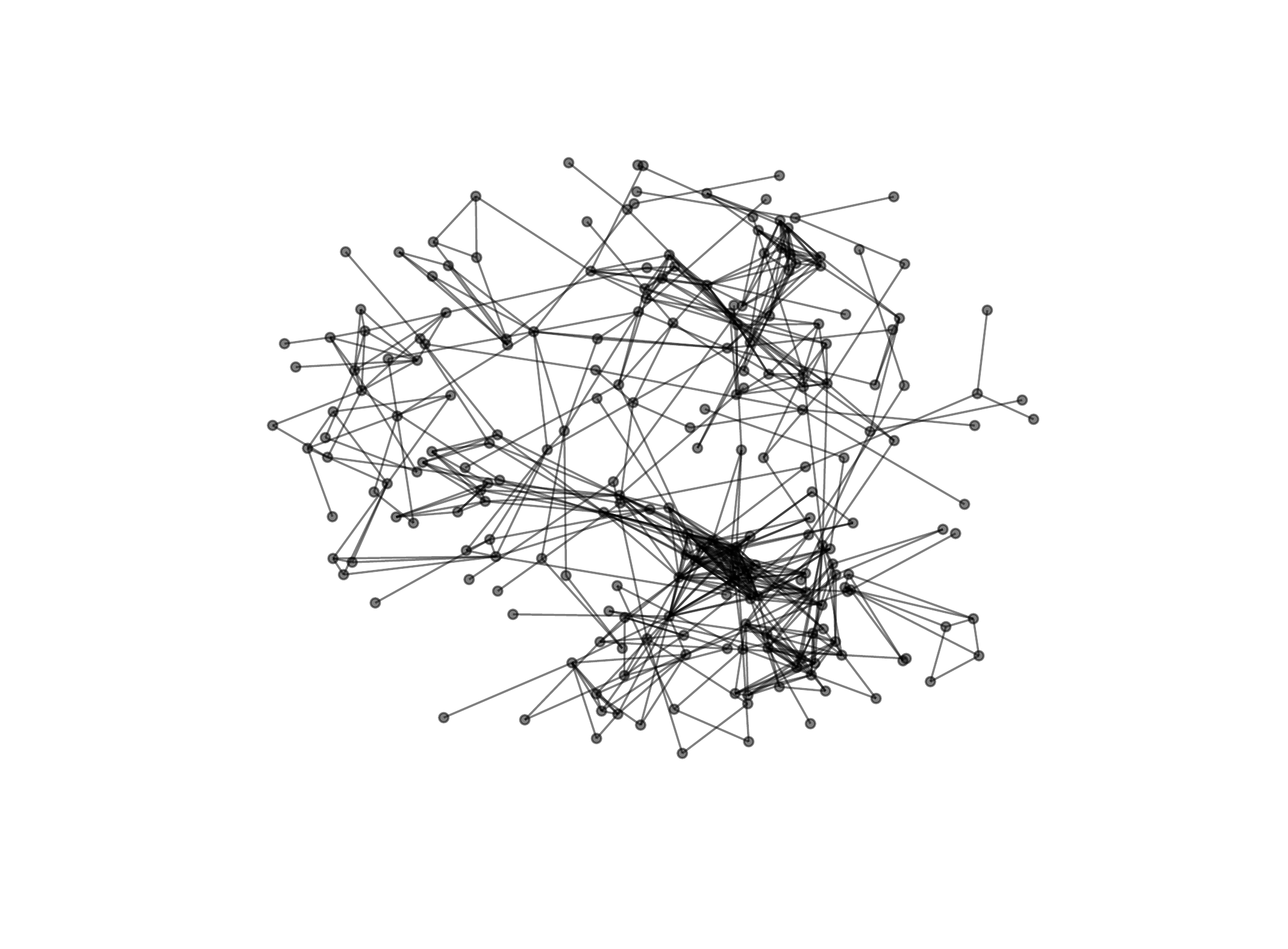}
\caption{A randomly generated distance-hereditary graph with $224$ vertices.}
\label{implementation-dhgraphlarge}
\end{center}
\end{figure}

To generate graphs of large size, we employ the technique of \emph{singular sampling}, described by Duchon~\etal\cite{duchon}, in which we sample from each class at the singularity of its corresponding generating function. Figures~\ref{implementation-dhgraph}, \ref{implementation-3lpgraph}, and \ref{implementation-dhgraphlarge} depict graphs that were drawn from a singular sampler.


\subsection{Details of interest}
\label{implementation-details}

\subsubsection{Oracles}
\label{implementation-details-oracles}

For each sum of two combinatorial classes that appears in the specifications of $\DHcp$ and $\TLPcp$, the corresponding Boltzmann sampler has a Bernoulli switch (\textit{cf.} Table~\ref{cyclepointing-polyaboltzmanntransfertable}) whose parameter depends on the values of the generating functions of the two classes that are being added, evaluated at the parameter $z$ that was passed into the sampler. To compute these values, we require an \emph{oracle} for each generating function of interest, which is a function that returns the value of the generating function at a particular input value.

Depending on the particular problem, oracles may be obtained by computing a closed form expression for the generating function, or by employing an iterative method such as Newton iteration. However since $\tt{combstruct}$ provides us access to the coefficients of each of the generating functions in question, we use these instead to build the necessary oracles. Specifically, for an OGF $\A(z)$, we build an approximate oracle by precomputing the exact values of the coefficients $\A_0, \A_1, \ldots, \A_{N}$ (we take $N = 2000$ in the code) and then, on input $z$, returning $$\sum_{i = 0}^{N}\A_iz^i.$$

\subsubsection{Radius of convergence}
\label{implementation-details-roc}

As mentioned above, we query our samplers at the singularity, or radius of convergence, of the corresponding generating function in order to generate graphs of large size. Finding the radius of convergence in Boltzmann sampling has traditionally been done using binary search \cite{pisaso12, darrasse} on a perfect oracle (or one that can be made infinitely precise) for the generating function; however, given that we have only finitely many coefficients of the generating function, we approach this question from a slightly different angle by directly using these coefficients.\footnote{At the time of writing, the NewtonGF package of Pivoteau~\etal\cite{newtongf} was not yet able to support cycle-pointed classes and specifications.}

Given the coefficients $\A_0, \ldots, \A_N$ of $\A(z)$ (for some fixed $N > 0$), a few methods of estimating the radius of convergence $\rhoA$ of $\A(z)$ immediately present themselves. We recall by the ratio and root convergence tests that the quantities $$\frac{\A_{n-1}}{\A_{n}}\qquad\text{and}\qquad\frac{1}{\sqrt[n]{\A_n}}$$ both converge to $\rhoA$ as $n\rightarrow\infty$ (assuming the limits exist), so we may estimate $\rhoA$ with the values $$\frac{\A_{N-1}}{\A_{N}}\qquad\text{and}\qquad\frac{1}{\sqrt[N]{\A_N}}.$$ However, we can obtain a more accurate estimate by taking into account a first-order approximation of the error between $\A_{N-1}/\A_N$ and $\rhoA$ as a function of $N$. Specifically, we know from Georgescu~\cite{georgescu}~that $$\frac{\A_n}{\A_{n-1}}\sim\frac{1}{\rhoA}\left(1 - \frac{C}{n}\right)$$ for some constant $C$ as $n\rightarrow\infty$, so the plot of $$\frac{\A_n}{\A_{n-1}}\qquad\text{vs.}\qquad\frac{1}{n}$$ is, in the limit $n\rightarrow\infty$, a straight line with slope $1/\rhoA$. We thus estimate $\rhoA$ by making such a plot (known as a Domb-Sykes plot) over the range $N/2\leq n\leq N$, using a linear interpolation to estimate its slope, and computing the reciprocal of this value.

To briefly compare the accuracy of these methods on a generating function whose radius of convergence is known, consider $$\A(z) = \frac{z}{(1-z)^2} = \sum_{n = 0}^{\infty}nz^n,$$ for which $\rhoA = 1$. Table~\ref{implementation-estimatestable} shows the estimates obtained by using the first $2000$ coefficients.
\begin{table}[!htb]
\begin{center}
\begin{tabular} {cccN}
\toprule
Estimation method & Value & Absolute error & \\[10pt]
\midrule
$\frac{\A_{1999}}{\A_{2000}}$ & $0.9995$ & $5\cdot10^{-4}$ & \\[20pt]
$\frac{1}{\sqrt[2000]{\A_{2000}}}$ & $0.996207$ & $3.793\cdot10^{-3}$ & \\[20pt]
Domb-Sykes & $0.999999999966895$ & $3.310\cdot10^{-11}$ & \\[10pt]
\bottomrule
\end{tabular}
\caption{Estimates of $\protect\rhoA$ for $\A(z) = \sum_{n=0}^{\infty}nz^n$.}
\label{implementation-estimatestable}
\end{center}
\end{table}

\noindent Using the Domb-Sykes method, we obtain the estimates $$\rhosub{\DH} = 0.137935\qquad\text{and}\qquad\rhosub{\TLP} = 0.259845.$$

\subsubsection{Sampling of random variables}
\label{implementation-details-rejectionsampling}

The one place where the implemented $\DH$ and $\TLP$ samplers employ some form of rejection is in order to sample certain pairs of complicated and correlated random variables.

For example, one of the subprocedures that is used is a Boltzmann sampler $\Gamma(\Set_{\geq2}(\A))(z)$, where $\A$ is an arbitrary class. According to the P\'{o}lya-Boltzmann sampler for $\Set_{\geq2}$ (which is a modification of the one for $\Set$ given in Figure~\ref{cyclepointing-setpolyaboltzmann}), this procedure requires sampling $$J\leftarrow\textsf{MAX\TextUnderscore{}INDEX}_{\geq1}(\A, z)\qquad\text{and}\qquad k_J\leftarrow\text{Pois}_{\geq1}\left(\frac{\A(z^J)}{J}\right)$$ such that either $J > 1$ or $k_J > 1$ (or both). We currently do so by sampling $J$ unconditionally using inversion sampling, then sampling $k_J$ given the value of $J$ (again using inversion sampling), and then checking if at least one of the values is greater than $1$ and repeating if not. A similar rejection is used in the Boltzmann samplers for $\Setcp_{\geq2}\sub\A$ and $\Setscp_{\geq3}\sub\A$.

We propose that these instances of rejection can be eliminated by sampling directly from the joint distributions on the pairs of variables. Using the same example, we wish to sample from the distribution $$\P[J = a, k_J = b\vb J > 1 \text{ or } k_J > 1].$$ The marginal of this distribution on $J$ is 
 \begin{displaymath}
   \P[J = a\vb J > 1 \text{ or } k_J > 1] = \left\{
     \begin{array}{lr}
       \frac{\P[J = 1] - \P[J = 1, k_J = 1]}{1 - \P[J = 1, k_J = 1]} & a = 1\\\\
       \frac{\P[J = a]}{1 - \P[J = 1, k_J = 1]} &  a > 1
     \end{array}
   \right.,
\end{displaymath}
and then the conditional on $k_J$ is 
 \begin{displaymath}
   \P[k_J = b\vb J = a \text{ and } (J > 1 \text{ or } k_J > 1)] = \left\{
     \begin{array}{lr}
       \P[k_1 = b\vb k_1 > 1] & a = 1\\\\
       \P[k_a = b] &  a > 1
     \end{array}
   \right..
\end{displaymath}
Since all terms on the right hand side of the above equalities except $\P[J = 1, k_J = 1]$ are already known from the respective individual inversion samplers on $J$ and $k_J$, and $\P[J = 1, k_J = 1]$ can be easily computed, we may sample from the joint distribution by first sampling $a$ from the marginal on $J$ using inversion, and then sampling $b$ from the conditional on $k_J$ using inversion once again. This will correctly sample the pair $(J, k_J)$ without using rejection.

\subsection{Empirical analysis}
\label{implementation-empirical}

\subsubsection{Chi-squared tests}
\label{implementation-empirical-chi2}
In order to analyze the accuracy of the samplers, we perform a Pearson's chi-squared test on the distribution over the possible sizes of the object generated by the sampler. Recall that for a class $\A$, the theoretical distribution of the size of an object generated by a Boltzmann sampler $\Gamma\A$ is $$\P_z[S = n] = \frac{\A_nz^n}{\A(z)},$$ so we can compute the theoretical size distributions for our samplers $\Gamma\DHcp$ and $\Gamma\TLPcp$ (recalling that, while the samplers are unbiased for $\DH$ and $\TLP$, they are only in fact Boltzmann samplers for $\DHcp$ and $\TLPcp$). 

To generate empirical size distributions, we sample $N = 1000$ objects from each class with parameter $z = 0.1$, and maintain bucket counts $O_i$ for $1\leq i\leq n-1$ ($n = 30$) of the number of trees of size $i$ that have been drawn, and a count $O_n$ of the number of trees of size at least $n$ that have been drawn. Then, the $\chi^2$ statistic is $$\chi^2 = \sum_{i = 1}^n\frac{\left(O_i - Np_i\right)^2}{Np_i}$$ where $p_i$ is the theoretical probability of being in bucket $i$~\cite{chi2calc}. 

Running this procedure for each class, we obtain values of $$\chi^2_{\DHcp} = 15.49\qquad\text{and}\qquad\chi^2_{\TLPcp} = 8.146.$$ Since the test has $n - 1$ degrees of freedom, the cutoff statistic value for $p = 0.05$ is $43.7$~\cite{nist}. As both of the computed statistics are (well) below this value, in both cases we fail to reject the null hypothesis that the sampler produces the correct size distribution.

\begin{figure}[!htb]
\begin{minipage}[b]{0.475\textwidth}
\begin{center}
\includegraphics[width=0.97\linewidth]{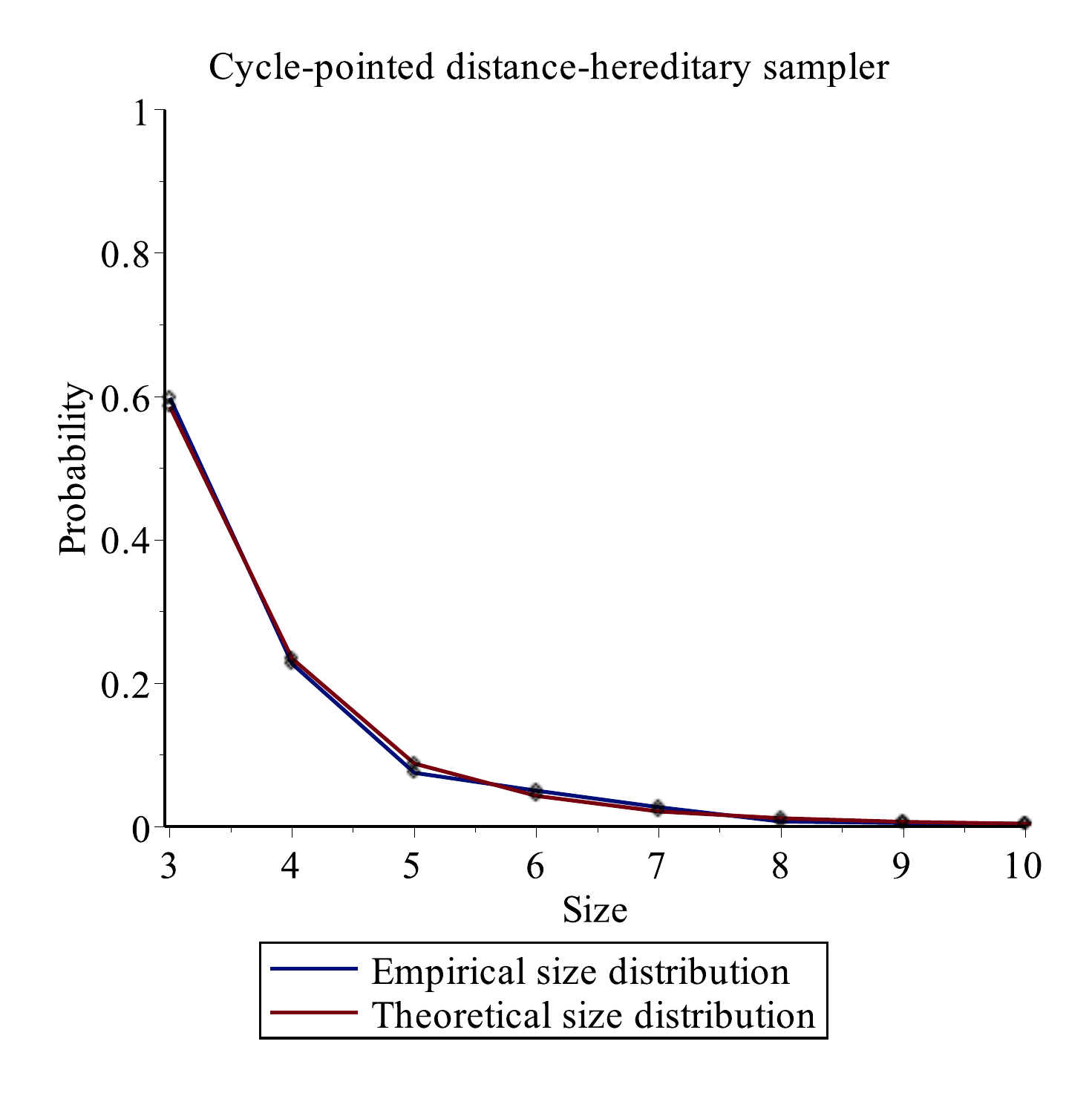}
\caption{Theoretical size distribution, \& empirical distribution from $1000$ samples of $\Gamma\DHcp(0.1)$.}
\label{implementation-dhsizedist}
\end{center}
\end{minipage}
\hfill
\begin{minipage}[b]{0.475\textwidth}
\begin{center}
\includegraphics[width=0.97\linewidth]{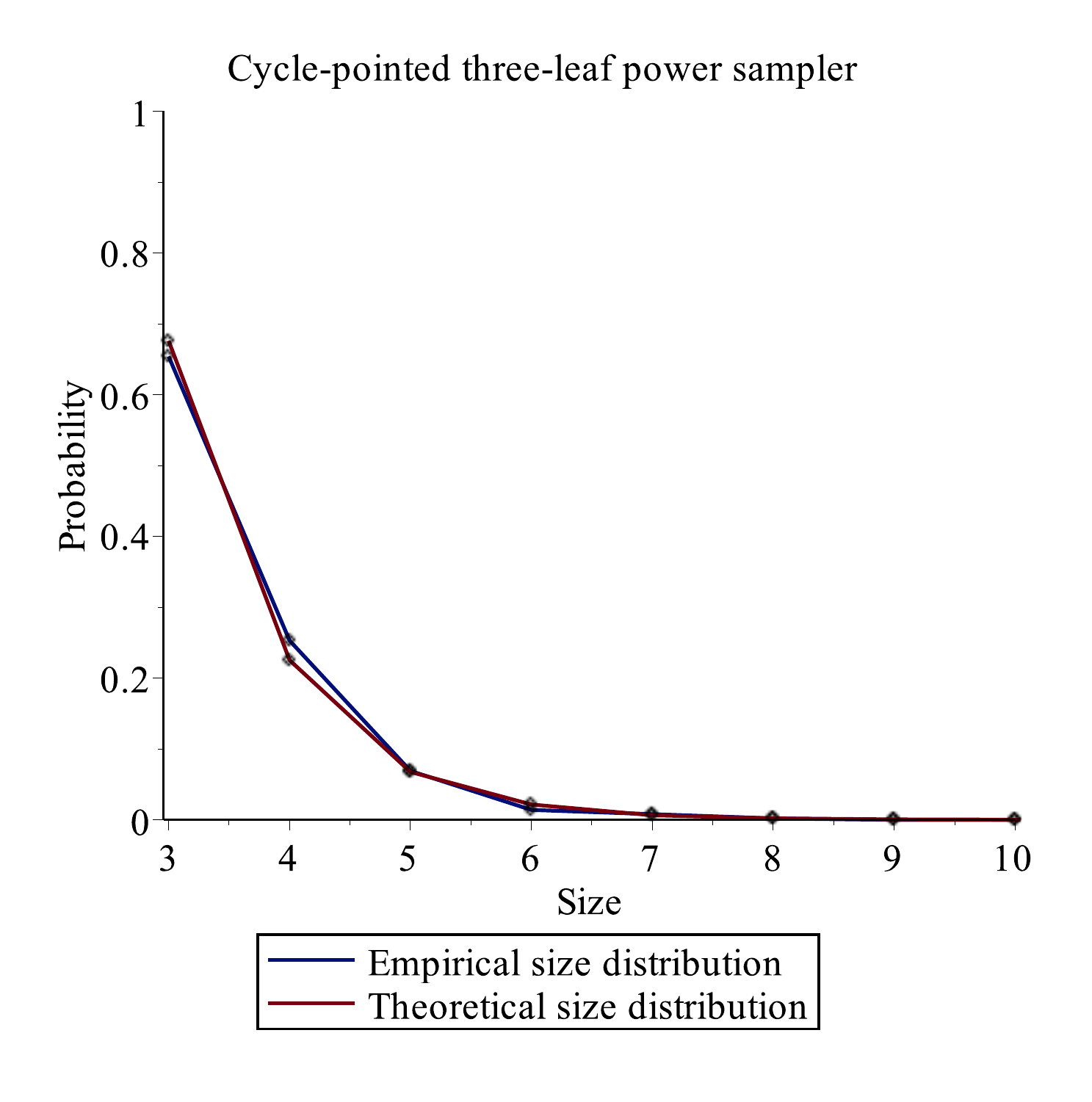}
\caption{Theoretical size distribution, \& empirical distribution from $1000$ samples of $\Gamma\TLPcp(0.1)$.}
\label{implementation-3lpsizedist}
\end{center}
\end{minipage}
\end{figure}

Figures~\ref{implementation-dhsizedist} and \ref{implementation-3lpsizedist} provide a graphical depiction of the agreement between the theoretical and empirical size distributions.

\subsubsection{Timing studies}
\label{implementation-empirical-timing}
We recall that standard Boltzmann samplers run in linear time in the size of the object that is returned. In order to study the running time of our cycle-pointed samplers, we repeatedly sample from them at a fixed parameter value, and for each sample we measure the size of the returned tree (\textit{i.e.} the number of leaves it has) and the time it took to generate. 

\begin{figure}[!htb]
\begin{minipage}[b]{0.475\textwidth}
\begin{center}
\includegraphics[width=\linewidth]{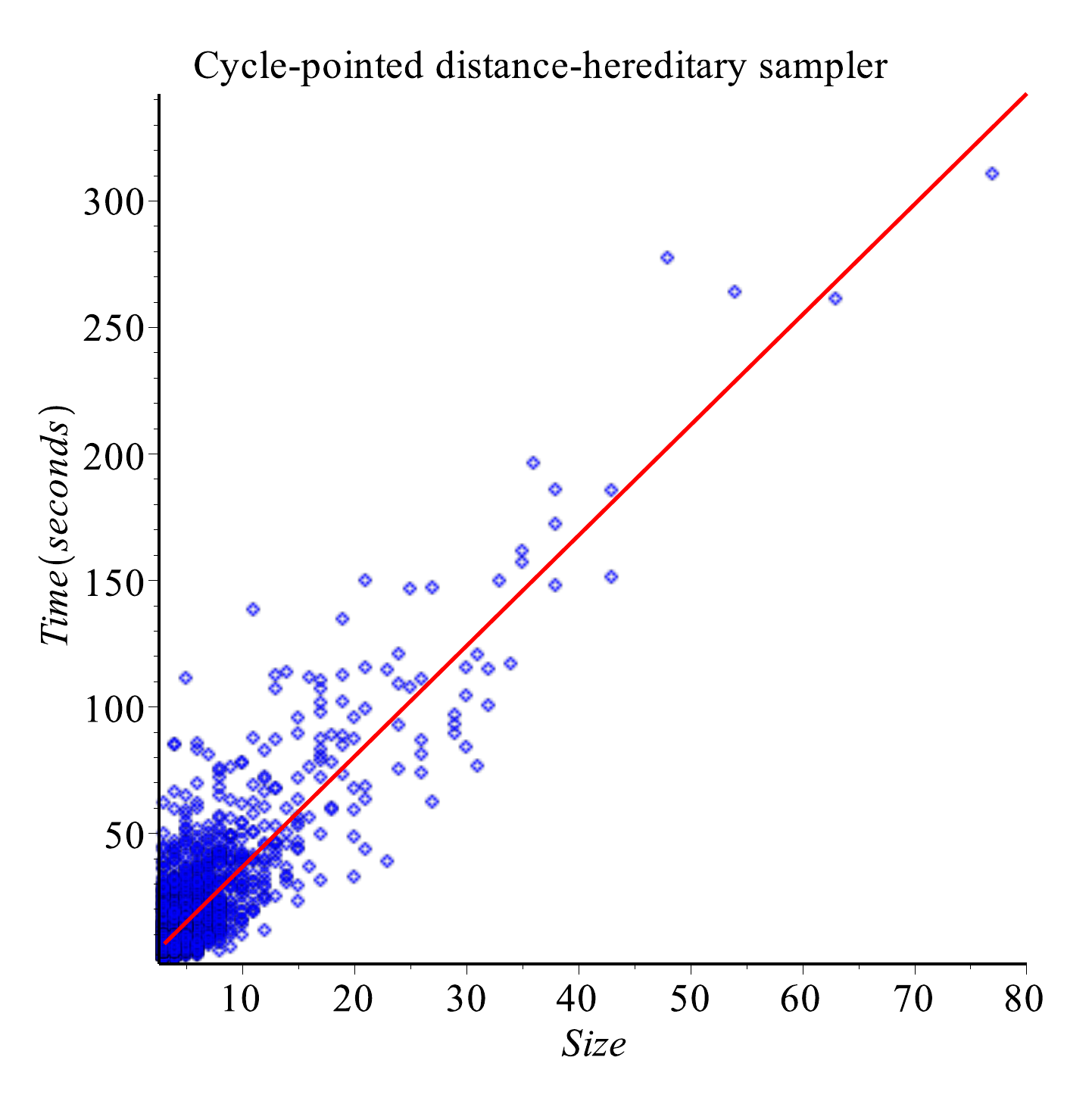}
\caption{Scatter plot of time vs. size for $2000$ samples of $\Gamma\DHcp(0.13)$, with line of best fit.}
\label{implementation-dhsizetimescatter}
\end{center}
\end{minipage}
\hfill
\begin{minipage}[b]{0.475\textwidth}
\begin{center}
\includegraphics[width=\linewidth]{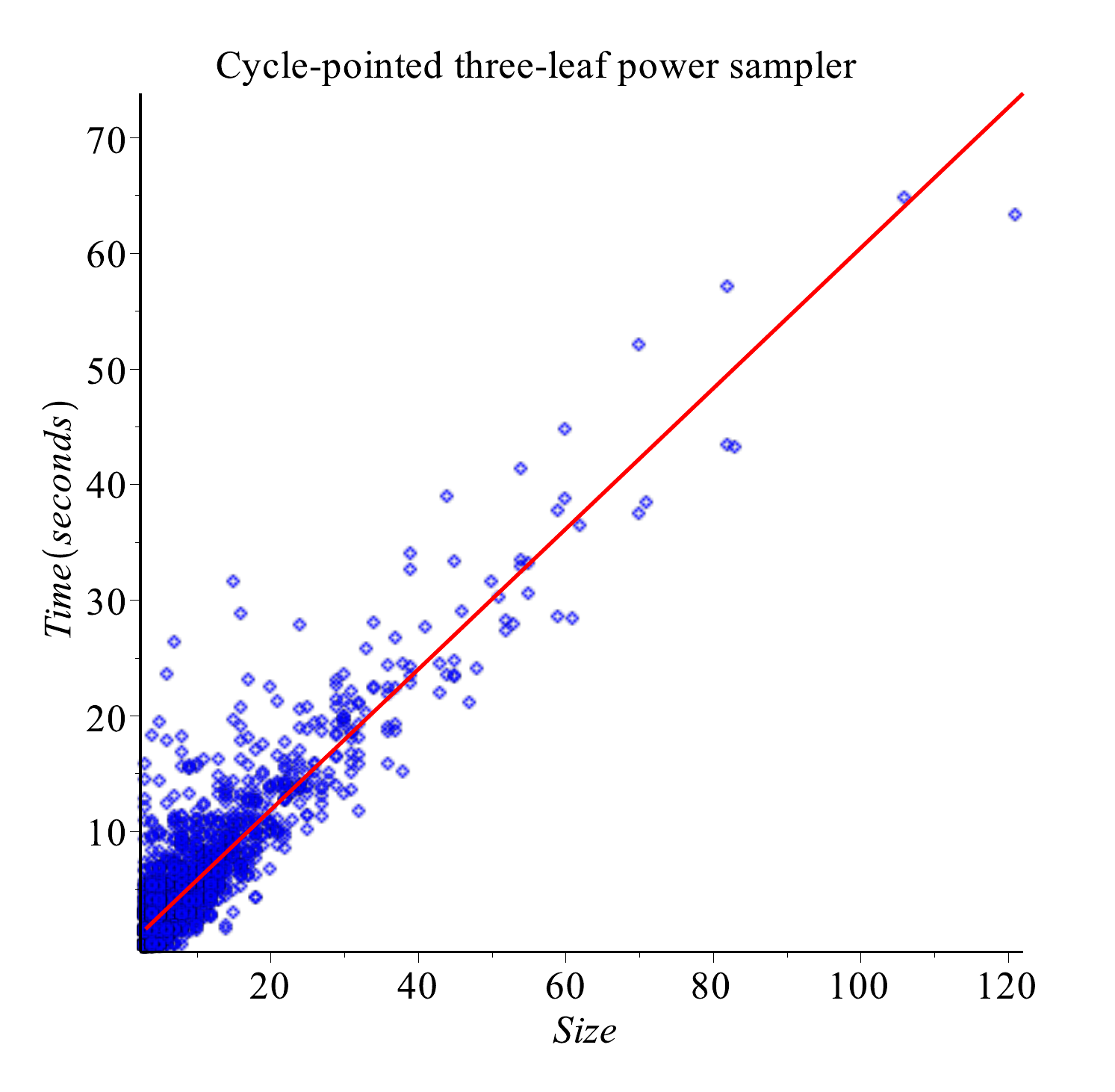}
\caption{Scatter plot of time vs. size for $2000$ samples of $\Gamma\TLPcp(0.25)$, with line of best fit.}
\label{implementation-3lpsizetimescatter}
\end{center}
\end{minipage}
\end{figure}

Figures~\ref{implementation-dhsizetimescatter} and \ref{implementation-3lpsizetimescatter} show scatter plots of time vs. size for 2000 graphs drawn from each Boltzmann sampler, with lines of best fit included. Both graphs show a distinct linear relationship between the two variables.
\clearpage

\section{Conclusion}

In this work, we study the problem of enumerating and sampling from combinatorial classes of unlabeled and unrooted graphs. We consider two techniques for addressing this problem: the dissymmetry theorem, which allows for the analysis of an unrooted class of trees by first analyzing three rooted counterparts; and cycle pointing, which marks certain cycles of a graph in such a way that there are exactly $n$ distinct pointed graphs for each unpointed graph of size $n$, thereby allowing for a straightforward translation of the analysis of the pointed class into an analysis of the unpointed class.

While at first glance the dissymmetry theorem provides only the enumeration of the unrooted class, we have shown how to build a Boltzmann sampler for an arbitrary combinatorial class specified by the dissymmetry theorem, assuming that there exist samplers for the corresponding vertex-rooted and undirected edge-rooted classes. Secondly, we have provided an exposition of cycle pointing that focuses on the enumeration and unbiased sampling of the underlying unpointed class, in the hope of elucidating this technique for future practitioners. Finally, we have applied the technique of cycle pointing to build the first unbiased samplers for the classes of distance-hereditary graphs and three-leaf power graphs

\begin{figure}[!htb]
\begin{minipage}[b]{0.475\textwidth}
\begin{center}
\includegraphics[width=0.95\linewidth]{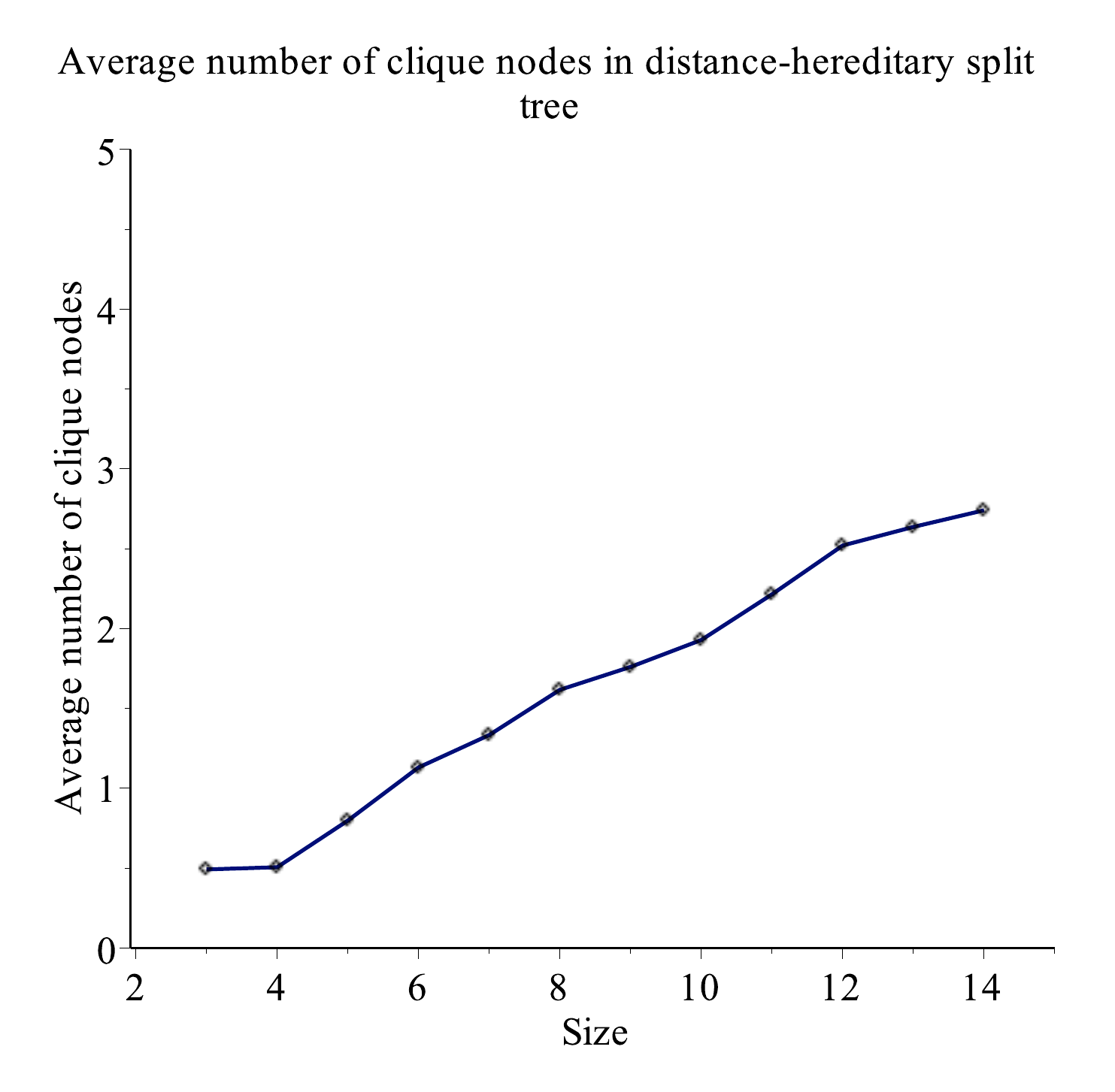}
\caption{Average number of clique nodes in a distance-hereditary split tree of size $n$ vs. $n$, from $2000$ samples of $\Gamma\cp{\DH}(0.137)$.}
\label{presentation-dhaveragecliques}
\end{center}
\end{minipage}
\hfill
\begin{minipage}[b]{0.475\textwidth}
\begin{center}
\includegraphics[width=0.95\linewidth]{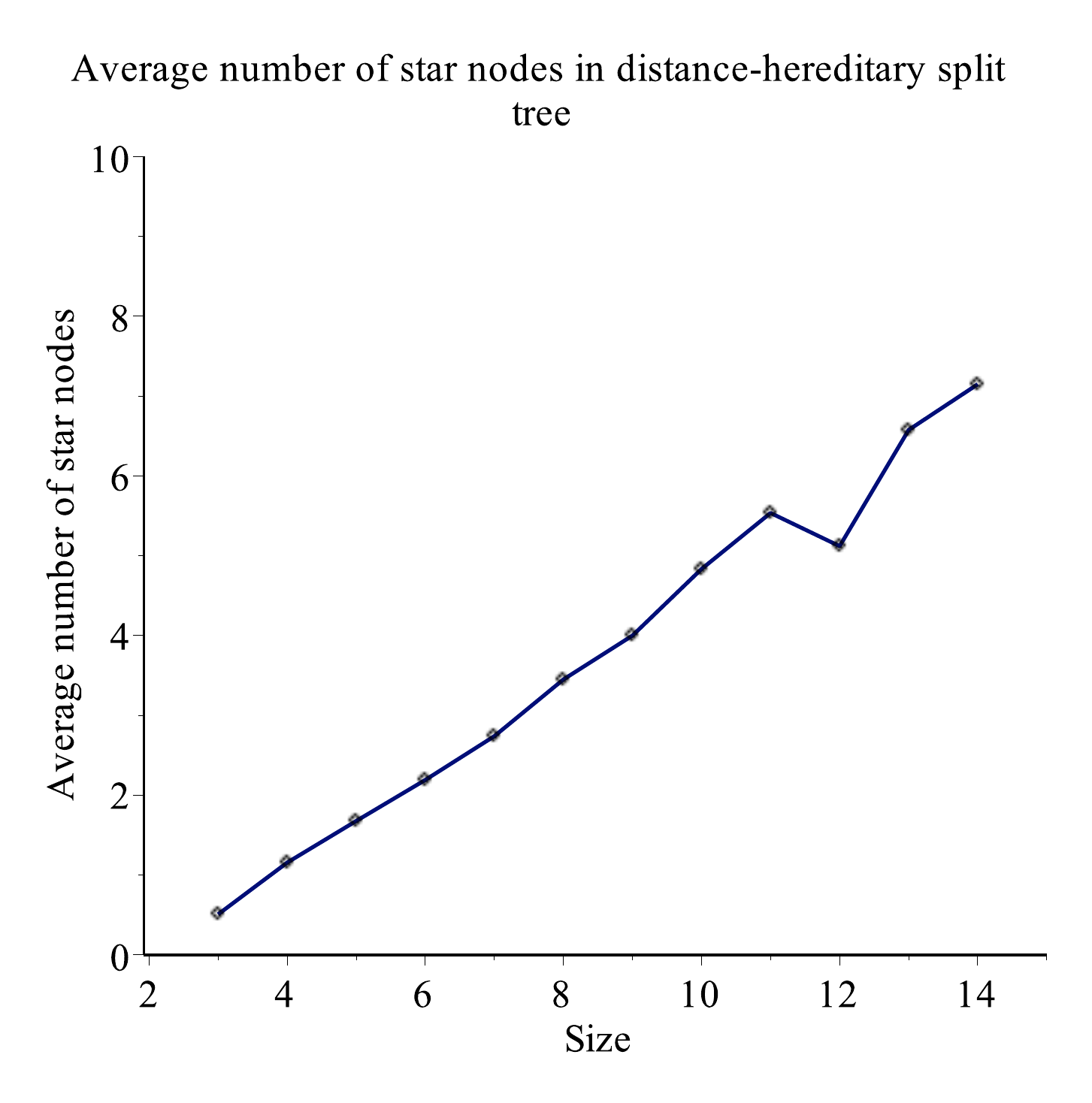}
\caption{Average number of star nodes in a distance-hereditary split tree of size $n$ vs. $n$, from $2000$ samples of $\Gamma\cp{\DH}(0.137)$.}
\label{presentation-dhaveragestars}
\end{center}
\end{minipage}
\end{figure}

Much further work remains. As a small point, we will make the changes to the implementation described in Section~\ref{implementation-details-rejectionsampling} so that the cycle-pointed samplers no longer use any form of rejection. Also, our distance-hereditary and three-leaf power samplers provide a fertile starting point for analyzing parameters of these graphs, whether parameters of the split trees or of the graphs themselves. For example, in Figures~\ref{presentation-dhaveragecliques} and \ref{presentation-dhaveragestars} we draw from $\Gamma\DHcp$ to estimate the average number of clique nodes and star nodes in a random distance-hereditary split tree of size $n$ as a function of $n$. We see that both of these parameters appear to grow linearly with $n$, and we conjecture that the number of clique nodes grows as approximately $\sim0.221n$ and the number of star nodes grows as approximately $\sim0.593n$. One interesting suggestion is that an understanding of the distribution of parameters in random graphs from these classes can be applied, for example, to see if the phylogenetic trees studied by Nishimura~\etal\cite{nishimura} appear to behave randomly. Finally, we propose that the techniques used in this work can be applied to many other classes of unlabeled and unrooted graphs which have not yet been analyzed in this manner.

\clearpage

\appendix
\appendixpage

\section{Commonly used notation}
\label{appendices-notation}

\begin{table}[!htb]
\begin{center}
\begin{tabular} {clN}
\toprule
Notation & Meaning & \\[10pt]
\midrule
$\Av$ & Vertex-rooted class of $\A$ & \\[10pt]
$\Au$ & Undirected edge-rooted class of $\A$ & \\[10pt]
$\Ad$ & Directed edge-rooted class of $\A$ & \\[10pt]
$\Sym(G)$ & Symmetries of $G$ & \\[10pt]
$\Rcsym((G, c))$ & Rooted $c$-symmetries of $(G, c)$ & \\[10pt]
$\Acp$ & Cycle-pointed class of $\A$ & \\[10pt]
$\Ascp$ & Symmetric cycle-pointed class of $\A$ & \\[10pt]
$\A\sub\B$ & Cycle-pointed substitution & \\[10pt]
$Z_{\A}$ & Cycle index sum of $\A$ & \\[10pt]
$\Gamma\A$ & Boltzmann sampler for $\A$ & \\[10pt]
$\Gamma\Z_{\A}$ & P\'{o}lya-Boltzmann sampler for $\A$ & \\[10pt]
\bottomrule
\end{tabular}
\caption{Commonly used notation in this work.}
\label{appendices-notation-notationtable}
\end{center}
\end{table}

\clearpage

\section{Unbiased sampler for the class of unlabeled, unrooted, non-plane 2-3 trees}
\label{appendices-twothreetrees}

We recall from Section~\ref{cyclepointing-decomposition-example} the decomposition of the class $\Tcp$ of unlabeled, unrooted, non-plane~trees: $$\Tcp = \Zcp\times\Set_{1, 3, 4}(\S) + \Z\times(\Setscp_{3, 4}\sub\S) + \Setscp_2\sub\S$$$$\S=\Z+\Z\times\Set_{2, 3}(\S).$$ To build Boltzmann samplers for these classes, we require Boltzmann samplers for $\Set_k(\A)$, \mbox{$\Setcp_k\sub\A$}, and $\Setscp_k\sub\A$ (for arbitrary $k$ and $\A$), which are shown in Figures~\ref{appendices-twothreetrees-setkboltzmann}, \ref{appendices-twothreetrees-setcpkboltzmann}, and \ref{appendices-twothreetrees-setscpkboltzmann}.\footnote{\textsf{PARTITION}$(\A, k, z)$ is a random generator over the tuples $(n_1, \ldots, n_k)$ of non-negative integers such that $\sum_{i = 1}^kin_i =~k$, for the distribution $$\P[(n_1, \ldots, n_k)] = \frac{(\A(z))^{n_1}(\A(z^2))^{n_2}\cdots(\A(z^k))^{n_k}\cdot[s_1^{n_1}s_2^{n_2}\cdots s_k^{n_k}]Z_{\Set_k}(s_1, s_2, \ldots)}{Z_{\Set_k}(\A(z), \A(z^2), \ldots)}.$$

\textsf{MARKED\TextUnderscore{}PARTITION}$(\A, k, z)$ is a random generator over the tuples $(\l, n_1, \ldots, n_k)$ of non-negative integers such that $\l + \sum_{i = 1}^kin_i = k$ and $\l\geq1$, for the distribution $$\P[(\l, n_1, \ldots, n_k)] = \frac{z^\l\A'(z^\l)(\A(z))^{n_1}(\A(z^2))^{n_2}\cdots(\A(z^k))^{n_k}\cdot[t_\l s_1^{n_1}s_2^{n_2}\cdots s_k^{n_k}]Z_{\Setcp_k}(s_1, s_2, \ldots; t_1, t_2, \ldots)}{Z_{\Setcp_k}(\A(z), \A(z^2), \ldots; z\A'(z), z^2\A'(z^2), \ldots)}.$$

\textsf{MARKED\TextUnderscore{}SYMM\TextUnderscore{}PARTITION}$(\A, k, z)$ is a random generator over the tuples $(\l, n_1, \ldots, n_k)$ of non-negative integers such that $\l + \sum_{i = 1}^kin_i = k$ and $\l\geq2$, for the distribution $$\P[(\l, n_1, \ldots, n_k)] = \frac{z^\l\A'(z^\l)(\A(z))^{n_1}(\A(z^2))^{n_2}\cdots(\A(z^k))^{n_k}\cdot[t_\l s_1^{n_1}s_2^{n_2}\cdots s_k^{n_k}]Z_{\Setscp_k}(s_1, s_2, \ldots; t_1, t_2, \ldots)}{Z_{\Setscp_k}(\A(z), \A(z^2), \ldots; z\A'(z), z^2\A'(z^2), \ldots)}.$$}

\begin{figure}[!htb]
\begin{center}
\begin{tabular} {|p{14cm}|N}
\hline
$\Gamma(\Set_k(\A))(z)$:\newline
\hspace*{4mm}$(n_1, \ldots, n_k)\leftarrow \textsf{PARTITION}(\A, k, z)$\newline
\hspace*{4mm}$S\leftarrow\textbf{null}$\newline
\hspace*{4mm}\textbf{for} $i$ \textbf{from} $1$ \textbf{to} $k$ \textbf{do}\newline
\hspace*{7mm}\textbf{for} $j$ \textbf{from} $1$ \textbf{to} $n_i$ \textbf{do}\newline
\hspace*{10mm}$\gamma\leftarrow\Gamma\A(z^i)$\newline
\hspace*{10mm}Add $i$ copies of $\gamma$ to $S$\newline
\hspace*{7mm}\textbf{end for}\newline
\hspace*{4mm}\textbf{end for}\newline
\hspace*{4mm}\textbf{return} $S$\newline
& \\[25pt]
\hline
\end{tabular}
\caption{Boltzmann sampler for $\protect\Set_k(\A)$.}
\label{appendices-twothreetrees-setkboltzmann}
\end{center}
\end{figure}
\clearpage
\begin{figure}[!htb]
\begin{center}
\begin{tabular} {|p{14cm}|N}
\hline
$\Gamma(\Setcp_k\sub\A)(z)$:\newline
\hspace*{4mm}$(\l, n_1, \ldots, n_k)\leftarrow\textsf{MARKED\TextUnderscore{}PARTITION}(\A, k, z)$\newline
\hspace*{4mm}$S\leftarrow\textbf{null}$\newline
\hspace*{4mm}\textbf{for} $i$ \textbf{from} $1$ \textbf{to} $k$ \textbf{do}\newline
\hspace*{7mm}\textbf{for} $j$ \textbf{from} $1$ \textbf{to} $n_i$ \textbf{do}\newline
\hspace*{10mm}$\gamma\leftarrow\Gamma\A(z^i)$\newline
\hspace*{10mm}Add $i$ copies of $\gamma$ to $S$\newline
\hspace*{7mm}\textbf{end for}\newline
\hspace*{4mm}\textbf{end for}\newline
\hspace*{4mm}$(\gamma, c)\leftarrow\Gamma\Acp(z^{\l})$\newline
\hspace*{4mm}Add $\l$ copies of $(\gamma, c)$ to $S$\newline
\hspace*{4mm}Let $c'$ be the cycle obtained by composing the $\l$ copies of $c$\newline
\hspace*{4mm}\textbf{return} $(S, c)$\newline
& \\[25pt]
\hline
\end{tabular}
\caption{Boltzmann sampler for $\protect\Setcp_k\sub\A$.}
\label{appendices-twothreetrees-setcpkboltzmann}
\end{center}
\end{figure}

\begin{figure}[!htb]
\begin{center}
\begin{tabular} {|p{14cm}|N}
\hline
$\Gamma(\Setscp_k\sub\A)(z)$:\newline
\hspace*{4mm}$(\l, n_1, \ldots, n_k)\leftarrow\textsf{MARKED\TextUnderscore{}SYMM\TextUnderscore{}PARTITION}(\A, k, z)$\newline
\hspace*{4mm}$S\leftarrow\textbf{null}$\newline
\hspace*{4mm}\textbf{for} $i$ \textbf{from} $1$ \textbf{to} $k$ \textbf{do}\newline
\hspace*{7mm}\textbf{for} $j$ \textbf{from} $1$ \textbf{to} $n_i$ \textbf{do}\newline
\hspace*{10mm}$\gamma\leftarrow\Gamma\A(z^i)$\newline
\hspace*{10mm}Add $i$ copies of $\gamma$ to $S$\newline
\hspace*{7mm}\textbf{end for}\newline
\hspace*{4mm}\textbf{end for}\newline
\hspace*{4mm}$(\gamma, c)\leftarrow\Gamma\Acp(z^{\l})$\newline
\hspace*{4mm}Add $\l$ copies of $(\gamma, c)$ to $S$\newline
\hspace*{4mm}Let $c'$ be the cycle obtained by composing the $\l$ copies of $c$\newline
\hspace*{4mm}\textbf{return} $(S, c)$\newline
& \\[25pt]
\hline
\end{tabular}
\caption{Boltzmann sampler for $\protect\Setscp_k\sub\A$.}
\label{appendices-twothreetrees-setscpkboltzmann}
\end{center}
\end{figure}

\clearpage

\noindent We then apply these rules to build Boltzmann samplers for $\S$, $\Scp$, and $\Tcp$, which are shown in Figures~\ref{appendices-twothreetrees-ssampler}, \ref{appendices-twothreetrees-scpsampler}, and \ref{appendices-twothreetrees-tcpsampler}. By Corollary~\ref{cyclepointing-introduction-correspondencecor}, running $\Gamma\Tcp(z)$ and forgetting the marked cycle provides an unbiased sampler for the class $\T$ of unlabeled, unrooted, non-plane 2-3 trees.

We use $\Zat$ to denote a node, $\Zcpat$ to denote a node with a marked singleton cycle, $\textsf{A} = (\textsf{B}_1, \ldots, \textsf{B}_k)$ to denote that $\textsf{B}_1, \ldots, \textsf{B}_k$ are neighbors of $\textsf{A}$, and $\textsf{e(A, B)}$ to denote that $\textsf{A}$ and $\textsf{B}$ are connected by an edge. Also for the sake of simplicity, we let $$\textsf{DRAW}(\alpha_1, \ldots, \alpha_k)$$ be a random generator that draws each integer $1\leq i\leq k$ with probability $$\frac{\alpha_i}{\alpha_1 + \ldots + \alpha_k}.$$

\begin{figure}[!htb]
\begin{center}
\begin{tabular} {|p{14cm}|N}
\hline
$\Gamma\S(z)$:\newline
\hspace*{4mm}$i\leftarrow\textsf{DRAW}\left(z, \frac{1}{2}z\S(z)^2, \frac{1}{2}z\S(z^2), \frac{1}{6}z\S(z)^3, \frac{1}{2}z\S(z)\S(z^2), \frac{1}{3}z\S(z^3)\right)$\newline
\hspace*{4mm}\textbf{switch} $i$\newline
\hspace*{7mm}\textbf{case} $1$\newline
\hspace*{10mm}\textbf{return} $\Zat$\newline
\hspace*{7mm}\textbf{case} $2$\newline
\hspace*{10mm}\textbf{return} $\Zat\textsf{(}\Gamma\S(z), \Gamma\S(z)\textsf{)}$\newline
\hspace*{7mm}\textbf{case} $3$\newline
\hspace*{10mm}$\gamma\leftarrow\Gamma\S(z^2)$\newline
\hspace*{10mm}\textbf{return} $\Zat\textsf{(}\gamma, \gamma\textsf{)}$\newline
\hspace*{7mm}\textbf{case} $4$\newline
\hspace*{10mm}\textbf{return} $\Zat\textsf{(}\Gamma\S(z), \Gamma\S(z), \Gamma\S(z)\textsf{)}$\newline
\hspace*{7mm}\textbf{case} $5$\newline
\hspace*{10mm}$\gamma\leftarrow\Gamma\S(z^2)$\newline
\hspace*{10mm}\textbf{return} $\Zat\textsf{(}\Gamma\S(z), \gamma, \gamma\textsf{)}$\newline
\hspace*{7mm}\textbf{case} $6$\newline
\hspace*{10mm}$\gamma\leftarrow\Gamma\S(z^3)$\newline
\hspace*{10mm}\textbf{return} $\Zat\textsf{(}\gamma, \gamma, \gamma\textsf{)}$\newline
& \\[25pt]
\hline
\end{tabular}
\caption{Boltzmann sampler for $\S$.}
\label{appendices-twothreetrees-ssampler}
\end{center}
\end{figure}

\begin{figure}[!htb]
\begin{center}
\begin{tabular} {|p{14cm}|N}
\hline
$\Gamma\Scp(z)$:\newline
\hspace*{4mm}$i\leftarrow\textsf{DRAW}\left(z, \frac{1}{2}z\S(z)^2, \frac{1}{2}z\S(z^2), \frac{1}{6}z\S(z)^3, \frac{1}{2}z\S(z)\S(z^2), \frac{1}{3}z\S(z^3), z^2\S'(z)\S(z),\right.$\newline\newline
\hspace*{25mm}$\left.z^3\S'(z^2), \frac{1}{2}z^2\S'(z)\S(z)^2, \frac{1}{2}z^2\S'(z)\S(z^2), z^3\S'(z^2)\S(z), z^4\S'(z^3)\right)$\newline
\hspace*{4mm}\textbf{switch} $i$\newline
\hspace*{7mm}\textbf{case} $1$\newline
\hspace*{10mm}\textbf{return} $\Zcpat$\newline
\hspace*{7mm}\textbf{case} $2$\newline
\hspace*{10mm}\textbf{return} $\Zcpat\textsf{(}\Gamma\S(z), \Gamma\S(z)\textsf{)}$\newline
\hspace*{7mm}\textbf{case} $3$\newline
\hspace*{10mm}$\gamma\leftarrow\Gamma\S(z^2)$\newline
\hspace*{10mm}\textbf{return} $\Zcpat\textsf{(}\gamma, \gamma\textsf{)}$\newline
\hspace*{7mm}\textbf{case} $4$\newline
\hspace*{10mm}\textbf{return} $\Zcpat\textsf{(}\Gamma\S(z), \Gamma\S(z), \Gamma\S(z)\textsf{)}$\newline
\hspace*{7mm}\textbf{case} $5$\newline
\hspace*{10mm}$\gamma\leftarrow\Gamma\S(z^2)$\newline
\hspace*{10mm}\textbf{return} $\Zcpat\textsf{(}\Gamma\S(z), \gamma, \gamma\textsf{)}$\newline
\hspace*{7mm}\textbf{case} $6$\newline
\hspace*{10mm}$\gamma\leftarrow\Gamma\S(z^3)$\newline
\hspace*{10mm}\textbf{return} $\Zcpat\textsf{(}\gamma, \gamma, \gamma\textsf{)}$\newline
\hspace*{7mm}\textbf{case} $7$\newline
\hspace*{10mm}\textbf{return} $\Zat\textsf{(}\Gamma\Scp(z), \Gamma\S(z)\textsf{)}$\newline
\hspace*{7mm}\textbf{case} $8$\newline
\hspace*{10mm}$\gamma\leftarrow\Gamma\Scp(z^2)$\newline
\hspace*{10mm}\textbf{return} $\Zat\textsf{(}\gamma, \gamma\textsf{)}$, with the cycles on the two copies of $\gamma$ composed\newline
\hspace*{7mm}\textbf{case} $9$\newline
\hspace*{10mm}\textbf{return} $\Zat\textsf{(}\Gamma\Scp(z), \Gamma\S(z), \Gamma\S(z)\textsf{)}$\newline
\hspace*{7mm}\textbf{case} $10$\newline
\hspace*{10mm}$\gamma\leftarrow\Gamma\S(z^2)$\newline
\hspace*{10mm}\textbf{return} $\Zat\textsf{(}\Gamma\Scp(z), \gamma, \gamma\textsf{)}$\newline
\hspace*{7mm}\textbf{case} $11$\newline
\hspace*{10mm}$\gamma\leftarrow\Gamma\Scp(z^2)$\newline
\hspace*{10mm}\textbf{return} $\Zat\textsf{(}\gamma, \gamma, \Gamma\S(z)\textsf{)}$, with the cycles on the two copies of $\gamma$ composed\newline
\hspace*{7mm}\textbf{case} $12$\newline
\hspace*{10mm}$\gamma\leftarrow\Gamma\Scp(z^3)$\newline
\hspace*{10mm}\textbf{return} $\Zat\textsf{(}\gamma, \gamma, \gamma\textsf{)}$, with the cycles on the three copies of $\gamma$ composed\newline
& \\[25pt]
\hline
\end{tabular}
\caption{Boltzmann sampler for $\Scp$.}
\label{appendices-twothreetrees-scpsampler}
\end{center}
\end{figure}

\clearpage
\newgeometry{top=1.5cm}
\thispagestyle{empty}
\begin{figure}[!htb]
\begin{center}
\begin{tabular} {|p{14cm}|N}
\hline
$\Gamma\Tcp(z)$:\newline
\hspace*{4mm}$i\leftarrow\textsf{DRAW}\left(z\S(z), \frac{1}{6}z\S(z)^3, \frac{1}{2}z\S(z)\S(z^2), \frac{1}{3}z\S(z^3), \frac{1}{24}z\S(z)^4, \frac{1}{4}z\S(z)^2\S(z^2), \frac{1}{8}z\S(z^2)^2,\right.$\newline\newline
\hspace*{25mm}$\left.\frac{1}{3}z\S(z)\S(z^3), \frac{1}{4}z\S(z^4), z^3\S'(z^2)\S(z), z^4\S'(z^3), \frac{1}{2}z^3\S'(z^2)\S(z)^2,\right.$\newline\newline
\hspace*{25mm}$\left.\frac{1}{2}z^3\S'(z^2)\S(z^2), z^4\S'(z^3)\S(z), z^5\S'(z^4), z^2\S'(z^2)\right)$\newline
\hspace*{4mm}\textbf{switch} $i$\newline
\hspace*{7mm}\textbf{case} $1$\newline
\hspace*{10mm}\textbf{return} $\Zcpat\textsf{(}\Gamma\S(z)\textsf{)}$\newline
\hspace*{7mm}\textbf{case} $2$\newline
\hspace*{10mm}\textbf{return} $\Zcpat\textsf{(}\Gamma\S(z), \Gamma\S(z), \Gamma\S(z)\textsf{)}$\newline
\hspace*{7mm}\textbf{case} $3$\newline
\hspace*{10mm}$\gamma\leftarrow\Gamma\S(z^2)$\newline
\hspace*{10mm}\textbf{return} $\Zcpat\textsf{(}\Gamma\S(z), \gamma, \gamma\textsf{)}$\newline
\hspace*{7mm}\textbf{case} $4$\newline
\hspace*{10mm}$\gamma\leftarrow\Gamma\S(z^3)$\newline
\hspace*{10mm}\textbf{return} $\Zcpat\textsf{(}\gamma, \gamma, \gamma\textsf{)}$\newline
\hspace*{7mm}\textbf{case} $5$\newline
\hspace*{10mm}\textbf{return} $\Zcpat\textsf{(}\Gamma\S(z), \Gamma\S(z), \Gamma\S(z), \Gamma\S(z)\textsf{)}$\newline
\hspace*{7mm}\textbf{case} $6$\newline
\hspace*{10mm}$\gamma\leftarrow\Gamma\S(z^2)$\newline
\hspace*{10mm}\textbf{return} $\Zcpat\textsf{(}\Gamma\S(z), \Gamma\S(z), \gamma, \gamma\textsf{)}$\newline
\hspace*{7mm}\textbf{case} $7$\newline
\hspace*{10mm}$\gamma\leftarrow\Gamma\S(z^2)$\newline
\hspace*{10mm}$\zeta\leftarrow\Gamma\S(z^2)$\newline
\hspace*{10mm}\textbf{return} $\Zcpat\textsf{(}\gamma, \gamma, \zeta, \zeta\textsf{)}$\newline
\hspace*{7mm}\textbf{case} $8$\newline
\hspace*{10mm}$\gamma\leftarrow\Gamma\S(z^3)$\newline
\hspace*{10mm}\textbf{return} $\Zcpat\textsf{(}\Gamma\S(z), \gamma, \gamma, \gamma\textsf{)}$\newline
\hspace*{7mm}\textbf{case} $9$\newline
\hspace*{10mm}$\gamma\leftarrow\Gamma\S(z^4)$\newline
\hspace*{10mm}\textbf{return} $\Zcpat\textsf{(}\gamma, \gamma, \gamma, \gamma\textsf{)}$\newline
\hspace*{7mm}\textbf{case} $10$\newline
\hspace*{10mm}$\gamma\leftarrow\Gamma\Scp(z^2)$\newline
\hspace*{10mm}\textbf{return} $\Zat\textsf{(}\gamma, \gamma, \Gamma\S(z)\textsf{)}$, with the cycles on the two copies of $\gamma$ composed\newline
\hspace*{7mm}\textbf{case} $11$\newline
\hspace*{10mm}$\gamma\leftarrow\Gamma\Scp(z^3)$\newline
\hspace*{10mm}\textbf{return} $\Zat\textsf{(}\gamma, \gamma, \gamma\textsf{)}$, with the cycles on the three copies of $\gamma$ composed\newline
\hspace*{7mm}\textbf{case} $12$\newline
\hspace*{10mm}$\gamma\leftarrow\Gamma\Scp(z^2)$\newline
\hspace*{10mm}\textbf{return} $\Zat\textsf{(}\gamma, \gamma, \Gamma\S(z), \Gamma\S(z)\textsf{)}$, with the cycles on the two copies of $\gamma$ composed\newline
\hspace*{7mm}\textbf{case} $13$\newline
\hspace*{10mm}$\gamma\leftarrow\Gamma\Scp(z^2)$\newline
\hspace*{10mm}$\zeta\leftarrow\Gamma\S(z^2)$\newline
\hspace*{10mm}\textbf{return} $\Zat\textsf{(}\gamma, \gamma, \zeta, \zeta\textsf{)}$, with the cycles on the two copies of $\gamma$ composed\newline
\hspace*{7mm}\textbf{case} $14$\newline
\hspace*{10mm}$\gamma\leftarrow\Gamma\Scp(z^3)$\newline
\hspace*{10mm}\textbf{return} $\Zat\textsf{(}\gamma, \gamma, \gamma, \Gamma\S(z)\textsf{)}$, with the cycles on the three copies of $\gamma$ composed\newline
\hspace*{7mm}\textbf{case} $15$\newline
\hspace*{10mm}$\gamma\leftarrow\Gamma\Scp(z^4)$\newline
\hspace*{10mm}\textbf{return} $\Zat\textsf{(}\gamma, \gamma, \gamma, \gamma\textsf{)}$, with the cycles on the four copies of $\gamma$ composed\newline
\hspace*{7mm}\textbf{case} $16$\newline
\hspace*{10mm}$\gamma\leftarrow\Gamma\Scp(z^2)$\newline
\hspace*{10mm}\textbf{return} $\textsf{e}\textsf{(}\gamma, \gamma\textsf{)}$, with the cycles on the two copies of $\gamma$ composed\newline
& \\[25pt]
\hline
\end{tabular}
\caption{Boltzmann sampler for $\Tcp$.}
\label{appendices-twothreetrees-tcpsampler}
\end{center}
\end{figure}

\restoregeometry

\bibliographystyle{plain}
\cleardoublepage
\bibliography{bib}

\end{document}